\documentclass[fleqn,preprint]{elsarticle}

\usepackage{booktabs}
\usepackage{longtable}
\usepackage[flushleft]{threeparttable}
\usepackage[flushleft]{threeparttablex}
\usepackage{multirow}
\usepackage{multicol}
\usepackage{amsmath,amssymb,amsfonts}
\usepackage{algorithmic}
\usepackage{algorithm}
\usepackage{array}
\usepackage{textcomp}
\usepackage{stfloats}
\usepackage{url}
\usepackage{verbatim}

\usepackage{eurosym}
\usepackage{acronym}
\usepackage[mathscr]{eucal}
\usepackage{xcolor}
\usepackage{subfigure}
\usepackage{enumitem}
\usepackage{accents}
\usepackage{amssymb}
\usepackage{natbib}
\usepackage{geometry}
\usepackage{graphicx}
\usepackage{txfonts}
\usepackage{hyperref}
\usepackage{float}
\usepackage{tcolorbox}
\tcbuselibrary{skins,breakable}

\biboptions{sort&compress}

\usepackage{subcaption} % For subfigure support
\usepackage{caption}    % For better control over captions
\usepackage{url}
\usepackage{breakurl}

\usepackage{makecell}   % Allows text wrapping in table cells

\usepackage{pifont} % Include in the preamble

\usepackage{rotating}

\newcommand{\ubar}[1]{\underaccent{\bar}{#1}}

\newcommand{\mR}{\mathscr{R}}

\usepackage{nomencl}
\newcommand{\nomunit}[1]{%
\renewcommand{\nomentryend}{\hspace*{\fill}#1}}
\makenomenclature

\usepackage{etoolbox}
\renewcommand\nomgroup[1]{%
  \item[\bfseries
  \ifstrequal{#1}{I}{Indexes and Sets}{%
  \ifstrequal{#1}{P}{Parameters}{%
  \ifstrequal{#1}{V}{\textcolor{black}{Variables}}{%
  \ifstrequal{#1}{W}{Binary Variables}{%
  \ifstrequal{#1}{Z}{Vectors}{}}}}}%
]}

\graphicspath{{figs/}} 

\acrodef{dam}[DAM]{Day Ahead Market}
\acrodef{idm}[IDM]{Intra-Day Market}
\acrodef{srm}[SRM]{Secondary Reserve Market}
\acrodef{asm}[ASM]{Ancillary Service Market}
\acrodef{bam}[BAM]{Balancing Market}
\acrodef{sr}[SR]{Secondary Reserve}
\acrodef{pr}[PR]{Primary Reserve}
\acrodef{rto}[RTO]{Real-Time Operation}
\acrodef{rt}[RT]{Real-Time}
\acrodef{ess}[ESS]{Energy Storage System}
\acrodef{bess}[BESS]{Battery Energy Storage System}
\acrodef{res}[RES]{Renewable Energy Source}
\acrodef{ndrs}[ND-RES]{Non-dispatchable Renewable Energy Sources}
\acrodef{csp}[CSP]{Concentrated Solar Power Plant}
\acrodef{vpp}[VPP]{Virtual Power Plant}
\acrodef{wf}[WF]{Wind Farm}
\acrodef{pv}[PV]{Photovoltaic}
\acrodef{opf}[OPF]{Optimal Power Flow}
\acrodef{pf}[PF]{Power Flow}
\acrodef{milp}[MILP]{Mixed Integer Linear Programming}
\acrodef{minlp}[MINLP]{Mixed Integer non-Linear Programming}
\acrodef{tso}[TSO]{Transmission System Operator}
\acrodef{so}[SO]{System Operator}
\acrodef{pcc}[PCC]{Point of Common Coupling}
\acrodef{ree}[REE]{Red El{\'e}ctrica de España}
\acrodef{afrr}[aFRR]{Automatic Frequency Restoration Reserve}
\acrodef{fcr}[FCR]{Frequency Containment Reserve}
\acrodef{rVPP}[RVPP]{Renewable-only Virtual Power Plant}
\acrodef{rvpp}[RVPP]{Renewable-only VPP}
\acrodef{gams}[GAMS]{General Algebraic Modeling System}
\acrodef{gdx}[.gdx]{GAMS Data eXchange}
\acrodef{picasso}[PICASSO]{Platform for the International Coordination of Automated Frequency Restoration and Stable System Operation}
\acrodef{so}[SO]{Stochastic Optimization}
\acrodef{ro}[RO]{Robust Optimization}
\acrodef{aro}[ARO]{Adaptive Robust Optimization}
\acrodef{sp}[SP]{Stochastic Programming}
\acrodef{saro}[SARO]{Stochastic Adaptive RO}
\acrodef{dro}[DRO]{Distributed RO}

\acrodef{omie}[OMIE]{Spanish Market Operator}
\acrodef{agc}[AGC]{Automatic Generation Control}

\acrodef{mo}[MO]{Market Operator}

\acrodef{ccg}[C\&CG]{Column \& Constraint Generation}
\acrodef{rtm}[RTM]{Real-time Market}
\acrodef{ngm}[NGM]{Natural Gas Market}
\acrodef{pdf}[PDF]{Probability Density Function}
\acrodef{ev}[EV]{Electric Vehicle}
\acrodef{ptc}[PTC]{Parabolic Trough Collector}
\acrodef{pb}[PB]{Power Block}
\acrodef{sf}[SF]{Solar Field}
\acrodef{ts}[TS]{Thermal Storage}
\acrodef{hpa}[HPA]{Heat Purchase Agreement}
\acrodef{sos-2}[SOS-2]{Special Ordered Set of type 2}

\acrodef{eh}[EH]{Electric Heater}
\acrodef{ed}[ED]{Electric Demand}
\acrodef{td}[TD]{Thermal Demand}

\acrodef{es}[ES]{Electrical Storage}

\acrodef{igdt}[IGDT]{Information Gap Decision Theory}
\acrodef{cvar}[CVaR]{Conditional Value-at-Risk}
\acrodef{vab}[VaB]{Value-at-Best}

\acrodef{pvt}[PVT]{Photovoltaic-Thermal}

\acrodef{caes}[CAES]{Compressed Air Energy Storage}
\acrodef{dni}[DNI]{Direct Normal Irradiance}

\journal{Journal of Energy Storage}

\ExplSyntaxOn
\NewDocumentCommand{\instringTF}{mmmm}
{\oleks_instring:nnnn { #1 } { #2 } { #3 } { #4 }}
\tl_new:N \l__oleks_instring_test_tl
\cs_new_protected:Nn \oleks_instring:nnnn
{
    \tl_set:Nn \l__oleks_instring_test_tl { #1 }
    \regex_match:nnTF { \u{l__oleks_instring_test_tl} } { #2 } { #3 } { #4 }
}
\ExplSyntaxOff
\let\mybibitem\bibitem
\renewcommand{\bibitem}[1]{%
    \instringTF{Refblack}{#1}
    {\color{black}\mybibitem{#1}}
    {\color{black}\mybibitem{#1}}%
}

\begin{document}

\begin{frontmatter}

\title{Integration of Concentrated Solar Power Plants in Renewable-Only VPP with Electrical and Thermal Demands: A Two-Stage Robust Bidding Approach}

\author{Hadi Nemati*}

%\affiliation{organization={Comillas Pontifical University ICAI School of Engineering, Institute for Research in Technology},%Department and Organization
%            state={Madrid},
%            country={Spain}} 
\author{Pedro S{\'a}nchez-Mart{\'i}n}
\author{{\'A}lvaro Ortega}
\author{Lukas Sigrist}
\author{Luis Rouco}

\address{Comillas Pontifical University ICAI School of Engineering, Institute for Research in Technology, Madrid, Spain}    

\address{*Corresponding author}
\address {E-mail address: hnemati@comillas.edu}

\begin{abstract}
This paper proposes the integration of \ac{csp} in the Renewable-only virtual power plant (RVPP) for bidding in the electricity day-ahead and secondary reserve markets, as well as trading thermal energy through a heat purchase agreement. A reformulated two-stage robust optimization approach is introduced to account for multiple uncertainties, including electricity prices, non-dispatchable renewable energy sources' electrical production, \ac{csp} thermal production, and uncertainties in electrical and thermal demand consumption. The provision of energy and reserve by the thermal storage of \ac{csp} is modeled using an adjustable approach, which allocates a share of energy for up and down reserves based on the profitability of the \acs{rvpp}. Simulations are conducted for several case studies to demonstrate the effectiveness and computational efficiency of the proposed approach under different \acs{rvpp} operator decisions against uncertain parameters and various trading strategies for electricity and thermal energy. The simulation results show that integrating \ac{csp} into \acs{rvpp} enhances \acs{rvpp} flexibility for both electrical and thermal trading. Furthermore, the results indicate that the profitability of the \acs{rvpp} increases when all trading options are considered, across different levels of conservatism adopted by the \acs{rvpp} operator in response to uncertain parameters.

\end{abstract}

\end{frontmatter}

\section*{Keywords}
Concentrated solar power plant, virtual power plant, thermal storage, electricity markets, heat purchase agreement, robust optimization
%Energy markets, renewable-only virtual power plant, reserve markets, robust optimization, stochastic sources.

%\section*{Nomenclature}
%This subsection presents the notation and nomenclature used in the remainder of the paper.

\vspace{.5cm}
\begin{tcolorbox}[breakable, colback=white, colframe=black, title=Abbreviations, width=\textwidth]
\begin{footnotesize}
\begin{tabular}{p{1.2cm} p{5.2cm} | p{1.2cm} p{5.2cm}}
\textbf{Acronym} & \textbf{Definition} & \textbf{Acronym} & \textbf{Definition} \\ \hline
aFRR  & Automatic Frequency Restoration Reserve  & PDF  & Probability Density Function  \\
ARO  & Adaptive Robust Optimization  & PTC  & Parabolic Trough Collectors  \\
ASM  & Ancillary Service Market  & PV  & Photovoltaic  \\
BAM  & Balancing Market  & PVT  & Photovoltaic-Thermal  \\
CAES  & Compressed Air Energy Storage  & RES  & Renewable Energy Source  \\
C\&CG  & Column \& Constraint Generation  & RO  & Robust Optimization  \\
CSP  & Concentrated Solar Power Plant  & RTM  & Real-Time Market  \\
CVaR  & Conditional Value-at-Risk  & RVPP  & Renewable-only Virtual Power Plant  \\
DAM  & Day-Ahead Market  & SF  & Solar Field  \\
DRO  & Distributed Robust Optimization  & SOS-2  & Special Ordered Set of Type 2  \\
ED  & Electric Demand  & SP  & Stochastic Programming  \\
EH  & Electric Heater  & SRM  & Secondary Reserve Market  \\
ES  & Electrical Storage  & TD  & Thermal Demand  \\
EV  & Electric Vehicle  & TS  & Thermal Storage  \\
HPA  & Heat Purchase Agreement  & TSO  & Transmission System Operator  \\
IDM  & Intra-Day Market  & VaB  & Value-at-Best  \\
IGDT  & Information Gap Decision Theory  & VPP  & Virtual Power Plant  \\
MILP  & Mixed Integer Linear Programming  & WF  & Wind Farm  \\
MO  & Market Operator  & ND-RES  & Non-Dispatchable Renewable Energy Sources  \\  
PB  & Power Block  &  &  \\  
\end{tabular}
\end{footnotesize}
\end{tcolorbox}

\vspace{.5cm}
\begin{tcolorbox}[breakable,colback=white, colframe=black,coltext=black,title=Nomenclature, width=\textwidth]
    
\begin{footnotesize}

\begin{comment}
    
\noindent \textbf{General Notation Concepts}
\begin{itemize}

    \item An uncertain parameter with a tilde symbol denotes the median value in the forecast distribution ($\tilde{A}$);
    \item an uncertain parameter with a tilde and hat/inverse hat symbol denotes the upper/lower bound in the forecast distribution ($\hat{\tilde{A}}$/$\tilde{\check{A}}$);
    \item the hat/inverse hat symbol on uncertain parameters signifies the positive/negative permitted deviation from the forecast's value ($\hat{A}$, $\check{A}$);
    \item parameters with an upper/lower bar represent their upper/lower bounds  ($\bar{A}$, $\ubar{A}$);
    \item upward/downward arrows indicate up/down direction of regulation in variables and parameters ($a^{\uparrow}$, $A^{\uparrow}$/$a^{\downarrow}$, $A^{\downarrow}$);
    \item positive/negative symbol on variables and parameters indicates the charging/discharging state of \acs{ts} ($a^{+}$, $A^{+}$/$a^{-}$, $A^{-}$).
\end{itemize}

%\item An uncertain parameter with a tilde/two tilde symbol denotes the median value/upper bound in the forecast distribution, representing a point where half of the samples are lower ($\tilde{A}$/$\tilde{\tilde{A}}$);

\end{comment}

\vspace{-.5cm}

\setlength{\nomitemsep}{0.05cm}

\nomenclature[I, 01]{$d \in \mathscr{D}$}{Set of \acsp{ed}/\acsp{td} \nomunit{}}
\nomenclature[I, 01]{$n \in \mathscr{N}$}{Set of segments of piecewise efficiency function of \acs{pb} of \acsp{csp} \nomunit{}}
\nomenclature[I, 01]{$r \in \mathscr{R}$}{Set of \acsp{ndrs} \nomunit{}}
\nomenclature[I, 01]{$u \in \mathscr{U}$}{Set of \acs{rvpp} units \nomunit{}}
\nomenclature[I, 01]{$t \in \mathscr{T}$}{Set of time periods \nomunit{}}
\nomenclature[I, 01]{$t \in \mathscr{T}^{DA}$}{Set of time periods in which the worst case of \acs{dam} price uncertainty occurs \nomunit{}}
\nomenclature[I, 01]{$t \in \mathscr{T}^{SR,\uparrow(\downarrow)}$}{Set of time periods in which the worst case of up (down) \acs{srm} price uncertainty occurs \nomunit{}}
\nomenclature[I, 01]{$t \in \mathscr{T}_{r}$}{Set of time periods in which the worst case of \acs{ndrs} $r$ electrical production uncertainty occurs \nomunit{}}
\nomenclature[I, 01]{$t \in \mathscr{T}_{\theta}$}{Set of time periods in which the worst case of \acs{sf} of \acs{csp} $\theta$ thermal production uncertainty occurs \nomunit{}}
\nomenclature[I, 01]{$t \in \mathscr{T}_{d}$}{Set of time periods in which the worst case of \acs{ed}/\acs{td} $d$ consumption uncertainty occurs \nomunit{}}

\nomenclature[I, 02]{$\theta \in {\Theta}$}{Set of \acsp{csp} \nomunit{}}
\nomenclature[I, 03]{$\Xi^{{DA+SR+HPA}}$}{Set of decision variables of \acs{dam}, \acs{srm}, and \acs{hpa} \nomunit{} \vspace{5pt}}
\nomenclature[I, 03]{$\Xi^{O}$/$\Xi^{C}$}{Set of decision variables of uncertainties in the objective function/constraints of the optimization problem \nomunit{} \vspace{5pt}}

\nomenclature[P, 01]{$C_{r(\theta)} \;$}{Operation and maintenance costs of \acs{ndrs} $r$ (\acs{csp} $\theta$) \nomunit{[€/MWh]}}

\nomenclature[P, 02]{$\ubar E_d/ \ubar Q_d \;$}{Minimum electrical/thermal energy consumption of \acs{ed}/\acs{td} $d$ throughout the operation horizon \nomunit{[MWh]}}
\nomenclature[P, 02]{$\bar E_{\theta}^{TS}$/$\ubar E_{\theta}^{TS}$}{Upper/lower bound of thermal energy of \acs{ts} of \acs{csp} $\theta$ \nomunit{[MWh]}}

\nomenclature[P, 03]{$\bar H_d$/$\ubar H_d$}{Upper/lower bound of thermal power consumption of \acs{td} $d$ \nomunit{[MW]}}
\nomenclature[P, 03]{$\tilde{\check{H}}_{d,t}$/$\hat H_{d,t}$}{Lower bound/positive deviation of the \acs{td} $d$ consumption forecast during period $t$ \nomunit{[MW]}}

\nomenclature[P, 04]{$K_{\theta}^{PB}$}{Start up electrical output multiplier of \acs{pb} of \acs{csp} \nomunit{[p.u.]}}
\nomenclature[P, 04]{$M \;$}{Big positive value \nomunit{[€] or [MW]}}
\nomenclature[P, 04]{$N^{OFF}_{\theta}$/$N^{ON}_{\theta}$}{Number of initial periods during which turbine of \acs{csp} $\theta$ must be offline/online \nomunit{[-]}}

\nomenclature[P, 05]{$P_u$}{Electrical power capacity of \acs{rvpp} unit $u$ \nomunit{[MW]}}
\nomenclature[P, 05]{$\bar P_d$/$\ubar P_d$}{Upper/lower bound of electrical power consumption of \acs{ed} $d$ \nomunit{[MW]}}
\nomenclature[P, 05]{$\bar P_{r(\theta)}$/$\ubar P_{r(\theta)}$}{Upper/lower bound of electrical power production of \acs{ndrs} $r$ (\acs{csp} $\theta$) \nomunit{[MW]}}
\nomenclature[P, 05]{$\hat{\tilde{P}}_{\theta,t}^{SF}$/$\check P_{\theta,t}^{SF}$}{Upper bound/negative deviation of the thermal power production forecast of \acs{sf} of \acs{csp} $\theta$ during period $t$ \nomunit{[MW]}}
\nomenclature[P, 05]{$\hat{\tilde{P}}_{r,t}$/$\check P_{r,t}$}{Upper bound/negative deviation of the \acs{ndrs} $r$ production forecast during period $t$ \nomunit{[MW]}}
\nomenclature[P, 05]{$\tilde{\check{P}}_{d,t}$/$\hat P_{d,t}$}{Lower bound/positive deviation of the \acs{ed} $d$ consumption forecast during period $t$ \nomunit{[MW]}}
\nomenclature[P, 05]{$P_{r,t}$}{\acs{ndrs} $r$ production forecast during period $t$ \nomunit{[MW]}}
\nomenclature[P, 05]{$P_{d,t}/H_{d,t}$}{\acs{ed}/\acs{td} $d$ consumption forecast during period $t$ \nomunit{[MW]}}
\nomenclature[P, 05]{$P_{\theta,t}^{SF}$}{Thermal power production forecast of \acs{sf} of \acs{csp} $\theta$ during period $t$ \nomunit{[MW]}}
\nomenclature[P, 05]{$\bar P_{\theta}^{PB}$/$\ubar P_{\theta}^{PB}$}{Upper/lower bound of thermal power of \acs{pb} of \acs{csp} $\theta$ \nomunit{[MW]}}
\nomenclature[P, 05]{$\bar P_{\theta}^{TS,+}$/$\ubar P_{\theta}^{TS,+}$}{Upper/lower bound of thermal charging power of \acs{ts} of \acs{csp} $\theta$ \nomunit{[MW]}}
\nomenclature[P, 05]{$\bar P_{\theta}^{TS,-}$/$\ubar P_{\theta}^{TS,-}$}{Upper/lower bound of thermal discharging power of \acs{ts} of \acs{csp} $\theta$ \nomunit{[MW]}}
\nomenclature[P, 05]{$P_{\theta,n}^{PB}$}{Thermal power of \acs{pb} of \acs{csp} $\theta$ in segment $n$ of piecewise function \nomunit{[MW]}}

\nomenclature[P, 06]{$\bar R_{r(\theta)}^{SR}$/$\ubar R_{r(\theta)}^{SR}$}{Up (down) secondary reserve ramp rate of \acs{ndrs} $r$ (\acs{csp} $\theta$) \nomunit{[MW/min]}}
\nomenclature[P, 06]{$\ubar R_{d}^{SR}$/$\bar R_{d}^{SR}$}{Up/down secondary reserve ramp rate of \acs{ed} $d$ \nomunit{[MW/min]}}

\nomenclature[P, 07]{$T^{SR} \;$}{Required time for secondary reserve action \nomunit{[min]}}
\nomenclature[P, 07]{$UT_{\theta}/DT_{\theta} \;$}{Minimum up/down time of turbine of \acs{csp} $\theta$ \nomunit{[-]}}

\nomenclature[P, 08]{$\ubar \beta_{d,t}$/$\bar \beta_{d,t}$}{Percentage of up/down flexibility of \acs{ed} $d$ during period $t$ \nomunit{[\%]}}
\nomenclature[P, 09]{$\Gamma^{DA}$}{Uncertainty budget of \acs{dam} electricity price \nomunit{[-]} }
\nomenclature[P, 09]{$\Gamma^{SR, \uparrow (\downarrow)}$}{Uncertainty budget of up (down) \acs{srm} electricity price \nomunit{[-]} }
\nomenclature[P, 09]{$\Gamma_{r(\theta)}$}{Uncertainty budget of \acs{ndrs} $r$ electrical (\acs{sf} of \acs{csp} $\theta$ thermal) production \nomunit{[-]}}
\nomenclature[P, 09]{$\Gamma_{d}$}{Uncertainty budget of consumption of \acs{ed}/\acs{td} $d$ \nomunit{[-]}}
\nomenclature[P, 10]{$\eta_{\theta}$}{Thermal power output efficiency of \acs{csp} $\theta$ \nomunit{[\%]}}
\nomenclature[P, 10]{$\eta_{\theta,n} \;$}{Conversion efficiency of thermal to electrical power in segment $n$ of piecewise function of \acs{pb} of \acs{csp} $\theta$ \nomunit{[\%]}}
\nomenclature[P, 10]{$\eta_{\theta}^{TS,+(-)} \;$}{Charging (discharging) thermal power efficiency of \acs{ts} of \acs{csp} $\theta$ \nomunit{[\%]}}
\nomenclature[P, 11]{$ \Delta{t} \;$}{Duration of periods \nomunit{[hour]}}
\nomenclature[P, 12]{$ \kappa \;$}{Percentage of reserve traded in the \acs{srm} relative to the power capacity of \acs{rvpp} \nomunit{[\%]}}
\nomenclature[P, 13]{$\lambda_t^{{DA}}$/$\lambda_t^{{HT}}$}{\acs{dam}/\acs{hpa} price during period $t$ \nomunit{[€/MWh]}}
\nomenclature[P, 13]{$\lambda_t^{{SR,\uparrow (\downarrow)}}$}{\acs{srm} price for up (down) reserve during period $t$ \nomunit{[€/MW]}}

\nomenclature[P, 14]{$\tilde \lambda_t^{{DA}}$/$\hat \lambda_t^{{DA}}$/$\check \lambda_t^{{DA}}$}{Median value/positive deviation/negative deviation of the \acs{dam} price forecast during period $t$ \nomunit{[€/MWh]}}
\nomenclature[P, 14]{$\hat{\tilde{\lambda}}{_t^{{SR,\uparrow}}}$/$\check \lambda_t^{{SR,\uparrow}}$}{Upper bound/negative deviation of the \acs{srm} price forecast for up reserve during period $t$ \nomunit{[€/MW]}}
\nomenclature[P, 14]{$\hat{\tilde{\lambda}}{_t^{{SR,\downarrow}}}$/$\check \lambda_t^{{SR,\downarrow}}$}{Upper bound/negative deviation of the \acs{srm} price forecast for down reserve during period $t$ \nomunit{[€/MW]}}

\nomenclature[V, 01]{$p_{r(\theta),t}$}{Electrical production of \acs{ndrs} $r$ (\acs{csp} $\theta$) during period $t$ \nomunit{[MW]}}
\nomenclature[V, 01]{$p_{d,t}/h_{d,t}$}{Electrical/thermal consumption of \acs{ed}/\acs{td} $d$ during period $t$ \nomunit{[MW]}}
\nomenclature[V, 01]{$h_{\theta,t}$}{Thermal power production of \acs{csp} $\theta$ during period $t$ \nomunit{[MW]}}
\nomenclature[V, 01]{$p_{\theta,t}^{SF}$}{Thermal power of \acs{sf} of \acs{csp} $\theta$ during period $t$ \nomunit{[MW]}}
\nomenclature[V, 01]{$p_{\theta,t}^{PB}$}{Thermal power of \acs{pb} of \acs{csp} $\theta$ during period $t$ \nomunit{[MW]}}
\nomenclature[V, 01]{$p_{\theta,t}^{TS,+(-)}$}{Thermal charging (discharging) power of \acs{ts} of \acs{csp} $\theta$ during period $t$ \nomunit{[MW]}}
\nomenclature[V, 01]{$x_{\theta,n,t}$}{Positive variable related to segment $n$ of piecewise function of \acs{pb} of \acs{csp} $\theta$ during period $t$ \nomunit{[-]}}
\nomenclature[V, 01]{$p_t^{DA}$}{Electrical power traded by \acs{rvpp} in the \acs{dam} during period $t$ (positive/negative for selling/buying) \nomunit{[MW]}}
\nomenclature[V, 01]{$h_t^{HT}$}{Thermal power purchased by \acs{rvpp} through \acs{hpa} during period $t$ \nomunit{[MW]}}
\nomenclature[V, 01]{$r_t^{SR,\uparrow(\downarrow)}$}{Up (down) reserve traded by \acs{rvpp} in the \acs{srm} during period $t$ \nomunit{[MW]}}
\nomenclature[V, 01]{$r_{r(\theta),t}^{\uparrow (\downarrow)}$}{Up (down) reserve provided by \acs{ndrs} $r$ (\acs{csp} $\theta$) during period $t$ \nomunit{[MW]}}
\nomenclature[V, 01]{$r_{u,t}^{\uparrow (\downarrow)}$}{Up (down) reserve provided by \acs{rvpp} unit $u$ during period $t$ \nomunit{[MW]}}

\nomenclature[V, 01]{$r_{d,t}^{\uparrow(\downarrow)}$}{Up (down) reserve provided by \acs{ed} $d$ during period $t$ \nomunit{[MW]}}
\nomenclature[V, 01]{$r_{\theta,t}^{TS,\uparrow(\downarrow)}$}{Up (down) reserve of \acs{ts} of \acs{csp} $\theta$ during period $t$ \nomunit{[MW]}}
\nomenclature[V, 01]{$r_{\theta,t}^{TS,+,\uparrow(\downarrow)}$}{Up (down) reserve of \acs{ts} of \acs{csp} $\theta$ in the charging state during period $t$ \nomunit{[MW]}}
\nomenclature[V, 01]{$r_{\theta,t}^{TS,-,\uparrow(\downarrow)}$}{Up (down) reserve of \acs{ts} of \acs{csp} $\theta$ in the discharging state during period $t$ \nomunit{[MW]}}
\nomenclature[V, 01]{$y_t^{DA}$}{Positive auxiliary variable of traded electrical energy in the \acs{dam} during period $t$ \nomunit{[MWh]}}
\nomenclature[V, 01]{$y_{r(\theta),t}$}{Positive auxiliary variable of \acs{ndrs} $r$ electrical (\acs{sf} of \acs{csp} $\theta$ thermal) production uncertainty during period $t$ \nomunit{[MW]}}
\nomenclature[V, 01]{$y_{d,t}$}{Positive auxiliary variable of \acs{ed}/\acs{td} $d$ consumption uncertainty during period $t$ \nomunit{[MW]}}
\nomenclature[V, 01]{$z_t^{DA}$}{Positive auxiliary variable of \acs{dam} electricity price uncertainty during period $t$ \nomunit{[-]}}
\nomenclature[V, 01]{$z_t^{SR, \uparrow (\downarrow)}$}{Positive auxiliary variable of up (down) \acs{srm} electricity price uncertainty during period $t$ \nomunit{[-]}}
\nomenclature[V, 01]{$z_{r(\theta),t}$}{Positive auxiliary variable of \acs{ndrs} $r$ electrical (\acs{sf} of \acs{csp} $\theta$ thermal) production uncertainty during period $t$ \nomunit{[-]}}
\nomenclature[V, 01]{$z_{d,t}$}{Positive auxiliary variable of \acs{ed}/\acs{td} $d$ consumption uncertainty during period $t$ \nomunit{[-]}}
\nomenclature[V, 01]{$e_{\theta,t}^{TS} \;$}{Thermal energy of \acs{ts} of \acs{csp} $\theta$ during period $t$ \nomunit{[MWh]}}

\nomenclature[V, 02]{$\sigma_{\theta}^{TS,\uparrow(\downarrow)} \;$}{Share of thermal energy capacity of \acs{ts} of \acs{csp} allocated to provide up (down) reserve \nomunit{[\%]}}
\nomenclature[V, 02]{$\zeta_t^{DA}$}{Dual variable to model the \acs{dam} price uncertainty during period $t$ \nomunit{[€]}}
\nomenclature[V, 02]{$\zeta_t^{SR, \uparrow (\downarrow)}$}{Dual variable to model the up (down) \acs{srm} price uncertainty during period $t$ \nomunit{[€]}}
\nomenclature[V, 02]{$\zeta_{r(\theta),t}$}{Dual variable to model the \acs{ndrs} $r$ electrical (\acs{sf} of \acs{csp} $\theta$ thermal) production uncertainty during period  $t$ \nomunit{[MW]}}
\nomenclature[V, 02]{$\zeta_{d,t}$}{Dual variable to model the \acs{ed}/\acs{td} $d$ consumption uncertainty during period $t$ \nomunit{[MW]}}
\nomenclature[V, 02]{$\phi^{DA}$}{Dual variable to model the \acs{dam} price uncertainty \nomunit{[€]}}
\nomenclature[V, 02]{$\phi^{SR,\uparrow (\downarrow)}$}{Dual variable to model the up (down) \acs{srm} price uncertainty \nomunit{[€]}}
\nomenclature[V, 02]{$\phi_{r(\theta)}$}{Dual variable to model the \acs{ndrs} $r$ electrical (\acs{sf} of \acs{csp} $\theta$ thermal) production uncertainty \nomunit{[MW]}}
\nomenclature[V, 02]{$\phi_{d}$}{Dual variable to model the \acs{ed}/\acs{td} $d$ consumption uncertainty \nomunit{[MW]}}

\nomenclature[W, 03]{$\varsigma_{\theta,t,n}$}{Binary variable that is 1 if segment $n$ of piecewise function of \acs{pb} of \acs{csp} $\theta$ is active during period $t$, and 0 otherwise \nomunit{[-]}}
\nomenclature[W, 01]{$u_{\theta,t}$}{Binary variable that is 1 if turbine of \acs{csp} $\theta$ is online during period $t$, and 0 otherwise \nomunit{[-]}}
\nomenclature[W, 01]{$v_{\theta,t}^{SU}$/$v_{\theta,t}^{SD}$}{Binary variable that is 1 if turbine of \acs{csp} $\theta$ starts up/shuts down at period $t$, and 0 otherwise \nomunit{[-]}}
\nomenclature[W, 01]{$u_{\theta,t}^{TS}$}{Binary variable that is 1 if charging state of \acs{ts} of \acs{csp} $\theta$ is active, and 0 otherwise \nomunit{[-]}}

\nomenclature[W, 03]{$\chi_{r(\theta),t}$}{Binary variable that is 1 if \acs{ndrs} $r$ electrical (\acs{sf} of \acs{csp} $\theta$ thermal) power production worst case occurs during period $t$, and 0 otherwise \nomunit{[-]}\vspace{5pt}}
\nomenclature[W, 03]{$\chi_{d,t}$}{Binary variable that is 1 if \acs{ed}/\acs{td} $d$ consumption worst case occurs during period $t$, and 0 otherwise \nomunit{[-]}\vspace{5pt}}

\nomenclature[Z, 01]{$\boldsymbol{r}_{t}^{SR} = \{r_{t}^{SR,\uparrow}, -r_{t}^{SR, \downarrow},0\}$}{\hspace{.4cm}Vector for possible reserve activation scenarios of \acs{rvpp}  \nomunit{[MW]}}
\nomenclature[Z, 01]{$\boldsymbol{r}_{r,t} = \{r_{r,t}^{\uparrow},-r_{r,t}^{\downarrow},0 \}$}{\hspace{.9cm}Vector for possible reserve activation scenarios of \acs{ndrs} $r$ \nomunit{[MW]}}
\nomenclature[Z, 01]{$\boldsymbol{r}_{\theta,t} = \{r_{\theta,t}^{\uparrow},-r_{\theta,t}^{\downarrow},0 \}$}{\hspace{.8cm}Vector for possible reserve activation scenarios of \acs{csp} $\theta$  \nomunit{[MW]}}
\nomenclature[Z, 01]{$\boldsymbol{r}_{d,t} = \{r_{d,t}^{\uparrow},-r_{d,t}^{\downarrow},0\}$}{\hspace{.8cm}Vector for possible reserve activation scenarios of \acs{ed} $d$  \nomunit{[MW]}}
\nomenclature[Z, 01]{$\boldsymbol{r}_{\theta,t}^{TS} = \{r_{\theta,t}^{TS,\uparrow},-r_{\theta,t}^{TS,\downarrow},0 \}$}{\hspace{.3cm}Vector for possible reserve activation scenarios of \acs{ts} of \acs{csp} $\theta$  \nomunit{[MW]}}
\nomenclature[Z, 01]{$\boldsymbol x_{\theta,t,n} = \{x_{\theta,t,n}^{\uparrow},x_{\theta,t,n}^{\downarrow}, x_{\theta,t,n} \}$}{Vector related to piecewise function of \acs{pb} of \acs{csp} $\theta$ for possible reserve activation scenarios \nomunit{[-]}}
\nomenclature[Z, 01]{$\boldsymbol \varsigma_{\theta,t,n}  = \{\varsigma_{\theta,t,n}^{\uparrow},\varsigma_{\theta,t,n}^{\downarrow}, \varsigma_{\theta,t,n} \}$}{\hspace{.05cm}Vector related to piecewise function of \acs{pb} of \acs{csp} $\theta$ for possible reserve activation scenarios \nomunit{[-]}}

\renewcommand{\nomname}{}
\printnomenclature[1.5cm]

\end{footnotesize} 
\end{tcolorbox}

\acresetall %resets all acronym definitions
\section{Introduction}
\subsection{Motivation}

{\color{black}\acp{csp} convert sunlight into heat energy, which can be stored in \ac{ts}, used for electricity generation, or directly applied in heating applications~\cite{dominguez2012optimal, zhao2021coordinated}. This thermal energy serves residential uses (e.g., water or space heating, absorption cooling) and industrial processes such as food, chemical, and textile production~\cite{USDOE_CSP}. Unlike photovoltaics, \ac{ptc} mirrors focus sunlight onto a fluid-filled tube that transfers heat to generate steam for turbines. \acp{csp} can store this heat for hours using \ac{ts} (e.g., molten salt), allowing power generation even after sunset~\cite{garcia2011performance}. This enhances their reliability over intermittent sources. \acp{csp} are increasingly applied for industrial decarbonization by supplying renewable process heat. Advances in materials, manufacturing, and design have reduced costs, making \acp{csp} more competitive with other renewables~\cite{miron2023cost}. Large-scale deployment is growing in high-irradiance regions such as Southern Europe, the Middle East, North Africa, Australia, and the United States~\cite{IRENA2025}. With improving storage technologies and falling costs, \acp{csp} are poised for further expansion. Their ability to provide reliable, high-temperature energy makes them well-suited for utility-scale and industrial use~\cite{khan2022progress}. Given their dual ability to generate both thermal and electrical energy—and their flexibility in providing ancillary services like secondary reserve via \ac{ts}~\cite{yao2023concentrated}—integrating \acp{csp} into the \ac{rvpp} framework has drawn increasing interest.} The \ac{rvpp} is defined as a set of \acp{res}, \acp{csp}, \acp{ed}, and \acp{td} aim of ensuring a safe and reliable operation by offering their combined flexibility (e.g., fast ramp-up/down capability for frequency control), the possibility to internally balance the stochastic \acp{res} fluctuations, and to sell their aggregate generation output in the wholesale market~\cite{alvaroPOSYFT}.

The primary marketplace for electric energy trading spans the full 24-hour period of the designated day, divided into 24 equal one-hour intervals. This market, known as the \ac{dam}, is usually settled 12 hours before the energy delivery period. In this market, generation and demand participants submit selling offers and purchasing bids for each time slot. The \ac{mo} then clears the market by evaluating these offers and bids, establishing electricity prices and determining the accepted energy quantities from each participant~\cite{conejo2010decision, naval2021virtual}. In addition to energy markets, \acp{asm} are in place to efficiently allocate resources and ensure the reliable operation of the power system. These markets primarily function by securing predetermined levels of power reserve. Reserve refers to the extra capacity that power plants must allocate in their generation plans to accommodate unforeseen changes in demand or generation shortfalls. The \ac{tso} is tasked with ensuring the reliability of the power system~\cite{fernandes2016participation}. The goal of secondary reserve, also known as \ac{afrr}, is to restore frequency and power exchange to their specified reference values. Any unit aiming to offer \ac{afrr} in the market must be certified by the \ac{tso}. Key qualifications include the ability to provide reserve for 15 minutes without interruption, within a resolution time of 1 hour, and a response time of 100 seconds~\cite{fernandes2016participation}. Due to the capabilities of \ac{rvpp} units—mainly stemming from the flexibility of \ac{csp}, and partly from the reserve capacity of flexible \acp{ed} and \ac{ndrs}—\ac{rvpp} can contribute to secondary reserve and thus participate in the \ac{srm} to maximize its profitability. Furthermore, enhancing the heat utilization of \ac{csp} can reduce operational costs and increase the efficiency of \ac{rvpp}~\cite{sun2022day}. The incorporation of thermal energy trading alongside electrical energy trading can boost the profitability of \ac{rvpp}~\cite{kong2020robust}. Thermal energy can be traded in various ways: through the thermal energy market~\cite{kong2020robust, ghasemi2022coordinated}, locally in the local market~\cite{foroughi2021bi}, or via contracts~\cite{wang2020non}. \ac{hpa} contracts for thermal energy are tailored agreements between two parties, enabling traders to negotiate terms outside a central market or organized exchange~\cite{hasni2023case, kircher2021heat}. The \ac{rvpp} can enter into an \ac{hpa} with a thermal energy service provider to satisfy its remaining \acp{td}. Additionally, the \ac{hpa} can act as a hedge against market price fluctuations and ensure the fulfillment of \acp{td} when the \ac{csp} has low thermal production or is unavailable.

Uncertainty characterization is one of the challenging aspects of \ac{rvpp} market participation~\cite{roald2023power, singh2022uncertainty}. Uncertainties in the \ac{rvpp} problem can influence multiple factors, including electricity prices in the \ac{dam} and \ac{srm}, the thermal output of \ac{csp}, the energy production from \ac{ndrs}, and the consumption patterns of \acp{ed}/\acp{td}~\cite{zhao2021coordinated, venegas2022review}. These uncertain parameters can have a substantial effect on the profitability of the \ac{rvpp} in the market. Thus, addressing this wide range of uncertainties, which characterize the behavior of \ac{rvpp} units, is crucial for enhancing competitiveness. On one hand, integrating various units into an \ac{rvpp} can help reduce uncertainties in both production and consumption. As the number of units within the \ac{rvpp} increases, the likelihood of significant deviations from expected production decreases~\cite{alvaroPOSYFT}. Additionally, the \ac{csp} can effectively manage uncertainty to maintain the balance between supply and demand within the \ac{rvpp}~\cite{xiong2024distributionally}. This is made possible by its \ac{ts} ability to provide energy when other units experience energy fluctuations. However, the \ac{rvpp} operator must still consider various uncertainties using appropriate optimization techniques. These techniques need to be computationally efficient so that the \ac{rvpp} can determine its optimal bids and offers before the energy and reserve markets’ gate closure or adjust the scheduling of its units accordingly~\cite{yu2019uncertainties}. Furthermore, the \ac{rvpp} operator must decide on the level of conservatism to apply to different uncertainties and conduct sensitivity analyses, as varying strategies for handling uncertainties can significantly affect profitability~\cite{nemati2025segan}. Optimization models should be developed to effectively address the various components of the \ac{rvpp} problem, including the complexities of different electricity markets, \ac{hpa} contracts, and inherent uncertainties. Consequently, formulating bidding strategies for \ac{rvpp} participation across multiple markets—while accounting for the technical characteristics of \ac{rvpp} units, uncertainty modeling, and market interactions—remains a critical area of focus for both \ac{rvpp} operators and researchers. This paper provides an in-depth examination of these critical aspects of \ac{rvpp} market participation.

\subsection{Literature Review}

The \ac{vpp} scheduling and bidding problem in the electrical energy market, with and without considering different uncertainties, is widely studied in the literature~\cite{foroughi2021bi, rahimi2021optimal, xiao2024windfall, kalantari2023strategic}. In~\cite{foroughi2021bi}, the optimal bidding strategy of a multi-carrier \ac{vpp} in energy markets is modeled by a bi-level nonlinear deterministic approach. In~\cite{rahimi2021optimal}, the optimal scheduling of a \ac{vpp} with both renewable and non-renewable energy sources is studied. The scenario-based \ac{sp} model is used to capture uncertainties in electricity price, \ac{ndrs} production, \ac{ed}, and \ac{td} consumption. The thermal market is adopted to improve the performance of multi-carrier \ac{vpp}, including \ac{res} and combined heat and power units. In~\cite{xiao2024windfall}, an industrial \ac{vpp}, including \ac{res}, \ac{es}, and material storage for supplying \ac{ed} and \ac{td} of industrial processes, and financial instruments, is developed. The best and worst-case scenarios of uncertain parameters related to electricity price and \ac{ndrs} production are modeled by scenario-based \ac{sp}, using \ac{vab} and \ac{cvar} criteria, respectively. The paper~\cite{kalantari2023strategic} studies the decision-making of a \ac{vpp} with energy storage in the \ac{dam} and \ac{idm}, considering \ac{res} uncertainties. The interactions between the \ac{vpp}, electricity market, and aggregators in the distribution network are taken into account by a rolling horizon optimization technique.
{\color{black} These studies predominantly focus on energy market participation, often neglecting \ac{asm}, which are critical for maintaining grid stability and reliability. Incorporating \ac{asm} participation can provide additional income sources and enhance the operational flexibility and economic performance of \acp{vpp}, representing a significant yet underexplored opportunity in these works.}

Some papers study different \acp{asm} in addition to energy markets, which can increase the profitability of \ac{vpp}. The paper~\cite{li2023robust} examines the peak-regulation \ac{asm} participation of a multi-energy \ac{vpp}, including electrical and thermal units. A two-stage \ac{ro} is proposed to account for the uncertainties in PV units production and \ac{ed} and \ac{td} consumption. In~\cite{gough2023bi}, a model is proposed for multiple technical \ac{vpp} in the distribution network, allowing for energy trading in the electricity market and energy trading among \acp{vpp}. The uncertainties related to \ac{res} and \ac{ed} are considered using a two-stage \ac{sp} model. The paper~\cite{NEMATI2025136421} proposes a single-level \ac{ro} approach for \ac{rvpp} participation in the sequential energy and reserve markets to find the worst-case profit of different uncertainties related to electricity price, \ac{ndrs} production, and \ac{ed} consumption.
In~\cite{nemati2025flexible}, this work is extended by incorporating the penalization costs of \ac{rvpp} in electricity markets using an economic risk analysis approach. Furthermore, carbon trading~\cite{yan2022two} and thermal energy trading~\cite{kong2020robust, ghasemi2022coordinated} are studied in the \ac{vpp} optimization problem. In~\cite{yan2022two}, to achieve low-carbon operation of a multi-energy \ac{vpp}, a carbon trading mechanism is proposed along with \ac{dam} participation of \ac{vpp}. A two-stage \ac{aro} approach, which is solved by the \ac{ccg} algorithm, is proposed to account for the uncertainties related to both the source and load sides. In~\cite{kong2020robust}, the optimal scheduling of a multi-energy \ac{vpp} with several uncertainties in electricity and thermal markets is studied. A two-stage \ac{ro}-\ac{sp} approach is implemented to capture the uncertainties of \ac{res} production and \ac{ed} and \ac{td} consumption. The paper~\cite{ghasemi2022coordinated} proposes a multi-objective scheduling of a multi-energy \ac{vpp} to maximize profit and minimize carbon emissions. Several uncertainties of electricity price, \ac{res} production (including PV, \ac{pvt}, and \ac{wf}), and \ac{ed} consumption are taken into account through scenarios. {\color{black}A common limitation of these studies is that they do not consider the joint participation of \acp{vpp} in electricity markets, \acp{asm}, and thermal energy trading, thus preventing fully leveraging their operational flexibility and maximizing economic benefits. In addition, while uncertainties related to renewable generation and demand are often considered, price uncertainty in both the energy market and \acp{asm} are typically neglected. Furthermore, the integration of \ac{csp} technologies is often overlooked, despite their growing importance in multi-energy \ac{vpp} systems. These limitations highlight the need for more holistic and realistic modeling approaches to support advanced market participation strategies.}

Given the flexibility of \ac{csp} through its \ac{ts} to provide both energy and ancillary services, some papers study \ac{csp} independent participation in different markets. In~\cite{dominguez2012optimal}, the offering curve of a price-taker \ac{csp} in the \ac{dam} is obtained. An \ac{ro}-\ac{sp} approach is proposed to consider the uncertainties of electricity price and \ac{csp} production. In~\cite{zhao2020mixed}, the optimal bidding of \ac{csp} in the \ac{dam} and \ac{rtm} is investigated. The uncertainties of electricity price and \ac{csp} production are modeled by a hybrid stochastic \ac{igdt} approach based on \ac{cvar}. The paper in~\cite{he2016optimal} studies the optimal offering strategy of \ac{csp} in the \ac{dam} energy, reserve market, and regulation markets. The market price uncertainties are considered by scenarios and solar generation uncertainties are modeled by \ac{ro}. Additionally, some research integrates \ac{csp} with \acp{wf}~\cite{xu2016coordinated, xiong2023dp, pousinho2014self, fang2020look} or biomass plants~\cite{khaloie2021day} to improve the performance of power producers in economic dispatch or market participation. In~\cite{xu2016coordinated}, the economic dispatch of \ac{csp} with \acp{eh} and \acp{wf} for the provision of energy and reserve is studied. A two-stage \ac{sp} model is used to capture the uncertainties of \ac{res}. The paper~\cite{xiong2023dp} suggests a multi-stage approach for coordinated \ac{wf}-\ac{csp} energy and reserve allocation in the \ac{dam} and \ac{idm}. The uncertainties of \ac{wf} and \ac{csp} production are taken into account by the \ac{aro} approach. In~\cite{pousinho2014self}, the self-scheduling of \ac{wf}-\ac{csp} power producers in the \ac{dam} energy and reserve market is studied. The uncertainties of \ac{wf} and \ac{csp} production are modeled by using the \ac{pdf} of these uncertain parameters. The paper in~\cite{fang2020look} studies the bidding approach of \acp{csp} containing \ac{eh} integrated with \acp{wf} for participation in the \ac{dam} and \ac{asm}. The uncertainties of electricity prices are considered by scenarios, while the uncertainties of \ac{csp} and \ac{wf} production are modeled by stochastic chance-constrained optimization. The paper~\cite{khaloie2021day} studies the \ac{dam} and \ac{idm} dispatch of integrated biomass-\ac{csp} by using a multi-objective risk-based approach. A two-stage \ac{sp} approach for market uncertainties and \ac{cvar}-\ac{igdt} for \ac{csp} uncertainties are used. {\color{black}However, these studies do not adopt a \ac{vpp} framework and instead consider separate participation of \ac{csp} units or their integration with \ac{wf} or biomass plants, missing the opportunity to fully exploit the operational and economic synergies of diverse assets within a coordinated \ac{vpp}. Moreover, they focus solely on electricity market participation, neglecting CSP’s potential role in thermal energy trading. }
%These limitations reduce the practical relevance of the proposed models for real-world multi-energy market applications.

Integration of \ac{csp} in \ac{vpp} has been studied in the literature. The paper~\cite{oladimeji2022optimal} examines the optimal participation of \ac{rvpp} including \ac{csp}, PV, \ac{wf}, and hydro plant in the \ac{dam} and \ac{idm} through a deterministic \ac{milp} problem. The paper~\cite{fang2023optimization} proposes the deterministic optimal scheduling of \ac{csp}-based \ac{vpp} in the electricity energy and carbon trading markets. The paper~\cite{xiong2024distributionally} suggests a coordinated energy management strategy for \ac{wf}-\ac{csp} integration in multiple \acp{vpp} in the \ac{dam}. A \ac{dro} approach using the \ac{ccg} algorithm is proposed to capture the uncertainties in \ac{csp} and \ac{wf} production. {\color{black}Despite the work done in this regard, the integration of \ac{csp} in \ac{vpp}, considering multiple uncertainties, electrical and thermal energy markets, and \ac{asm}, has not been fully studied in the literature.} The paper~\cite{nemati2025segan} proposes a flexible \ac{ro} approach for \ac{rvpp} participation in electrical markets, including sequential energy and reserve markets. {\color{black}However, thermal energy trading is neglected, and a detailed model for energy conversion is not provided in this paper.} In~\cite{zhao2021coordinated}, a coordinated strategy for \ac{vpp}, including \ac{csp}, solar PVs, and \acp{ed}/\acp{td}, is proposed for scheduling in the \ac{dam} and \ac{bam}. \ac{sp}, through scenario generation and considering \ac{cvar}, is used to model various uncertainties in electricity prices, \ac{csp} and \ac{pv} production, and \acp{ed}/\acp{td} consumption. The paper~\cite{sun2022day} proposes a model for the heat utilization of \ac{csp} in a \ac{vpp} composed of integrated \ac{caes}, \ac{ts}, combined heat and power, and \ac{csp}. The uncertainty in solar irradiation and \acp{ed}/\acp{td} is addressed by using four scenarios for different seasons. {\color{black} Again, the above two references do not take into account \ac{asm}.}

Considering the gaps in the literature, this paper investigates the integration of \ac{csp} in \ac{vpp} for the electrical \ac{dam} market and \ac{asm} (through \ac{srm}), along with thermal energy trading via \ac{hpa} contracts. A two-stage \ac{ro} approach is developed to account for various uncertainties related to \ac{dam} and \ac{srm} electricity prices, the electrical production of solar PVs, \acp{wf} production, and thermal production of \ac{csp}, as well as uncertainties in \ac{ed} and \ac{td} consumption. This work extends the \ac{ro} framework in~\cite{bertsimas04} by incorporating time-dependent selection of uncertain parameters, allowing for a more practical approach to handling uncertainty on both the source and load sides in \ac{vpp} market bidding applications. Unlike previous formulations~\cite{dominguez2012optimal, he2016optimal, xiong2023dp}, which often define uncertainty sets for each time period, our method integrates temporal considerations into the protection function of uncertain parameters, (the term in the objective function that defines the necessary protection against deviations of uncertain parameters). The procedure for obtaining the refined \ac{milp} mathematical formulation from the protection function of uncertainties is also provided. Additionally, for the first time, the reserve provision of \ac{csp} is modeled by considering the energy limitation of its \ac{ts} to guarantee the safe energy and reserve provision of \ac{csp}. This is achieved by proposing a mathematical model that assigns an adjustable share of \ac{ts} exclusively for the provision of reserve. The optimization problem determines the optimal value of the assigned share for reserve provision based on the profitability of the \ac{rvpp}. Overall, this paper proposes a more comprehensive model for the integration of \ac{csp} in \ac{rvpp} by considering multiple uncertainties and electricity and thermal energy trading options, which is more advanced than existing literature. To support this claim, Table~\ref{table:Literature} compares the key aspects of the reviewed literature with the proposed approach in this paper.
%[h]

\begin{sidewaystable}
%\begin{table*}
  \centering
  \caption{Comparison of proposed approach in this paper and literature.}
%  \tiny
\scriptsize
  \setlength{\tabcolsep}{2.1pt} % Adjust 3pt to a smaller value if needed
  \renewcommand{\arraystretch}{1.5} % Adjust the factor (default is 1.0)
  \vspace{-.5em}
  \begin{threeparttable}
  \begin{tabular}{@{}ccccccccccccccccc@{}}  % Removes extra space
    \toprule
    & \multicolumn{4}{c}{\textbf{Components}} 
    & \multicolumn{3}{c}{\textbf{Market (contract)}} 
    & \multicolumn{4}{c}{\textbf{Uncertainty}} 
    & \multicolumn{1}{c}{\textbf{\makecell{VPP\\concept}}} 
    & \multicolumn{1}{c}{\textbf{{\makecell{Reserve\\concept}}}} 
    & \multicolumn{1}{c}{\textbf{\makecell{Method\\ \& solution}}} \\

    \cmidrule(lr){2-5} \cmidrule(lr){6-8} \cmidrule(lr){9-12} 
    \textbf{Ref.} & \textbf{\ac{csp}} & \textbf{\ac{res}} & \textbf{Storage} & \textbf{Load}  
    & \textbf{Energy} & \textbf{\ac{asm}} & \textbf{Thermal}  
    & \textbf{Price} & \textbf{\ac{csp}} & \textbf{\ac{res}} & \textbf{Load}  
    & \textbf{} & \textbf{} & \textbf{} \\

    \cmidrule{1-15}

    \cite{dominguez2012optimal} & \ding{51} & $\times$ & \ac{ts} & $\times$ & \ac{dam} & $\times$ & $\times$ & \ac{dam} & \ding{51} & $\times$ & $\times$ & $\times$ & $\times$ & \ac{sp}, \ac{ro} \\ [0.2em]

    \cite{zhao2021coordinated} & \ding{51} & PV & \ac{ts} & \ac{ed}/\ac{td}  
    & \ac{dam}/\ac{bam} & $\times$ & $\times$  
    & \ac{dam}/\ac{bam} & \ding{51} & PV & \ac{ed}/\ac{td}  
    & \ding{51} & $\times$ & \ac{sp}, \ac{cvar} \\  

    \cite{sun2022day} & \ding{51} & $\times$ & \ac{es}/\ac{ts} & \ac{ed}/\ac{td}  
    & \ac{dam} & $\times$ & Thermal  
    & $\times$ & \ding{51} & $\times$ & \ac{ed}/\ac{td}  
    & \ding{51} & $\times$ & Scenario-based, \ac{milp} \\

    \cite{kong2020robust}  & $\times$  & PV/\ac{wf} & \ac{es}/\ac{ts} & \ac{ed}/\ac{td} & \ac{dam} & $\times$ &  Thermal  & $\times$ &  $\times$  & PV/\ac{wf}  & \ac{ed}/\ac{td} & \ding{51} & $\times$ & \makecell{Two-stage \ac{ro}-\ac{sp},\\  \ac{ccg} } \\ [0.2em]

    \cite{ghasemi2022coordinated}  & $\times$ & PV/\ac{pvt}/\ac{wf} & \ac{es}/\ac{ts} & \ac{ed}/\ac{td} & \ac{dam} & Reserve &  Thermal  & \ac{dam} &  $\times$  & PV/\ac{pvt}/\ac{wf}  & \ac{ed} & \ding{51} & Spinning reserve & \makecell{Scenario-based \ac{sp},\\ Fuzzy} \\ [0.2em]

    \cite{foroughi2021bi}  & $\times$ & PV/\ac{wf} & $\times$ & \ac{ed}/\ac{td} & \ac{dam} & $\times$ &  Thermal  & $\times$ &  $\times$  & $\times$  & $\times$ & \ding{51} & $\times$ & \makecell{Deterministic,\\ Bi-level, Nonlinear} \\ [0.2em]

    \cite{xiong2024distributionally} & \ding{51} & \ac{wf} & \ac{es}/\ac{ts} & \ac{ed} & \ac{dam} & $\times$ & $\times$  
    & $\times$ & \ding{51} & \ac{wf} & $\times$ & \ding{51} & $\times$ & \ac{dro}, \ac{ccg}  \\  

    \cite{nemati2025segan}  & \ding{51}  & PV/\ac{wf} & $\times$ & \ac{ed} & \ac{dam}/\ac{idm} &  \ac{srm} &  $\times$  & \ac{dam}/\ac{idm}/\ac{srm} &  \ding{51}  & PV/\ac{wf}  & \ac{ed} & \ding{51} & Up/down reserve & \ac{ro}, \ac{milp} \\ [0.2em]

    \cite{rahimi2021optimal}  & $\times$ & PV/\ac{pvt}/\ac{wf} & \ac{es} & \ac{ed}/\ac{td} & \ac{dam} & $\times$ &  $\times$  & \ac{dam} &  $\times$  & PV/\ac{pvt}/\ac{wf}  & \ac{ed}/\ac{td} & \ding{51} & $\times$ & \makecell{Scenario-based \ac{sp},\\ \ac{milp}} \\ [0.2em]

    \cite{xiao2024windfall}  & $\times$  & \ac{wf} & \ac{es} & \ac{ed}/\ac{td} & \ac{dam}/\ac{rtm} &  $\times$ &  $\times$  & \ac{dam}/\ac{rtm} &  $\times$  & \ac{wf}  & $\times$ & \ding{51} & $\times$ & \makecell{Scenario-based \ac{sp},\\  \ac{cvar}, \ac{vab}} \\ [0.2em]

    \cite{kalantari2023strategic}  & $\times$ & PV/\ac{wf} & \ac{es}/\ac{ts} & \ac{ed}/\ac{td} & \ac{dam}/\ac{idm} & $\times$ &  $\times$  & $\times$ &  $\times$  & PV/\ac{wf}  & $\times$ & \ding{51} & $\times$ & \makecell{Rolling horizon,\\ Nonlinear} \\ [0.2em]

    \cite{li2023robust}  & $\times$  & PV & \ac{es}/\ac{ts} & \ac{ed}/\ac{td} & \ac{dam} &  Peak regulation &  $\times$  & $\times$ &  $\times$  & PV  & \ac{ed}/\ac{td} & \ding{51} & $\times$ & \makecell{Two-stage \ac{ro},\\  \ac{ccg} } \\ [0.2em]

    \cite{gough2023bi}  & $\times$ & PV/\ac{wf} & \ac{es} & \ac{ed} & \ac{dam} & \makecell{Congestion\\management}  &  $\times$  & $\times$ &  $\times$  & PV/\ac{wf}  & \ac{ed} & \ding{51} & $\times$ & \makecell{Two-stage \ac{sp},\\ \ac{milp}} \\ [0.2em]

    \cite{NEMATI2025136421}  & $\times$  & PV/\ac{wf} & $\times$ & \ac{ed} & \ac{dam} &  \ac{srm} &  $\times$  & \ac{dam}/\ac{srm} &  $\times$  & PV/\ac{wf}  & \ac{ed} & \ding{51} & Up/down reserve & \ac{ro}, \ac{milp} \\ [0.2em]

    \cite{yan2022two}  & $\times$  & PV/\ac{wf} & \ac{es} & \ac{ed}/\ac{td} & \ac{dam} & $\times$ &  $\times$  & $\times$ &  $\times$  & PV/\ac{wf}  & \ac{ed}/\ac{td} & \ding{51} & $\times$ & \makecell{Two-stage \ac{aro},\\  \ac{ccg} } \\ [0.2em]

    \cite{zhao2020mixed} & \ding{51} & $\times$ & \ac{ts} & $\times$  
    & \ac{dam}/\ac{rtm} & $\times$ & $\times$  
    & \ac{dam}/\ac{rtm} & \ding{51} & $\times$ & $\times$  
    & $\times$ & $\times$ & \makecell{\ac{sp}-\ac{igdt},\\ \ac{cvar}} \\ 

    \cite{he2016optimal} & \ding{51} & $\times$ & \ac{ts} & $\times$  
    & \ac{dam} & \makecell{Reserve,\\Regulation} & $\times$  
    & \ac{dam}/\ac{asm} & \ding{51} & $\times$ & $\times$  
    & $\times$ & \makecell{Responsive reserve,\\Regulation up/down}  
    & \ac{sp}, \ac{ro} \\  

    \cite{xu2016coordinated}  & \ding{51}  & \ac{wf} & \ac{ts} & $\times$ & $\times$ &  $\times$  &  $\times$  & $\times$ &  \ding{51}  & \ac{wf}  & $\times$ & $\times$ & Reserve scheduling  & Two-stage \ac{sp} \\ [0.2em]  

    \cite{xiong2023dp} & \ding{51} & \ac{wf} &  \ac{ts} & $\times$  
    & \ac{dam}/\ac{idm} & $\times$ & $\times$  
    & $\times$ & \ding{51} & $\times$ & $\times$  
    & $\times$ & Reserve scheduling & Multi-stage \ac{aro} \\ 

    \cite{pousinho2014self} & \ding{51} & \ac{wf} & \ac{ts} & $\times$  
    & \ac{dam} & Reserve & $\times$  
    & $\times$ &   \ding{51} & \ac{wf} & $\times$  
    & $\times$ & Spinning reserve & \ac{pdf}, \ac{milp} \\

    \cite{fang2020look} & \ding{51} & \ac{wf} & \ac{ts} & $\times$  
    & \ac{dam} & \makecell{Reserve,\\Regulation} & $\times$  
    & \ac{dam}/\ac{asm} & \ding{51} & \ac{wf} & $\times$  
    & $\times$ & \makecell{Responsive reserve,\\Regulation up/down}  
    & \makecell{\ac{sp},\\Chance constrained} \\  

    \cite{khaloie2021day} & \ding{51} & Biomass & \ac{ts} & $\times$  
    & \ac{dam}/\ac{idm} & $\times$ & $\times$  
    & \ac{dam}/\ac{idm} & \ding{51} & $\times$ & $\times$  
    & $\times$ & $\times$ & \makecell{Two-stage \ac{sp},\\ \ac{cvar}-\ac{igdt}} \\

    \cite{oladimeji2022optimal} & \ding{51} & PV/\ac{wf}/Hydro & \ac{ts} & \ac{ed} & \ac{dam}/\ac{idm} & $\times$ & $\times$ & $\times$ & $\times$ & $\times$ & $\times$ & \ding{51} & $\times$ & Deterministic, \ac{milp} \\ [0.2em]

    \cite{fang2023optimization} & \ding{51} & PV/\ac{wf} & \ac{es} & \ac{ed}/\ac{td}  
    & \ac{dam} & $\times$ & $\times$  
    & $\times$ & $\times$ & $\times$ & $\times$  
    & \ding{51} & $\times$ & Master-slave game \\

    \cmidrule{1-15}
    \makecell{\textbf{This}\\ \textbf{paper}} & \ding{51} & PV/\ac{wf} & \ac{ts} & \ac{ed}/\ac{td}  
    & \ac{dam} & \ac{srm} & \ac{hpa} contract  
    & \ac{dam}/\ac{srm} & \ding{51} & PV/\ac{wf} & \ac{ed}/\ac{td}  
    & \ding{51} & \makecell{Up/down reserve,\\ \ac{ts} reserve allocation}  
    & \makecell{Two-stage \ac{ro},\\ \ac{milp}} \\  

    \bottomrule
  \end{tabular}
  \end{threeparttable}
  \label{table:Literature}
%\end{table*}
\end{sidewaystable}

\subsection{Paper Contributions}
Considering the research gaps in the literature, the main contributions of this paper are outlined as follows:

\begin{itemize}

    \item {\textit{To consider different markets and contracts to trade electrical and thermal energy for the integration of \ac{csp} in \ac{rvpp}:} This paper models the participation of \ac{csp}-based \ac{rvpp} in the \ac{dam} and \ac{srm}, and considers the possibility of trading thermal energy through \ac{hpa}. The participation of \ac{rvpp} in different combinations of these markets and the use of \ac{hpa} are presented, and the profitability of \ac{rvpp} is compared.}

    \item {\textit{To develop a refined two-stage \ac{ro} formulation that considers different uncertainties in the \ac{rvpp} problem:} In this paper, several uncertainties related to the \ac{dam} price, \ac{srm} price, \ac{ndrs} electrical production, \ac{csp} thermal production, and \ac{ed} and \ac{td} consumption are considered using a reformulated two-stage \ac{ro} model. For the first time, the refined \ac{milp} mathematical formulation, incorporating temporal considerations to obtain the worst-case uncertainty for both the source and load sides, is presented.}

    \item {\textit{Energy and reserve provision of \ac{csp} is modeled by considering the energy limitation of its \ac{ts}:} The mathematical model in this paper considers an adjustable share of energy from the energy-limited device \ac{ts} for the provision of up and down reserve. This share of energy is assigned by the optimization problem to maximize \ac{rvpp} profitability.}

    \item {\textit{The proposed two-stage approach for \ac{csp} integration in \ac{vpp} provides high computational efficiency:} Considering several features in the \ac{csp}-based \ac{rvpp} market bidding problem can increase the computational time of the optimization. Some of these features include the number of \ac{rvpp} units, various uncertainties, trading options, reserve provision, and time-coupling constraints of \ac{ts} for \ac{rvpp}. Despite taking into account these features, the proposed \ac{milp} two-stage model maintains the computational efficiency, allowing for multiple economic analyses to be performed before determining the final market bid.}
    
\end{itemize}

\section{Problem Description}
\label{sec:Problem_description}

\begin{figure}[ht!]
    \centering
    \includegraphics[width=1\linewidth]{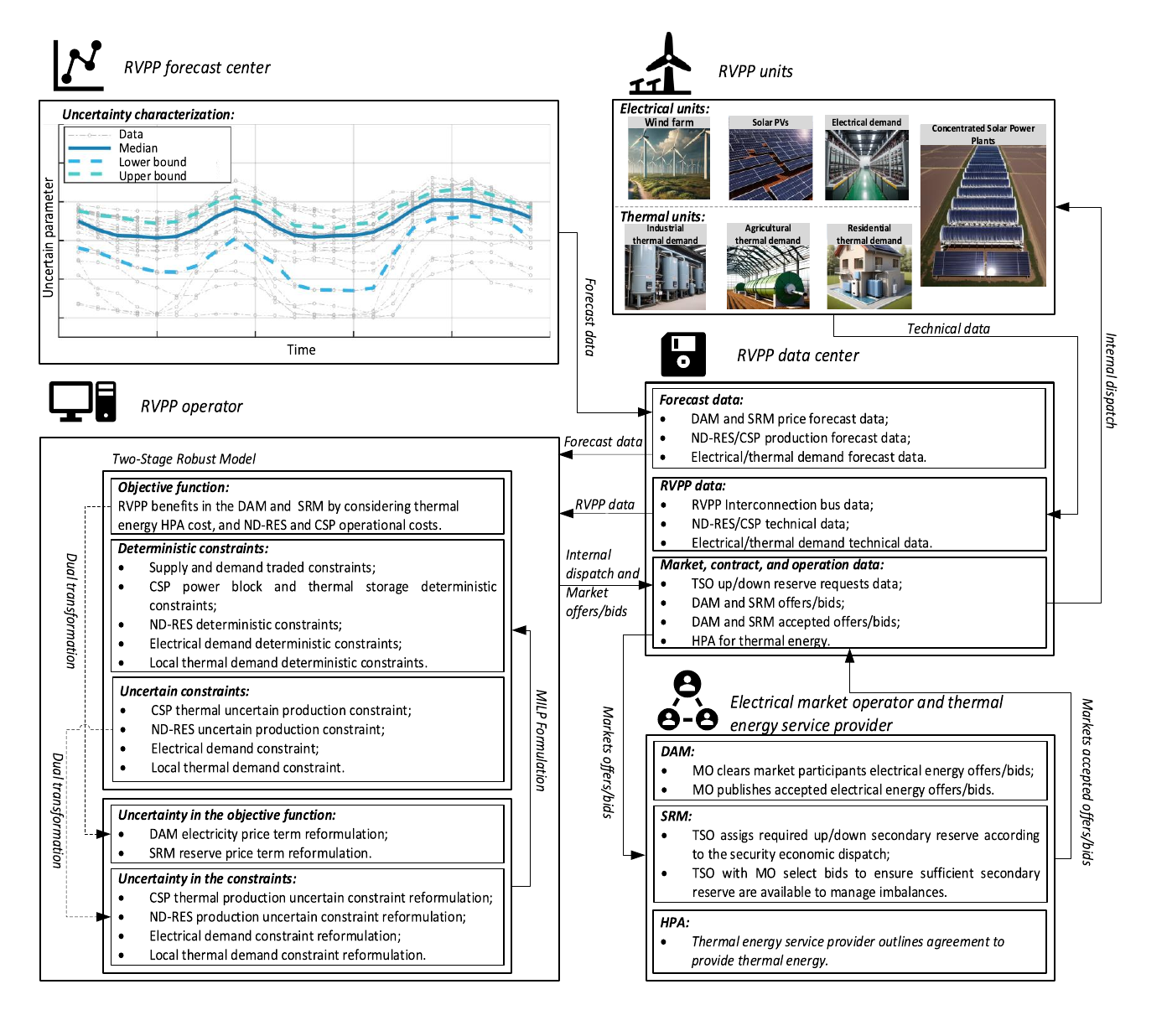}
    \vspace{-6mm}
    \caption{The scheme of considered \ac{rvpp} for participation in electricity markets, as well as the establishment of \ac{hpa} contracts.}
     \label{fig:Scheme_RVPP}
\end{figure}

{\color{black}\acp{rvpp} play a key role in integrating \ac{ndrs} into electricity markets while reliably meeting \ac{ed}/\ac{td} demand, thus maintaining operational and economic viability. This is achieved by operating its units across multiple markets—such as the \ac{dam} and \ac{srm}—and by trading thermal energy, while managing uncertainties from market prices, renewable output, and demand. The main challenge is to determine optimal bidding strategies that maximize \ac{rvpp} profit while ensuring technical feasibility and robustness. Figure~\ref{fig:Scheme_RVPP} presents the structure of the proposed \ac{rvpp} model, showing its components and optimization framework. The \ac{rvpp} may include solar PVs, \acp{wf}, and \acp{csp}, along with local \acp{ed}/\acp{td}. It is connected to the electrical power network to trade electricity in the \ac{dam} and sell reserves in the \ac{srm}. \ac{rvpp} thermal demand can be met locally by the \ac{csp} or externally via the \ac{hpa}, which offers fixed time-based pricing~\cite{hasni2023case}, helping hedge against price fluctuations.}

The electricity \ac{mo} is responsible for clearing the \ac{dam} based on the offers and bids submitted by the \ac{rvpp} and other participants in the electricity markets~\cite{conejo2010decision}. The \ac{tso} assigns the required secondary reserve according to the security economic dispatch to maintain network security. Additionally, the \ac{tso} and \ac{mo} determine the final accepted reserve offers in the \ac{srm} to ensure that sufficient reserves are available to manage real-time imbalances~\cite{fernandes2016participation}. It is assumed that the \ac{rvpp} is a price taker in the market, offering electricity at zero price in the \ac{dam} and \ac{srm} and bidding at a high price for purchasing electricity. This approach increases the likelihood of the \ac{rvpp}'s offers and bids being accepted. Since all \ac{rvpp} units are renewable, with zero production costs and low operational costs, and given that the \ac{rvpp} is relatively small compared to the overall network, this assumption is reasonable and aligns with the integration of renewable resources into the grid.

%\vspace{-3mm}

{\color{black}The \ac{rvpp} forecast center analyzes historical data on key uncertain parameters—including \ac{dam}/\ac{srm} prices, the output of \acp{wf}, PVs, and the \ac{sf} of \ac{csp}, and the consumption of \acp{ed}/\acp{td}—to determine forecast bounds. These bounds are passed to the \ac{rvpp} data center. The \ac{rvpp} operator uses forecasted prices, generation, load, and technical data to manage energy flows by solving the proposed optimization problem. The goal is to maximize the value of electricity and reserve trading in the \ac{dam} and \ac{srm}, while accounting for \ac{hpa} thermal energy costs and the operational costs of \acp{ndrs} and \acp{csp}. Deterministic constraints govern production, consumption, and energy balances, while uncertain constraints cover generation from \acp{ndrs}/\acp{csp} and load forecasts. Price uncertainty in the objective function is modeled using additional constraints.}

Since uncertain parameters can negatively impact the objective function of the \ac{rvpp} operator, a two-stage \ac{ro} model is used to account for their effects in the optimization problem~\cite{li2023robust}. The first stage aims to maximize the benefits of the \ac{rvpp}, while the second stage considers the flexible worst cases of uncertain parameters. The flexibility of the two-stage \ac{ro} model is achieved by introducing an input parameter called the \textit{uncertainty budget} for each source of uncertainty~\cite{bertsimas04}. This parameter determines the number of hours within the entire time period during which uncertain parameters deviate to their worst-case values. Therefore, the \ac{rvpp} operator can adjust the required protection against uncertainty by modifying this parameter. After solving the optimization problem, the \ac{rvpp} operator provides the internal dispatch of its units, the offers/bids for electrical energy and reserve in the \ac{dam} and \ac{srm}, as well as the amount of thermal energy purchased through the \ac{hpa}, to the \ac{rvpp} data center. This information is then transmitted to the relevant entities, including \ac{rvpp} units, \ac{mo} and \ac{tso}, and the thermal energy service provider.

\section{Two-stage Robust Optimization Problem Formulation}
\label{sec:Problem_formulation}

To effectively manage the participation of the \ac{rvpp} in both electricity and reserve markets, as well as the assignment of thermal energy through \ac{hpa} contracts, a comprehensive mathematical formulation is required. The optimization framework must account for the technical characteristics of the units, market mechanisms, and multiple sources of uncertainty. The two-stage \ac{ro} approach effectively captures uncertainty by structuring decisions into two sequential stages: the first stage determines the \ac{rvpp}'s initial market bids and operational plans, while the second stage adjusts to the flexible worst-case realization of uncertain parameters, ensuring that the solution remains feasible and robust against adverse conditions~\cite{li2023robust}. Therefore, this section develops a two-stage \ac{ro} optimization framework to model the \ac{rvpp}'s decision-making process under uncertainty, aiming to maximize its profitability while satisfying operational and market constraints. The deterministic formulation by considering a single value for different uncertain parameters is presented in Section~\ref{subsec:Deterministic_Formulation}. The two-stage robust approach to take into account different uncertainties in the objective function and constraints of optimization problem is provided in Section~\ref{subsec:Robust_Formulation}.

%The final formulation of the optimization problem as an \ac{milp} problem is presented in Section~\ref{subsec:MILP_Formulation}.

%Section~\ref{subsec:Problem_Description} describes the structure of the \ac{rvpp}.

%This section provides the mathematical formulation for the \ac{rvpp} participation in the electricity energy and reserve markets and assigning thermal energy \ac{hpa} contract.

\subsection{Deterministic Formulation}
\label{subsec:Deterministic_Formulation}
The deterministic formulation accounts for the operation and bidding of the \ac{rvpp} in the electrical energy and reserve markets, assuming the \textit{exact} value of uncertain parameters. This section presents the deterministic objective function and constraints of the optimization problem.

\subsubsection{Objective Function}
\label{subsubsec:Objective_Function}
The deterministic objective function, outlined in~\eqref{RVPP: Obj_Deterministic}, aims to maximize the \ac{rvpp}'s benefits in the electrical \ac{dam} and \ac{srm}, while accounting for thermal energy \ac{hpa} costs. The first term of~\eqref{RVPP: Obj_Deterministic} represents the expected revenues from the \ac{rvpp}'s bids in the \ac{dam}, as well as upward and downward \ac{srm}. The second term accounts for the expected costs from purchasing thermal energy through the \ac{hpa}, while the third and fourth terms calculate the operational costs associated with \acp{ndrs} and \acp{csp}.

\vspace{-1em}
\begingroup
\allowdisplaybreaks
\begin{align} \label{RVPP: Obj_Deterministic}
&\mathop {\max }\limits_{{\Xi ^{DA+SR+HPA}}} \sum\limits_{t \in \mathscr{T}} {\left[ {\lambda _t^{DA}p_t^{DA}\Delta t +{\lambda _t^{{SR, \uparrow}}r_t^{SR,\uparrow} } +{\lambda _t^{{SR, \downarrow}}r_t^{SR,\downarrow} }  } \right]}
-\sum\limits_{t \in \mathscr{T}} {{\lambda _t^{HT}h_t^{HT}\Delta t} } \nonumber \\&
- \sum\limits_{t \in \mathscr{T}} {\sum\limits_{r \in \mathscr{R}} {C_rp_{r,t}\Delta t} } - \sum\limits_{t \in \mathscr{T}} {\sum\limits_{\theta \in \Theta} {C_{\theta}(p_{\theta,t}+h_{\theta,t})\Delta t} } 
\end{align}
\endgroup
\vspace{-1em}

\subsubsection{Supply \& Demand Traded Constraints}
\label{subsubsec:Supply_Demand_Constraints}

 The equality constraint governing the supply-demand balance of electrical energy and reserve for \ac{rvpp} units is defined in~\eqref{cons: Supply-Demand1}. This accounts for all possible reserve activation scenarios in real-time, including upward reserve activation, downward reserve activation, and no reserve activation. To model these scenarios, vectors $\boldsymbol{r}_{t}^{SR}=\{r_{t}^{SR,\uparrow}, -r_{t}^{SR, \downarrow},0\}$; $\boldsymbol{r}_{r,t}= \{r_{r,t}^{\uparrow},-r_{r,t}^{\downarrow},0 \}$; $\boldsymbol{r}_{\theta,t}= \{r_{\theta,t}^{\uparrow},-r_{\theta,t}^{\downarrow},0 \}$; and $\boldsymbol{r}_{d,t}= \{r_{d,t}^{\uparrow},-r_{d,t}^{\downarrow},0\}$ are defined for \ac{rvpp}, \ac{ndrs}, \ac{csp}, and \ac{ed}/\ac{td}. 
 Hence,~\eqref{cons: Supply-Demand1} conforms a set of three equations. Equation~\eqref{cons: Supply-Demand1_Heat} defines the equality constraint for thermal energy, where the heat generated by the \ac{csp}, along with the thermal energy purchased through the \ac{hpa}, meets the demand. The upper and lower bounds for the total electrical energy and reserve traded by the \ac{rvpp} are governed by equations~\eqref{cons: Power-Traded1} and~\eqref{cons: Power-Traded2}, respectively. The traded up and down reserve of the \ac{rvpp} is limited to a proportion of its maximum electrical production minus its \acp{ed} capacity, as outlined in constraints~\eqref{cons: Power-Traded3} and~\eqref{cons: Power-Traded4}~\cite{nemati2025segan}.

 %To model these scenarios, general variables $\boldsymbol{r}_{t}^{SR} = \{r_{t}^{SR,\uparrow}, -r_{t}^{SR, \downarrow},0\}$, $\boldsymbol{r}_{r,t} = \{r_{r,t}^{\uparrow},-r_{r,t}^{\downarrow},0 \}$, $\boldsymbol{r}_{\theta,t} = \{r_{\theta,t}^{\uparrow},-r_{\theta,t}^{\downarrow},0 \}$, and $\boldsymbol{r}_{d,t} = \{r_{d,t}^{\uparrow},-r_{d,t}^{\downarrow},0\}$ are defined. 
 %The constraint~\eqref{cons: Power-Traded3} restricts the down-to-up reserve requested by the \ac{tso}, while the traded up reserve of the \ac{rvpp} is capped as a proportion of its maximum production capacity, as outlined in constraint~\eqref{cons: Power-Traded4}.

 %Similarly, the limits on traded heat for the \ac{rvpp} are defined by the maximum heat production capacity of the \ac{csp} and the maximum heat demand, as specified in equations~\eqref{cons: Power-Traded_Heat1} and~\eqref{cons: Power-Traded_Heat2}.

{\color{black}
\vspace{-1em}
\begingroup
\allowdisplaybreaks
\begin{subequations}
\begin{align}
    &\sum\limits_{r \in \mathscr{R}} \left[ p_{r,t} + \boldsymbol{r}_{r,t} \right] + \sum\limits_{\theta \in \Theta} \left[ p_{\theta,t} + \boldsymbol{r}_{\theta,t} \right] - \sum\limits_{d \in \mathscr{D}} \left[ p_{d,t} - \boldsymbol{r}_{d,t} \right] = p_{t}^{DA}+ \boldsymbol{r}_{t}^{SR}~;
    & 
    \forall t \in \mathscr{T} \label{cons: Supply-Demand1} \\
    & \sum\limits_{\theta \in \Theta} h_{\theta,t} +h_{t}^{HT}  = \sum\limits_{d \in \mathscr{D}} h_{d,t} ~;
    & 
    \forall t \in \mathscr{T} \label{cons: Supply-Demand1_Heat} \\
    & p_{t}^{DA} + r_{t}^{SR,\uparrow} \le \sum\limits_{r \in \mathscr{R}} {\bar P_r} + \sum\limits_{\theta \in \Theta} {\bar P_{\theta}}~; 
    & 
    \forall t \in \mathscr{T} \label{cons: Power-Traded1} \\
    & -\sum\limits_{d \in \mathscr{D}} {\bar P_d} \le p_{t}^{DA} - r_{t}^{SR,\downarrow}~; 
    & 
    \forall t \in \mathscr{T} \label{cons: Power-Traded2} \\
%
%    & h_{t}^{HT} \le \sum\limits_{\theta \in \Theta} {\bar H_{\theta}}~; 
%    & 
%    \forall t \label{cons: Power-Traded_Heat1} \\
%
%    & -\sum\limits_{d \in \mathscr{D}} {\bar H_d} \le h_{t}^{HT}~; 
%    & 
%    \forall t \label{cons: Power-Traded_Heat2} \\
%
%    & r_{t}^{SR,\uparrow} = \varrho_t r_{t}^{SR,\downarrow}~; 
%    &
%    \forall t \label{cons: Power-Traded3} \\
%
    & r_{t}^{SR,\uparrow} \le \kappa \left (\sum\limits_{r \in \mathscr{R}} {\bar P_r} + \sum\limits_{\theta \in \Theta} {\bar P_{\theta}} - \sum\limits_{d \in \mathscr{D}} {\bar P_{d}} \right) ~;  
    & 
    \forall t \in \mathscr{T} \label{cons: Power-Traded3}\\
    & r_{t}^{SR,\downarrow} \le \kappa \left (\sum\limits_{r \in \mathscr{R}} {\bar P_r} + \sum\limits_{\theta \in \Theta} {\bar P_{\theta}} - \sum\limits_{d \in \mathscr{D}} {\bar P_{d}} \right) ~;  
    & 
    \forall t \in \mathscr{T} \label{cons: Power-Traded4}
\end{align}
\label{RVPP: Supply-Demand}
\end{subequations}
\endgroup
%\vspace{-1em}
}

\vspace{1em}
\subsubsection{Concentrated Solar Power Plant}
\label{subsubsec:STU}

\begin{figure}[t!]
    \centering
    \includegraphics[width=0.69\linewidth]{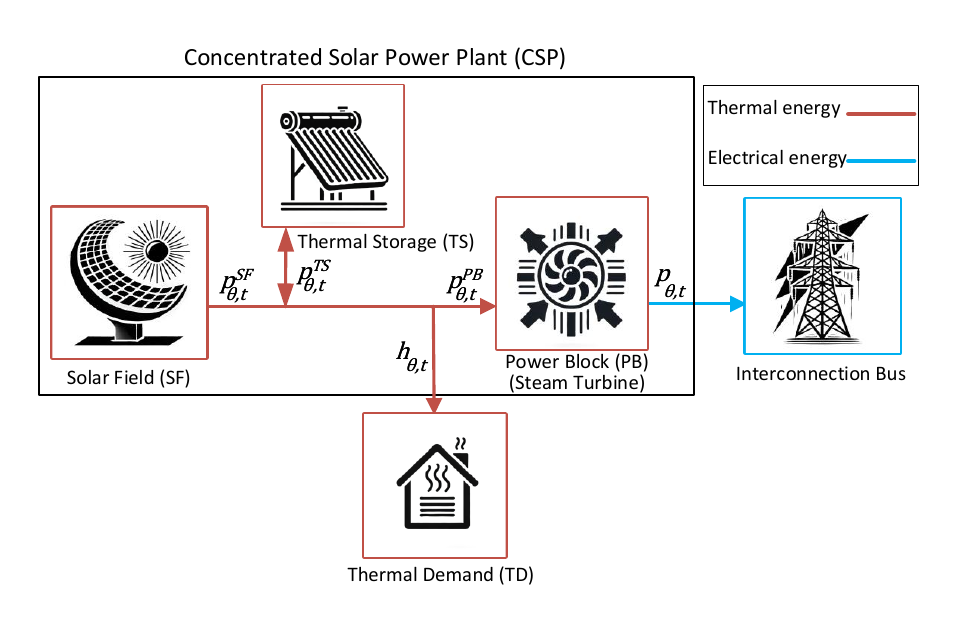}
\vspace{-5mm}
    \caption{Scheme of the \ac{csp} providing electrical and thermal energy.}
     \label{fig:Scheme_STU}
\end{figure}

\acp{csp} are synchronous power plants that harness sunlight to produce thermal energy, which can be utilized for electricity generation, residential heating, or industrial applications. \acp{ptc} consist of long, curved mirrors that concentrate sunlight onto a tube containing a heat-transfer fluid, which absorbs the heat and facilitates energy transfer in the \ac{sf}. The conversion of thermal energy into electrical power occurs in the \ac{pb} of the \ac{csp}, where steam generated by the heated fluid drives a turbine. Additionally, molten salt \ac{ts} enable \ac{csp} to store heat for several hours, allowing for electricity production or heat supply even after sunset~\cite{garcia2011performance}. As a note, the \ac{ts} can only be charged if energy is available at the \ac{sf}, i.e., it cannot be charged by consuming (\textit{purchasing}) energy from the grid. Figure~\ref{fig:Scheme_STU} illustrates the \ac{csp} configuration proposed in this paper, which can deliver both electricity (including energy and reserve) and thermal energy. Compared to \acp{ndrs}, \acp{csp} introduce additional complexities in uncertainty modeling, as solar resource variability affects thermal energy production within the \ac{sf}. Furthermore, the conversion process from thermal to electrical energy requires explicit modeling while considering reserve provision. Additionally, the presence of \ac{ts} adds further complexity due to the need to account for different states of charging and discharging. The model must ensure that reserve provision does not compromise the availability of stored energy in the \ac{ts} for later use. These interdependencies make the modeling of \acp{csp} significantly more intricate than that of \ac{ndrs} assets.

\vspace{2mm}
\textit{A. \ac{csp} Power Block Constraints}

{\color{black}The formulation for converting thermal energy into electrical energy in the \ac{pb} of the \ac{csp}, with explicit consideration of reserve provision, is provided in~\eqref{Deterministic: STU}~\cite{alvaroPOSYFT}. This set of constraints models the interactions between the thermal subsystem (\ac{sf}, \ac{ts}, and turbine) and the electrical output. Specifically, constraint~\eqref{Deterministic: STU1} defines the operational bounds of the thermal power output from the \ac{sf}, which depend on a fixed realization of the uncertain \ac{sf} power (${P}_{\theta,t}^{SF}$). Constraint~\eqref{Deterministic: STU2} captures the core thermal power balance feeding into the \ac{pb}, linking it to several components, as illustrated in Figure~\ref{fig:Scheme_STU}. These components include the thermal power supplied by the \ac{sf} ($p_{\theta,t}^{SF}$), the charging and discharging power of the \ac{ts} ($p_{\theta,t}^{TS,+}$ and $p_{\theta,t}^{TS,-}$), the thermal power allocated to local \acp{td} ($h_{\theta,t}$), adjusted by the efficiency of thermal energy provision to \acp{td} ($\eta_{\theta}$), and the thermal losses associated with turbine startup ($K_{\theta}^{PB}$). This constraint demonstrates the first key coupling between thermal inflows and their effect on the net thermal energy available for electricity generation. The thermal power and reserve supplied to the \ac{pb} of the \ac{csp} are constrained by the \ac{pb}'s upper and lower power capacity ($\bar P_{\theta}^{PB}$ and $\ubar P_{\theta}^{PB}$) and the turbine's commitment status, represented by the binary variable $u_{\theta,t}$, as outlined in~\eqref{Deterministic: STU3}–\eqref{Deterministic: STU4}. These bounds ensure that the turbine’s operation aligns with its physical limitations and operational status. The commitment status of the \ac{pb} and its associated constraints, including the minimum up/down time requirements, are detailed in the supplementary constraints provided in Appendix A. The electrical power ($p_{\theta,t}$) and reserve output ($\boldsymbol{r}_{\theta,t}$) from the \ac{pb} are calculated in~\eqref{Deterministic: STU5}, which uses a piecewise linear function to model thermal-to-electric conversion efficiency. If the total thermal power and reserve ($p_{\theta,t}^{PB} + \boldsymbol{r}_{\theta,t}^{TS}$) injected into the \ac{pb} falls below a specified threshold, the turbine cannot turn on, resulting in zero output, as defined by the first term on the right-hand side of~\eqref{Deterministic: STU5}. Once the injected thermal energy exceeds this threshold, the turbine begins generating electricity according to the piecewise linear segments represented by the second term on the right-hand side of~\eqref{Deterministic: STU5}. The efficiency $\eta_{\theta,n}$ of each segment in the piecewise linear function varies based on the thermal power and reserve injected into the \ac{pb}. The real-time reserve activation scenarios for the \ac{ts}, encompassing upward reserve activation, downward reserve activation, and no reserve activation, are represented by the vector $\boldsymbol{r}_{\theta,t}^{TS}= \{r_{\theta,t}^{TS,\uparrow},-r_{\theta,t}^{TS,\downarrow},0 \}$. The vector $\boldsymbol{r}_{\theta,t}= \{r_{\theta,t}^{\uparrow},-r_{\theta,t}^{\downarrow},0 \}$ represents the different states of reserve activation for the final reserve supplied by the \ac{csp} (see \eqref{cons: Supply-Demand1}). This formulation illustrates the second major thermal-electrical coupling, where thermal reserve activation directly impacts potential electrical reserve output. Constraints~\eqref{Deterministic: STU6}–\eqref{Deterministic: STU7} set upper limits on the electrical upward and downward reserves ($r_{\theta,t}^{\uparrow}$ and $r_{\theta,t}^{\downarrow}$) the \ac{csp} can supply to the system. Finally, the binary nature of the turbine commitment variables $u_{\theta,t}$ and $v_{\theta,t}^{SU}$ is enforced in~\eqref{Deterministic: STU8}, supporting the logical consistency of thermal operation states. Collectively, the set of constraints~\eqref{Deterministic: STU} ensures that the thermal decision variables—whether related to thermal power or reserve—are internally consistent with the electrical output and market participation framework, thereby establishing a valid thermal-electrical coupling within the model.}

{\color{black}
\begingroup
\allowdisplaybreaks
\begin{subequations}
\begin{align}
    & 0 \leq p_{\theta,t}^{SF} \leq {P}_{\theta,t}^{SF}~; 
    & \forall \theta, t \in {\Theta},\mathscr{T} \label{Deterministic: STU1} \\
    &p_{\theta,t}^{PB} %+\boldsymbol{r}_{\theta,t}^{PB} 
    = p_{\theta,t}^{SF} - \frac{h_{\theta,t}}{\eta_{\theta}} + p_{\theta,t}^{TS, -} - p_{\theta,t}^{TS, +} %+\boldsymbol{r}_{\theta,t}^{TS}
    - K_{\theta}^{PB} v_{\theta,t}^{SU} \bar P_{\theta}^{PB}~; 
    &\forall \theta, t \in {\Theta},\mathscr{T} \label{Deterministic: STU2} \\
    &p_{\theta,t}^{PB} + r_{\theta,t}^{TS,\uparrow} \leq \bar P_{\theta}^{PB} u_{\theta,t}~; 
    &\forall \theta, t \in {\Theta},\mathscr{T} \label{Deterministic: STU3} \\
    &\ubar P_{\theta}^{PB} u_{\theta,t} \leq p_{\theta,t}^{PB} - r_{\theta,t}^{TS,\downarrow}~; 
    &\forall \theta, t \in {\Theta},\mathscr{T} \label{Deterministic: STU4} \\
    & p_{\theta,t} + \boldsymbol{r}_{\theta,t} =
    \begin{cases} 
        \begin{aligned}
         &0,  \qquad  \qquad \quad \quad if \quad  0 \leq p_{\theta,t}^{PB} + \boldsymbol{r}_{\theta,t}^{TS} \leq P_{\theta,n=1}^{PB}~; \qquad  \forall \theta, t \in {\Theta},\mathscr{T} \\
         &\eta_{\theta,n} (p_{\theta,t}^{PB} + \boldsymbol{r}_{\theta,t}^{TS}),  \: \: if \quad   P_{\theta,n-1}^{PB} \leq p_{\theta,t}^{PB} + \boldsymbol{r}_{\theta,t}^{TS} \leq P_{\theta,n}^{PB}~; \quad \forall \theta, t, n \in {\Theta},\mathscr{T},\mathscr{N}, n\backslash\{0,1\} \hspace{-2cm} 
        \end{aligned}
    \end{cases} \label{Deterministic: STU5} \\
    & r_{\theta,t}^{\uparrow} \le T^{SR} \bar R_{\theta}^{SR}~;
    & 
    \forall \theta,t \in {\Theta},\mathscr{T}  \label{Deterministic: STU6} \\
    & r_{\theta,t}^{\downarrow} \le T^{SR} \ubar R_{\theta}^{SR}~;
    & 
    \forall \theta,t \in {\Theta},\mathscr{T}  \label{Deterministic: STU7}  \\
    & u_{\theta,t}, v_{\theta,t}^{SU} \in \{0,1\}~; 
    &\forall \theta, t \in {\Theta},\mathscr{T} \label{Deterministic: STU8}
    \end{align} 
\label{Deterministic: STU}
\end{subequations}
\endgroup
}

{\color{black}The conditional piecewise function in~\eqref{Deterministic: STU5} must be reformulated as a set of linear constraints to ensure compatibility with the \ac{milp} programming framework. To achieve this, \ac{sos-2} constraints are introduced in~\eqref{STU_SOS}, which allow the piecewise linear relationship to be modeled accurately and efficiently within the optimization problem. Constraint~\eqref{STU_SOS1} ensures that the input thermal power and reserve variables lie within the breakpoints of each piecewise segment of the \ac{pb}, thereby defining the convex combination of these breakpoints. Constraint~\eqref{STU_SOS2} defines the output electrical power and reserve variables as weighted sums of the corresponding efficiencies and breakpoints. Constraints~\eqref{STU_SOS3}–\eqref{STU_SOS6} enforce the \ac{sos-2} property, guaranteeing that only two adjacent breakpoints can have positive weights at any time, which preserves the piecewise linearity and ensures solution feasibility. The vectors $\boldsymbol x_{\theta,t,n} = \{x_{\theta,t,n}^{\uparrow},x_{\theta,t,n}^{\downarrow}, x_{\theta,t,n} \}$ and $\boldsymbol \varsigma_{\theta,t,n}  = \{\varsigma_{\theta,t,n}^{\uparrow},\varsigma_{\theta,t,n}^{\downarrow}, \varsigma_{\theta,t,n} \}$ represent variables associated with the different reserve activation scenarios. Finally, constraints~\eqref{STU_SOS7} and~\eqref{STU_SOS8} define the nature of the continuous and binary variables, respectively, ensuring the correct formulation of the \ac{sos-2} structure.}

%The conditional piecewise function~\eqref{Deterministic: STU5} must be expressed as mathematical constraints to be compatible with the \ac{milp} programming framework. To achieve this, \ac{sos-2} constraints~\eqref{STU_SOS} are introduced. Constraint~\eqref{STU_SOS1} ensures that the input variables related to the thermal energy and reserve lie within the breakpoints of each piecewise segment of \ac{pb}. Constraint~\eqref{STU_SOS2} defines the output variables related to the electrical energy and reserve of \ac{csp}. Constraints~\eqref{STU_SOS3}-\eqref{STU_SOS6} enforce the \ac{sos-2} property by ensuring that the solution falls between two consecutive breakpoints. The vectors $\boldsymbol x_{\theta,t,n} = \{x_{\theta,t,n}^{\uparrow},x_{\theta,t,n}^{\downarrow}, x_{\theta,t,n} \}$ and $\boldsymbol \varsigma_{\theta,t,n}  = \{\varsigma_{\theta,t,n}^{\uparrow},\varsigma_{\theta,t,n}^{\downarrow}, \varsigma_{\theta,t,n} \}$ are defined for the different reserve activation scenarios. Constraints~\eqref{STU_SOS7}-\eqref{STU_SOS8} specify the nature of the positive and binary variables, respectively.

%The vectors $\boldsymbol x_{\theta,t,n}$ and $\boldsymbol \varsigma_{\theta,t,n}$ are defined for the different reserve activation scenarios. Constraints~\eqref{STU_SOS7}-\eqref{STU_SOS8} specify the nature of the positive and binary variables, respectively.

{\color{black}
\begingroup
\allowdisplaybreaks
\begin{subequations}
\begin{align}
    & p_{\theta,t}^{PB} + \boldsymbol{r}_{\theta,t}^{TS} = \sum_{n \in \mathscr{N}} P_{\theta,n}^{PB} \boldsymbol x_{\theta,t,n}~; 
    &\forall \theta, t \in {\Theta},\mathscr{T} \label{STU_SOS1} \\
    & p_{\theta,t} + \boldsymbol{r}_{\theta,t} = \sum_{n \in \mathscr{N}} \eta_{\theta,n} P_{\theta,n}^{PB} \boldsymbol x_{\theta,t,n}~; 
    &\forall \theta, t \in {\Theta},\mathscr{T} \label{STU_SOS2} \\
    & \sum_{n \in \mathscr{N}} \boldsymbol x_{\theta,t,n} = 1~; 
    &\forall \theta, t \in {\Theta},\mathscr{T} \label{STU_SOS3} \\
    & \sum_{n \in \mathscr{N}} \boldsymbol \varsigma_{\theta,t,n} \leq 2~; 
    &\forall \theta, t \in {\Theta},\mathscr{T} \label{STU_SOS4} \\
    & \boldsymbol x_{\theta,t,n} \leq \boldsymbol \varsigma_{\theta,t,n}~; 
    &\forall \theta, t, n \in {\Theta},\mathscr{T},\mathscr{N} \label{STU_SOS5} \\
    & \boldsymbol \varsigma_{\theta,t,n} + \boldsymbol \varsigma_{\theta,t,n^\prime} \leq 1~; 
    &\forall \theta, t, n \in {\Theta},\mathscr{T},\mathscr{N}, n \leq |\mathscr{N}| - 3, n^{\prime} \geq n + 2 \label{STU_SOS6} \\
    & \boldsymbol x_{\theta,t,n} \geq 0~; 
    &\forall \theta, t, n \in {\Theta},\mathscr{T},\mathscr{N} \label{STU_SOS7} \\
    & \boldsymbol \varsigma_{\theta,t,n} \in \{0,1\}~; 
    &\forall \theta, t, n \in {\Theta},\mathscr{T},\mathscr{N} \label{STU_SOS8} 
\end{align}
\label{STU_SOS}
\end{subequations}
\endgroup
}

\textit{B. \ac{csp} Thermal Storage Constraints}

{\color{black}The \ac{ts} system of the \ac{csp} unit is modeled to simultaneously support both thermal energy delivery and reserve provision, using a set of operational and energy balance constraints defined in~\eqref{Deterministic: TS}. To mitigate potential infeasibility issues in the operation of the \ac{ts}, the model—based on the approach in~\cite{alvaroPOSYFT}—allocates a specific portion of the \ac{ts}'s energy capacity exclusively for upward $(\uparrow)$ and downward $(\downarrow)$ reserve provision. Constraints~\eqref{Deterministic: TS1}-\eqref{Deterministic: TS4} define the operational bounds for thermal charging $(+)$ and discharging $(-)$, while incorporating the corresponding upward and downward reserve contributions. The \ac{ts} can operate in only one mode at a time—either charging (storing thermal energy) or discharging (release stored energy)—as enforced by the binary variable $u_{\theta,t}^{TS}$, which takes the value 1 for charging and 0 for discharging. Specifically, constraints~\eqref{Deterministic: TS1}-\eqref{Deterministic: TS2} define the allowable charging power range $p_{\theta,t}^{TS,+}$, adjusted for the upward or downward reserve allocated during charging (i.e., $r_{\theta,t}^{TS,+,\uparrow}$, $r_{\theta,t}^{TS,+,\downarrow}$). Similarly, constraints~\eqref{Deterministic: TS3}-\eqref{Deterministic: TS4} define limits for discharging power $p_{\theta,t}^{TS,-}$, accounted for both upward and downward reserve provision in that mode (i.e., $r_{\theta,t}^{TS,-,\uparrow}$, $r_{\theta,t}^{TS,-,\downarrow}$). These constraints ensure that the \ac{ts} does not exceed its power capacity limits during either mode of operation, even when reserves are provided. Because the \ac{ts} may contribute reserves in both charging and discharging states, the total upward and downward reserves $r_{\theta,t}^{TS,\uparrow}$ and $r_{\theta,t}^{TS,\downarrow}$ are expressed as the sum of the contributions from both modes. Constraint~\eqref{Deterministic: TS5} shows that total upward reserve is the sum of reserve available during discharging and charging. Constraint~\eqref{Deterministic: TS6} is similarly presented for provided total downward reserve by \ac{ts}. This flexible formulation allows the \ac{ts} to participate in reserve ancillary services regardless of its charging state, enhancing its system value. Constraints~\eqref{Deterministic: TS7}–\eqref{Deterministic: TS8} ensure thermal energy balance and proper cycling within the storage tank. Constraint~\eqref{Deterministic: TS7} captures the energy balance over time, where the energy level is updated based on the charging power ($p_{\theta,t}^{TS,+}$), discharging power ($p_{\theta,t}^{TS,-}$), their respective efficiencies ($\eta_\theta^{TS,+}$ and $\eta_\theta^{TS,-}$), and the time step size ($\Delta t$). To complement~\eqref{Deterministic: TS7}, equation~\eqref{Deterministic: TS8} ensures that the state of charge is equal at the initial and final periods. This constraint is relevant for \ac{dam} operation and scheduling of representative periods, as it maintains the \ac{ts} energy level at the same state at the start of each cycle, avoiding any bias in the schedule caused by depleting or overfilling the tank near the end of the horizon. To prevent conflicts between energy commitments and reserve activation, a share of the thermal energy capacity is explicitly reserved. Constraints~\eqref{Deterministic: TS9} and~\eqref{Deterministic: TS10} limit the total energy that can be used for upward and downward reserve over the horizon to a fraction of the usable capacity. These fractions are represented by the variables $\sigma_{\theta}^{TS,\uparrow}$ and $\sigma_{\theta}^{TS,\downarrow}$, and they prevent overcommitting reserves without sufficient energy backup. Based on the energy allocated for reserve provision, the operational energy limits of the \ac{ts} are established in~\eqref{Deterministic: TS11}, ensuring that energy allocated for reserves remains available by effectively separating energy used for power and reserve. Finally, constraint \eqref{Deterministic: TS12} enforces the binary nature of the operating mode variable $u_{\theta,t}^{TS}$, which is essential to ensure mutually exclusive charging and discharging states.}

{\color{black}
\begingroup
\allowdisplaybreaks
\begin{subequations}
\begin{align}
    &\ubar {P}_{\theta}^{TS,+} u_{\theta,t}^{TS} \leq p_{\theta,t}^{TS,+} - r_{\theta,t}^{TS,+,\uparrow}~; 
    &\forall \theta, t \in {\Theta},\mathscr{T} \label{Deterministic: TS1} \\
    & p_{\theta,t}^{TS,+} + r_{\theta,t}^{TS,+,\downarrow} \leq \bar {P}_{\theta}^{TS,+} u_{\theta,t}^{TS}~;
    &\forall \theta, t \in {\Theta},\mathscr{T} \label{Deterministic: TS2} \\
    &p_{\theta,t}^{TS,-} + r_{\theta,t}^{TS,-,\uparrow} \leq \bar {P}_{\theta}^{TS,-} \left( 1 - u_{\theta,t}^{TS} \right)~; 
    &\forall \theta, t \in {\Theta},\mathscr{T} \label{Deterministic: TS3} \\
    &\ubar {P}_{\theta}^{TS,-} \left( 1 - u_{\theta,t}^{TS} \right) \leq p_{\theta,t}^{TS,-} - r_{\theta,t}^{TS,-,\downarrow}~; 
    &\forall \theta, t \in {\Theta},\mathscr{T} \label{Deterministic: TS4} \\
    &r_{\theta,t}^{TS,\uparrow} = r_{\theta,t}^{TS,+,\uparrow} + r_{\theta,t}^{TS,-,\uparrow}~; 
    &\forall \theta, t \in {\Theta},\mathscr{T} \label{Deterministic: TS5} \\
    &r_{\theta,t}^{TS,\downarrow} = r_{\theta,t}^{TS,+,\downarrow} + r_{\theta,t}^{TS,-,\downarrow}~; 
    &\forall \theta, t \in {\Theta},\mathscr{T} \label{Deterministic: TS6} \\
    & e_{\theta,t}^{TS} = e_{\theta,t-1}^{TS} + p_{\theta,t}^{TS,+} \eta_\theta^{TS,+} \Delta t - \frac{p_{\theta,t}^{TS,-} \Delta t}{\eta_{\theta}^{TS,-}}~; 
    &\forall \theta ,t \in {\Theta},\mathscr{T}, t \backslash\{1\} \label{Deterministic: TS7} \\
    & e_{\theta,1}^{TS} = e_{\theta,t=T}^{TS}~;% = \alpha_\theta^{TS}, 
    &\forall \theta \in {\Theta} \label{Deterministic: TS8} \\
    & \sum_{t \in \mathscr{T}} \frac{r_{\theta,t}^{TS,\uparrow} \Delta t} {\eta_\theta^{TS,-}} \leq \sigma_{\theta}^{TS,\uparrow} \left( \bar E_\theta^{TS} - \ubar E_\theta^{TS} \right)~; 
    &\forall \theta \in {\Theta} \label{Deterministic: TS9} \\
    & \sum_{t \in \mathscr{T}} r_{\theta,t}^{TS,\downarrow} \eta_\theta^{TS,+} \Delta t \leq \sigma_{\theta}^{TS,\downarrow} \left( \bar E_\theta^{TS} - \ubar E_\theta^{TS} \right)~; 
    &\forall \theta \in {\Theta} \label{Deterministic: TS10} \\
    & \ubar E_\theta^{TS} + \sigma_\theta^{TS,\downarrow} \left( \bar E_\theta^{TS} - \ubar E_\theta^{TS} \right) \leq e_{\theta,t}^{TS} \leq \bar E_{\theta}^{TS} - \sigma_{\theta}^{TS,\downarrow} \left( \bar E_{\theta}^{TS} - \ubar E_{\theta}^{TS} \right)~; 
    &\forall \theta, t \in {\Theta},\mathscr{T} \label{Deterministic: TS11} \\
    & u_{\theta,t}^{TS} \in \{0,1\}~; 
    &\forall \theta, t \in {\Theta},\mathscr{T} \label{Deterministic: TS12}
\end{align}
\label{Deterministic: TS}
\end{subequations}
\endgroup
}

\subsubsection{ND-RES Constraints}
\label{subsubsec:ND-RES_Constraints}

The constraints related to \acp{ndrs} are formulated in~\eqref{RVPP: NDRES}. The upper and lower bounds for the output energy and reserve of \acp{ndrs}, using the precise value for the uncertain parameter, are defined in constraints~\eqref{cons: NDRES1} and~\eqref{cons: NDRES2}, respectively. It is important to highlight that \acp{ndrs}, such as \acp{wf} and solar PVs, have the potential to provide reserve in the \ac{srm}, subject to their technical capabilities~\cite{zhang2022frequency, yin2021state}. Upward reserve can be supplied when \acp{ndrs} are operated at a curtailed level and then increase their generation during reserve activation. In contrast, downward reserve is provided by reducing their output power from the existing operational level. The up and down reserve provision capabilities by \acp{ndrs} are constrained by equations~\eqref{cons: NDRES3} and~\eqref{cons: NDRES4}, respectively.

{\color{black}
\vspace{-1em}
\begingroup
\allowdisplaybreaks
\begin{subequations}
\begin{align}
    & p_{r,t}+r_{r,t}^{\uparrow} \leq P_{r,t}~; 
    & 
    \forall r,t \in \mathscr{R},\mathscr{T} \label{cons: NDRES1} \\ 
    & \ubar P_{r} \le p_{r,t}-r_{r,t}^{\downarrow}~; 
    & 
    \forall r,t \in \mathscr{R},\mathscr{T}  \label{cons: NDRES2} \\ 
    & r_{r,t}^{\uparrow} \le T^{SR} \ubar R_{r}^{SR}~;
    & 
    \forall r,t \in \mathscr{R},\mathscr{T} \label{cons: NDRES3} \\
    & r_{r,t}^{\downarrow} \le T^{SR} \bar R_{r}^{SR}~;
    & 
    \forall r,t \in \mathscr{R},\mathscr{T} \label{cons: NDRES4}
    \end{align}
\label{RVPP: NDRES}
\end{subequations}
\endgroup
\vspace{-1em}
}

\subsubsection{Demand Constraints}
\label{subsubsec:Demand_Constraints}
The constraints for flexible \acp{ed} and local \acp{td} are provided in~\eqref{RVPP: Demand}. Constraints~\eqref{cons: Demand1} and~\eqref{cons: Demand_Heat1} set fixed values for the uncertainties associated with \acp{ed}/\acp{td}, respectively. The lower and upper bounds for flexible \acp{ed}, considering reserve provision, are defined in~\eqref{cons: Demand2} and~\eqref{cons: Demand3}, respectively. The bound for \acp{td} are determined by the capacity of the thermal pipe connected to the demand, as specified in~\eqref{cons: Demand_Heat2}. In this study, \acp{ed} are allowed a certain percentage of flexibility relative to its predefined profile, which is allocated for up and down reserve provision according to~\eqref{cons: Demand4} and~\eqref{cons: Demand5}, respectively~\citep{oladimeji2022modeling}. The up and down reserves provided by \acp{ed} are further constrained by the demand's ability to provide reserve, as specified in~\eqref{cons: Demand6} and~\eqref{cons: Demand7}. Additionally, the minimum energy consumption over the time horizon for both \acp{ed}/\acp{td} is constrained by~\eqref{cons: Demand8} and~\eqref{cons: Demand_Heat3}, respectively.

{\color{black}
\vspace{-1em}
\begingroup
\allowdisplaybreaks
\begin{subequations}
\begin{align}
    & p_{d,t} \geq P_{d,t}~;
    & 
    \forall d,t \in \mathscr{D},\mathscr{T}  \label{cons: Demand1}\\
    & h_{d,t} \geq H_{d,t}~;
    & 
    \forall d,t \in \mathscr{D},\mathscr{T}  \label{cons: Demand_Heat1}\\
    & \ubar P_{d} \le p_{d,t} - r_{d,t}^{\uparrow}~;
    &
    \forall d,t \in \mathscr{D},\mathscr{T} \label{cons: Demand2} \\
    & p_{d,t} + r_{d,t}^{\downarrow} \le \bar P_{d}~;
    & 
    \forall d,t \in \mathscr{D},\mathscr{T} \label{cons: Demand3}\\
    & %\ubar H_{d} \le
    h_{d,t}  \le \bar H_{d}~;
    & 
    \forall d,t \in \mathscr{D},\mathscr{T} \label{cons: Demand_Heat2}\\
    & r_{d,t}^{\uparrow} \le \ubar\beta_{d,t} p_{d,t}~;
    &
    \forall d,t \in \mathscr{D},\mathscr{T} \label{cons: Demand4} \\
    & r_{d,t}^{\downarrow} \le \bar\beta_{d,t} p_{d,t}~;
    & 
    \forall d,t \in \mathscr{D},\mathscr{T} \label{cons: Demand5} \\
    & r_{d,t}^{\uparrow} \le T^{SR} \ubar R_{d}^{SR}~;
    & 
    \forall d,t \in \mathscr{D},\mathscr{T} \label{cons: Demand6} \\
    & r_{d,t}^{\downarrow} \le T^{SR} \bar R_{d}^{SR}~;
    & 
    \forall d,t \in \mathscr{D},\mathscr{T} \label{cons: Demand7} \\
    & \ubar E_{d} \le \sum\limits_{t \in \mathscr{T}} \left[{p_{d,t}\Delta t} - r_{d,t}^{\uparrow} \right]~;
    &
    \forall d \in \mathscr{D} \label{cons: Demand8}\\
    & \ubar Q_{d} \le \sum\limits_{t \in \mathscr{T}} {h_{d,t}\Delta t}~;
    &
    \forall d \in \mathscr{D} \label{cons: Demand_Heat3}
%
%    & u_{d,p} \in \left\{ {0,1} \right\}~; 
%    &
%    \forall d,p \label{cons: Demand9} 
%    & \sum\limits_{p \in \mP} {u_{d,p}} = 1~; 
%    &
%    \forall d\phantom{,t} \label{cons: Demand2}\\
\end{align}
\label{RVPP: Demand}
\end{subequations}
\endgroup
\vspace{-1em}
}

\subsubsection{Remarks}
The optimization problem presented in~\eqref{RVPP: Obj_Deterministic}-\eqref{RVPP: Demand} is a deterministic model for \ac{rvpp} participation in the electrical energy and reserve markets, as well as for establishing a thermal \ac{hpa} contract. In the next section, the proposed formulation is developed to incorporate various uncertainties into the optimization problem.

\subsection{Robust Formulation}
\label{subsec:Robust_Formulation}

The deterministic formulation presented in Section~\ref{subsec:Deterministic_Formulation} neglects the variation of uncertain parameters. However, uncertainty in electricity energy and reserve market prices can negatively (or positively) affect the \ac{rvpp} profit. Additionally, uncertainties in the production and demand of \ac{rvpp} units can lead to reduced production or increased demand, thus impacting \ac{rvpp} market profits. Therefore, the \ac{rvpp} operator must account for the effect of various uncertainties in its decision-making process. In the following section, the optimization problem of the \ac{rvpp} is extended to incorporate these variations.

\subsubsection{Two-Stage Robust Model}
\label{subsubsec:Two-Stage_Robust}
A two-stage \ac{ro} problem is formulated in~\eqref{Two-Stage_Robust} as a max-min problem to account for various uncertain parameters in the objective function and constraints of the optimization problem. In the first stage, the \ac{rvpp} operator maximizes its objective function~\eqref{Robust1: Obj_Robust}, which is analogous to~\eqref{RVPP: Obj_Deterministic}. The set of first-stage decision variables, ${\Xi ^{DA+SR+HPA}}$, is also similar to those in the deterministic model. In the second stage, uncertainty adversely impacts the electricity energy and reserve prices in the objective function~\eqref{Robust1: Obj_Robust}. This is modeled through the minimization in the inner problem, which is defined for the objective function's uncertainty set $\Xi ^{O} = \{ \lambda _t^{DA}, \lambda _t^{{SR, \downarrow}}, \lambda _t^{{SR, \uparrow}} \}$. Additionally, uncertainty is modeled to potentially decrease the thermal production of the \ac{sf} of \acp{csp} and the electrical production of \acp{ndrs}, while increasing the consumption of \acp{ed}/\acp{td}, as reflected in constraints~\eqref{Robust1: STU1_Robust}-\eqref{Robust1: Demand_Heat1_Robust}. The decision variables associated with uncertainty in these constraints are defined by the set $\Xi ^{C} = \{ {P}_{\theta,t}^{SF}, {P}_{r,t}, {P}_{d,t}, {H}_{d,t} \}$. It is important to note that, unlike the deterministic problem, the new sets $\Xi ^{O}$ and $\Xi ^{C}$ include second-stage decision variables, which were parameters in the deterministic problem. The constraints of the deterministic problem described in Section~\ref{subsec:Deterministic_Formulation} that are unaffected by uncertainty are defined in~\eqref{Robust1: Other_Cons_Det}.

{\color{black}
\vspace{-1em}
\begingroup
\allowdisplaybreaks
\begin{subequations}
\begin{align} \label{Robust1: Obj_Robust}
&\mathop {\max}\limits_{{\Xi ^{DA+SR+HPA}}} \scalebox{1.5}{\Bigg\{ } \mathop {\min}\limits_{{\Xi ^{O}}} \scalebox{1.2}{\Bigg\{ } \sum\limits_{t \in \mathscr{T}} {\left[ {\lambda _t^{DA}p_t^{DA}\Delta t +{\lambda _t^{{SR, \uparrow}}r_t^{SR,\uparrow} } +{\lambda _t^{{SR, \downarrow}}r_t^{SR,\downarrow} }  } \right]} \nonumber \\ & 
-\sum\limits_{t \in \mathscr{T}} {{\lambda _t^{HT}h_t^{HT}\Delta t} }
- \sum\limits_{t \in \mathscr{T}} {\sum\limits_{r \in \mR} {{C_r}p_{r,t}\Delta t} } - \sum\limits_{t \in \mathscr{T}} {\sum\limits_{\theta \in \Theta} {C_{\theta}(p_{\theta,t}+h_{\theta,t})\Delta t} } \scalebox{1.2}{\Bigg\} } \scalebox{1.5}{\Bigg\} } \\
    \nonumber\text{st.} \\
    & p_{\theta,t}^{SF} \leq \mathop {\min}\limits_{{{P}_{\theta,t}^{SF}}} \left\{ {P}_{\theta,t}^{SF} \right\}~;
    \hspace{22.8em} \forall \theta, t \in {\Theta},\mathscr{T} \label{Robust1: STU1_Robust} \\
    & p_{r,t}+r_{r,t}^{\uparrow} \leq \mathop {\min}\limits_{{{P}_{r,t}}} \left\{ P_{r,t} \right\}~;  
    \hspace{21.5em} \forall r,t \in \mathscr{R},\mathscr{T} \label{Robust1: NDRES1_Robust} \\ 
    & p_{d,t} \geq - \mathop {\min}\limits_{{{P}_{d,t}}} \left\{ -P_{d,t} \right\}~; 
    \hspace{21.8em} \forall d,t \in \mathscr{D},\mathscr{T}   \label{Robust1: Demand1_Robust}\\
    & h_{d,t} \geq - \mathop {\min}\limits_{{{H}_{d,t}}} \left\{ -H_{d,t} \right\}~; 
    \hspace{21.8em} \forall d,t \in \mathscr{D},\mathscr{T}   \label{Robust1: Demand_Heat1_Robust}\\
    &\eqref{RVPP: Supply-Demand}, \eqref{Deterministic: STU2}-\eqref{Deterministic: STU4}, \eqref{Deterministic: STU6}-\eqref{Deterministic: STU8}, \eqref{STU_SOS}, \eqref{Deterministic: TS}, \eqref{cons: NDRES2}-\eqref{cons: NDRES4}, \eqref{cons: Demand2}-\eqref{cons: Demand_Heat3}~; 
    &   
    \label{Robust1: Other_Cons_Det}
\end{align}
\label{Two-Stage_Robust}
\end{subequations}
\endgroup
\vspace{-1em}    
}

In this section, a flexible risk-averse strategy is developed by extending the theory of \ac{ro} presented in~\cite{bertsimas04} and the forward-backward asymmetric \ac{ro} approach in~\cite{chen2007robust} to model the behavior of the \ac{rvpp} operator under uncertain parameters. {\color{black}Flexibility in addressing uncertainty is achieved through the use of the \textit{uncertainty budget} parameter, which is applied to both the objective function (i.e., $\Gamma^{DA}$, $\Gamma^{SR,\uparrow}$, and $\Gamma^{SR,\downarrow}$) and the constraints ($\Gamma_{\theta}$, $\Gamma_{r}$, and $\Gamma_{d}$) of the optimization problem. Each uncertainty budget is defined as an integer value ranging from 0 to 24 over the full scheduling horizon (i.e., 24 hours), and it determines the level of conservatism adopted by the \ac{rvpp} operator. Mathematically, the uncertainty budget specifies the number of time periods during which uncertain parameters (such as market prices or \ac{ndrs} generation) are allowed to deviate from their nominal (forecasted) values. A lower uncertainty budget leads to a more optimistic (less robust) solution, while a higher budget increases conservatism by accounting for more potential deviations. For instance, electricity market price uncertainties are modeled using asymmetric uncertainty bounds, defined as ${\lambda}_t^{DA} {\in} \left[\tilde{\lambda}_t^{DA}-\check{\lambda}_t^{DA}, \tilde{\lambda}_t^{DA} + \hat{\lambda}_t^{DA} \right]$, where in general, $\check{\lambda}_t^{DA} \neq \hat{\lambda}_t^{DA}$.} In the defined bound, the worst-case electricity prices are determined based on the direction of the \ac{rvpp}'s traded energy in a given period; i.e., the minimum price represents the worst case when the \ac{rvpp} is selling energy, and the maximum price represents the worst case when the \ac{rvpp} is buying energy. For the uncertainty of up/down reserve prices $\left({\lambda}_t^{{SR, \uparrow}} {\in} \left[ \hat{\Tilde{\lambda}}_t^{{SR, \uparrow}}-\check{\lambda}_t^{{SR, \uparrow}}, \hat {\Tilde{\lambda}}_t^{{SR, \uparrow}} \right]\right.$ and $\left.{\lambda}_t^{{SR, \downarrow}} {\in} \left[ \hat{\Tilde{\lambda}}_t^{{SR, \downarrow}}-\check{\lambda}_t^{{SR, \downarrow}}, \hat{\Tilde{\lambda}}_t^{{SR, \downarrow}} \right]\right)$, thermal production of \ac{sf} of \acp{csp} $\left({P}_{\theta,t}^{SF} {\in} \left[ \hat{\Tilde{P}}_{\theta,t}^{SF}-\check{P}_{\theta,t}^{SF}, \hat{\Tilde{P}}_{\theta,t}^{SF} \right]\right)$, and electrical production of \acp{ndrs} $\left({P}_{r,t} {\in} \left[ \hat{\Tilde{P}}_{r,t}-\check{P}_{r,t}, \hat{\Tilde{P}}_{r,t} \right]\right)$, only negative deviations are considered, while for \acp{ed}/\acp{td} $\left({P}_{d,t} {\in} \left[\tilde{\check{P}}_{d,t},\tilde{\check{P}}_{d,t}+\hat{P}_{d,t}\right]\right.$ and ${H}_{d,t} {\in} \left.\left[\tilde{\check{H}}_{d,t},\tilde{\check{H}}_{d,t}+\hat{H}_{d,t}\right]\right)$, only positive deviations are taken into account. This ensures that the worst-case scenario, according to the strategy or level of conservatism chosen by the \ac{rvpp} operator through the uncertainty budgets, is appropriately represented in the optimization problem, as deviations in opposite directions would result in higher profits for the \ac{rvpp}.

%For the uncertainty of up/down reserve prices $\left({\lambda}_t^{{SR, \uparrow}} {\in} \left[\tilde{\lambda}_t^{{SR, \uparrow}}, \tilde{\lambda}_t^{{SR, \uparrow}}-\check{\lambda}_t^{{SR, \uparrow}}\right]\right.$ and $\left.{\lambda}_t^{{SR, \downarrow}} {\in} \left[\tilde{\lambda}_t^{{SR, \downarrow}},\tilde{\lambda}_t^{{SR, \downarrow}}-\check{\lambda}_t^{{SR, \downarrow}}\right]\right)$, thermal production of \ac{sf} of \acp{csp} $\left({P}_{\theta,t}^{SF} {\in} \left[\tilde{P}_{\theta,t}^{SF},\tilde{P}_{\theta,t}^{SF}-\check{P}_{\theta,t}^{SF}\right]\right)$, and electrical production of \acp{ndrs} $\left({P}_{r,t} {\in} \left[\tilde{P}_{r,t},\tilde{P}_{r,t}-\check{P}_{r,t}\right]\right)$, only negative deviations are considered, while for \acp{ed}/\acp{td} $\left({P}_{d,t} {\in} \left[\tilde{P}_{d,t},\tilde{P}_{d,t}+\hat{P}_{d,t}\right]\right.$ and ${H}_{d,t} {\in} \left.\left[\tilde{H}_{d,t},\tilde{H}_{d,t}+\hat{H}_{d,t}\right]\right)$, only positive deviations are taken into account.

The objective function~\eqref{Uncertainty: Obj} is formulated by considering the defined bounds for different uncertain parameters in the objective function of the problem. The inner maximization problem in~\eqref{Uncertainty: Obj} identifies the worst-case scenario for \ac{rvpp} profit across different market participation. This is achieved by introducing new time period sets $\mathscr{T}^{DA}$, $\mathscr{T}^{SR,\uparrow}$, and $\mathscr{T}^{SR,\downarrow}$, which represent the periods where price deviations from the median value or higher bound have the greatest negative impact on the \ac{rvpp}'s profit. Since the cost of purchased thermal energy through \ac{hpa}, the operation costs of \ac{ndrs} and \acp{csp}, as well as the \ac{rvpp}'s profit for median values of \ac{dam} price and upper bound of \ac{srm} price of uncertain parameters, are not influenced by these uncertainties, the corresponding terms can be moved to the outer-layer problem. Additionally, a new positive auxiliary variable ${y_t}^{DA}$ is introduced to represent the traded electrical energy in~\eqref{Uncertainty: Obj}. This variable is bounded by constraint~\eqref{Uncertainty: Asymetric_DAprice}, effectively modeling the asymmetric behavior of electricity price uncertainties based on the direction of traded energy in the market.

%\footnote{When not explicitly indicated in~\eqref{Uncertainty_ALL}, the time set considered for $t$ is $\mathscr{T}$.}

%Equations~\eqref{Uncertainty: STU}-\eqref{Uncertainty: Demand_Heat} address the uncertainty in the thermal production of \ac{sf} of \acp{csp}, electrical production of \acp{ndrs}, and \acp{ed}/\acp{td} consumption, respectively. These constraints are developed by incorporating the described uncertainty bounds and the newly defined worst-case sets $\mathscr{T}_{\theta}$, $\mathscr{T}_{r}$, and $\mathscr{T}_{d}$ for each time period. For example, in constraint~\eqref{Uncertainty: STU}, the worst-case deviation of thermal power due to uncertainty, $\check{P}_{\theta,t^\prime}^{SF}$, is reduced from the upper bound of forecast value $\hat{\Tilde{P}}_{\theta,t}^{SF}$ only when $t^\prime$ corresponds to period $t$ and belongs to the set $\mathscr{T}_{\theta}$ (i.e., ${t^\prime \in \mathscr{T}_{\theta}, t^\prime = t}$). The positive nature of the auxiliary variables is established in~\eqref{Uncertainty: Auxillary_price}, while constraint~\eqref{Uncertainty: Other_Cons_Det} is identical to~\eqref{Robust1: Other_Cons_Det}.

{\color{black}Equations~\eqref{Uncertainty: STU}-\eqref{Uncertainty: Demand_Heat} address the uncertainty in the thermal production of \ac{sf} of \acp{csp}, electrical production of \acp{ndrs}, and \acp{ed}/\acp{td} consumption, respectively. These constraints are developed by incorporating the described uncertainty bounds and the newly defined worst-case sets $\mathscr{T}_{\theta}$, $\mathscr{T}_{r}$, and $\mathscr{T}_{d}$ for each time period. For example, in constraint~\eqref{Uncertainty: STU}, the worst-case deviation of thermal power due to uncertainty, $\check{P}_{\theta,t^\prime}^{SF}$, is reduced from the upper bound of forecast value $\hat{\Tilde{P}}_{\theta,t}^{SF}$ only when $t^\prime$ corresponds to period $t$ and belongs to the set $\mathscr{T}_{\theta}$ (i.e., ${t^\prime \in \mathscr{T}_{\theta}, t^\prime = t}$). Physically, $\hat{\Tilde{P}}_{\theta,t}^{SF}$ represents the nominal upper bound of thermal power expected under ideal solar conditions, while $\check{P}_{\theta,t^\prime}^{SF}$ captures the maximum possible reduction from this value due to adverse effects such as cloud cover or irradiance variability. This constraint ensures that the available thermal power of the \ac{sf} is robustly limited by accounting for possible underproduction within a predefined uncertainty budget $\Gamma_{\theta}$. The formulation effectively captures a flexible adversarial scenario in which the actual thermal power output from the \ac{sf} is reduced to account for forecast errors and unforeseen fluctuations in solar irradiation. This uncertain constraint has a direct impact on the downstream operation of the \ac{ts} and \ac{pb} systems, as outlined by constraints~\eqref{Deterministic: STU}-\eqref{Deterministic: TS} in Section~\ref{subsubsec:STU}. Specifically, a reduction in $p_{\theta,t}^{SF}$ due to solar forecast uncertainty can limit the thermal input to \ac{ts} and \ac{pb}, thereby constraining the electrical energy and reserve that the \ac{csp} can commit. The positive nature of the auxiliary variables is established in~\eqref{Uncertainty: Auxillary_price}, while constraint~\eqref{Uncertainty: Other_Cons_Det} is identical to~\eqref{Robust1: Other_Cons_Det}.}

{\color{black}
\vspace{-1em}
\begingroup
\allowdisplaybreaks
\begin{subequations}
\begin{align} \label{Uncertainty: Obj}
&\mathop {\max}\limits_{{\Xi ^{DA+SR+HPA}}} \scalebox{1.5}{\Bigg\{} -\sum\limits_{t \in \mathscr{T}} {{\lambda _t^{HT}h_t^{HT}\Delta t} } - \sum\limits_{t \in \mathscr{T}} {\sum\limits_{r \in \mathscr{R}} {C_rp_{r,t}\Delta t} } - \sum\limits_{t \in \mathscr{T}} {\sum\limits_{\theta \in \Theta} {C_{\theta}(p_{\theta,t}+h_{\theta,t})\Delta t} } \nonumber \\& + \sum\limits_{t \in \mathscr{T}} {\left[ {\tilde{\lambda}_t^{DA}p_t^{DA}\Delta t +{\hat{\tilde{\lambda}}_t^{{SR, \uparrow}}r_t^{SR,\uparrow} } +{\hat{\tilde{\lambda}}_t^{{SR, \downarrow}}r_t^{SR,\downarrow} }  } \right]} \nonumber
\\ & - \mathop{\max}\limits_{\substack{
    \left\{
    \substack{
    \mathscr{T}^{DA},\\
    %\mathscr{T}^{HT},\\
    \mathscr{T}^{SR,\uparrow},\\
    \mathscr{T}^{SR,\downarrow} \,
    }
    \middle| \; 
    \substack{
    \left| \mathscr{T}^{DA} \right| = \Gamma^{DA}, \\
    %\left| \mathscr{T}^{HT} \right| = \Gamma^{HT}, \\
    \; \left| \mathscr{T}^{SR,\uparrow} \right| = \Gamma^{SR,\uparrow}, \\
    \; \left| \mathscr{T}^{SR,\downarrow} \right| = \Gamma^{SR,\downarrow}
    }
    \right\}
    }}
    \left\{ \sum\limits_{t \in {\mathscr{T}^{DA}}} {\check{\lambda}_t^{DA} {y_t}^{DA} } + \sum\limits_{t \in {\mathscr{T}^{SR,\uparrow}}} {\check {\lambda}_t^{{SR, \uparrow}}r_t^{SR,\uparrow} } + \sum\limits_{t \in {\mathscr{T}^{SR,\downarrow}}} {\check {\lambda}_t^{{SR, \downarrow}}r_t^{SR,\downarrow} }  
\right\}   \scalebox{1.5}{\Bigg\}} 
\\
    \nonumber\text{st.} \\
    &- \frac{\check{\lambda}_t^{DA}}{\hat{\lambda}_t^{DA}} {y_t}^{DA} \le {p_t}^{DA} \Delta t \le {y_t}^{DA}~; 
    \hspace{17.7em}   
    \forall t \in \mathscr{T} \label{Uncertainty: Asymetric_DAprice} \\
    %
%    &- \frac{\check{\lambda}_t^{HT}}{\hat{\lambda}_t^{DA}} {y_t}^{HT} \le {p_t}^{HT} \Delta t \le {y_t}^{HT}~; 
%    \hspace{14.9em}   
%    \forall t \in \mathscr{T}^{HT} \label{Uncertainty: Asymetric_Heatprice} \\
    %
    & p_{\theta,t}^{SF} \leq \hat{\Tilde{P}}_{\theta,t}^{SF}-  \mathop {\max}\limits_{\left\{ \mathscr{T}_{\theta} \middle| \; \;  \left| \mathscr{T}_{\theta} \right| = \Gamma_{\theta} \right\}} \left\{ \sum_{t^\prime \in \mathscr{T}_{\theta}, t^\prime = t} \check{P}_{\theta,t^\prime}^{SF} \right\}~;
    \hspace{11.8em} \forall \theta, t \in {\Theta},\mathscr{T} \label{Uncertainty: STU} \\
    & p_{r,t}+r_{r,t}^{\uparrow} \leq \hat{\Tilde{P}}_{r,t} - \mathop {\max}\limits_{ \left\{ \mathscr{T}_{r} \middle| \;  \; \left| \mathscr{T}_{r} \right| = \Gamma_{r} \right\} } \left\{ \sum_{t^\prime \in \mathscr{T}_{r}, t^\prime = t } \check{P}_{r,t'} \right\}~;  
    \hspace{10.5em} \forall r,t \in \mathscr{R},\mathscr{T} \label{Uncertainty: NDRES} \\ 
    & p_{d,t} \geq \tilde{\check{P}}_{d,t} + \mathop {\max}\limits_{ \left\{ \mathscr{T}_{d} \middle| \; \; \left| \mathscr{T}_{d} \right| = \Gamma_{d} \right\} } \left\{ \sum_{t^\prime \in \mathscr{T}_{d}, t^\prime = t} \hat{P}_{d,t^\prime}  \right\}~; 
    \hspace{11.9em} \forall d,t  \in \mathscr{D},\mathscr{T} \label{Uncertainty: Demand}\\
    & h_{d,t} \geq \tilde{\check{H}}_{d,t} + \mathop {\max}\limits_{ \left\{ \mathscr{T}_{d} \middle| \; \; \left| \mathscr{T}_{d} \right| = \Gamma_{d} \right\} } \left\{ \sum_{t^\prime \in \mathscr{T}_{d}, t^\prime = t} \hat{H}_{d,t^\prime} \right\}~; 
    \hspace{11.9em} \forall d,t  \in \mathscr{D},\mathscr{T} \label{Uncertainty: Demand_Heat}\\
    & {y_t}^{DA} \geq 0~; 
    \hspace{26.2em} \forall t \in \mathscr{T}  \label{Uncertainty: Auxillary_price}\\
    &\eqref{RVPP: Supply-Demand}, \eqref{Deterministic: STU2}-\eqref{Deterministic: STU4}, \eqref{Deterministic: STU6}-\eqref{Deterministic: STU8}, \eqref{STU_SOS}, \eqref{Deterministic: TS}, \eqref{cons: NDRES2}-\eqref{cons: NDRES4}, \eqref{cons: Demand2}-\eqref{cons: Demand_Heat3}~; 
    &   
    \label{Uncertainty: Other_Cons_Det}
\end{align}
\label{Uncertainty_ALL}
\end{subequations}
\endgroup
\vspace{-1em}    
}

\subsubsection{Inner Problems Reformulation}
\label{subsubsec:Inner_Reformulation}

Although the maximization term in the last part of the objective function~\eqref{Uncertainty: Obj} (the protection function) captures the worst-case scenario for the uncertain parameters, the process of selecting values based on the defined sets can be expressed in a linear manner. To achieve this, assuming the optimal values (indicated with superscript $^*$) of the upper-level variables ${y_t}^{DA^*}$, $r_t^{SR,\uparrow^{*}}$, and $r_t^{SR,\downarrow^{*}}$ and using Proposition 1 from~\cite{bertsimas04}, linear problem~\eqref{Protection_Function_obj} is equivalent to the protection function in~\eqref{Uncertainty: Obj}. Constraints~\eqref{Protection_obj: con1}-\eqref{Protection_obj: con4} limit the summation of the new auxiliary variables ${z_t}^{DA}$, ${z_t}^{SR,\uparrow}$, and ${z_t}^{SR,\downarrow}$ for each uncertain parameter in the objective function of the problem to the corresponding uncertainty budgets $\Gamma^{DA}$, $\Gamma^{SR,\uparrow}$, and $\Gamma^{SR,\downarrow}$, respectively. These constraints ensure that the positive auxiliary variables ${z_t}^{DA}$, ${z_t}^{SR,\uparrow}$, and ${z_t}^{SR,\downarrow}$ are less or equal than 1, thereby achieving the same optimal value for the objective function in~\eqref{Protection_obj: obj} and the protection function in~\eqref{Uncertainty: Obj}. The dual variables of each constraint are defined in these equations, which are then used in Section~\ref{subsubsec:MILP_Formulation} to derive the final \ac{milp} formulation.

\begingroup
\allowdisplaybreaks
\begin{subequations}
\begin{align}
    & %\beta^{O} = %\nonumber \\ & 
    {\max} \left\{         
        \sum\limits_{t \in {\mathscr{T}^{DA}}} {\check{\lambda}_t^{DA} {y_t}^{DA^*} {z_t}^{DA} } + \sum\limits_{t \in {\mathscr{T}^{SR,\uparrow}}} {\check {\lambda}_t^{{SR, \uparrow}} r_t^{SR,\uparrow^{*}} {z_t}^{SR,\uparrow} } + \sum\limits_{t \in {\mathscr{T}^{SR,\downarrow}}} {\check {\lambda}_t^{{SR, \downarrow}} r_t^{SR,\downarrow^*} {z_t}^{SR,\downarrow} } \right\}
    \label{Protection_obj: obj} \\
    \nonumber\text{st.} \\
    & \sum_{t \in {\mathscr{T}^{DA}}} {z_t}^{DA} \le \Gamma^{DA}: \phi^{DA}~;   
    \label{Protection_obj: con1} \\
    %
%    & \sum_{t \in {\mathscr{T}^{HT}}} {z_t}^{HT} \le \Gamma^{HT}: \phi^{HT}~;    
%    \label{Protection_obj: con2} \\
        %
    & \sum_{t \in {\mathscr{T}^{SR,\uparrow}}} {z_t}^{SR,\uparrow} \le \Gamma^{SR,\uparrow}: \phi^{SR,\uparrow}~;    
    \label{Protection_obj: con3} \\
    & \sum_{t \in {\mathscr{T}^{SR,\downarrow}}} {z_t}^{SR,\downarrow} \le \Gamma^{SR,\downarrow}: \phi^{SR,\downarrow}~;    
    \label{Protection_obj: con4} \\
    & 0 \le {z_t}^{DA} \le 1: \zeta_{t}^{DA}~; 
    \hspace{16.3em} 
    \forall {t \in {\mathscr{T}^{DA}}}   \label{Protection_obj: con5} \\
        %
%    & 0 \le {z_t}^{HT} \le 1: \zeta_{t}^{HT}~; 
%    \hspace{15.8em} 
%    \forall {t \in {\mathscr{T}^{HT}}}  \label{Protection_obj: con6} \\
        %
    & 0 \le {z_t}^{SR,\uparrow} \le 1: \zeta_{t}^{SR,\uparrow}~; 
    \hspace{15em} 
    \forall {t \in {\mathscr{T}^{SR,\uparrow}}} \label{Protection_obj: con7} \\
    & 0 \le {z_t}^{SR,\downarrow} \le 1: \zeta_{t}^{SR,\downarrow}~; 
    \hspace{15em} 
    \forall {t \in {\mathscr{T}^{SR,\downarrow}}} \label{Protection_obj: con8}
\end{align}
\label{Protection_Function_obj}
\end{subequations}
\endgroup

The equivalent linear formulation for selecting the worst-case scenarios of thermal production uncertainty of \ac{sf} of \ac{csp} in constraint~\eqref{Uncertainty: STU} is presented as linear formulation~\eqref{Protection_Function_STU}. The objective function~\eqref{Protection_STU: obj} maximizes the thermal production deviation caused by uncertainty in the corresponding period $t$ ($t^\prime \in \mathscr{T}_{\theta}$, $t^\prime = t$). The summation of the auxiliary positive variables $z_{\theta,t^\prime}$ for the corresponding period $t$ and other worst-case periods, which are fixed at their optimal values, is constrained to be less than the uncertainty budget $\Gamma_{\theta}$. The bound for the auxiliary variable $z_{\theta,t^\prime}$ is defined in~\eqref{Protection_STU: con2}. Worth noting that the linear formulation~\eqref{Protection_Function_STU} considers the temporal constraints of the uncertainty parameter over the entire operation period, rather than defining the worst case of the uncertain parameter in a single period, as done in~\cite{dominguez2012optimal, he2016optimal, xiong2023dp}. This feature allows the \ac{rvpp} to adjust a single uncertain parameter for the entire period instead of multiple parameters for each time period and uncertain parameter.

 The equivalent linear formulations for worst-case selection in constraints~\eqref{Uncertainty: NDRES}-\eqref{Uncertainty: Demand_Heat}, addressing \ac{ndrs} production, and \acp{ed}/\acp{td} uncertainty, respectively, can be derived similarly to~\eqref{Protection_Function_STU}. These are omitted here for brevity.

\begingroup
\allowdisplaybreaks
\begin{subequations}
\begin{align}
    & %\beta_{\theta, t}^{C} =
    {\max} \sum_{t^\prime \in \mathscr{T}_{\theta}, t^\prime = t} \check{P}_{\theta,t^\prime}^{SF} z_{\theta,t^\prime}~; & \forall \theta, t \in {\Theta},\mathscr{T}      
    &   
    \label{Protection_STU: obj} \\
    \nonumber\text{st.} \\
    & \sum_{t^\prime \in \mathscr{T}_{\theta}, t^\prime = t} z_{\theta,t^\prime} + \sum_{t^\prime \in \mathscr{T}_{\theta}, t^\prime \neq t} z_{\theta,t^\prime}^{*} \le \Gamma_{\theta}: \phi_{\theta}~; 
    &   
    \label{Protection_STU: con1} \\
    & 0 \le z_{\theta,t^\prime} \le 1: \zeta_{\theta,t^\prime}~; 
    & 
    \forall t^\prime \in \mathscr{T}_{\theta}, t^\prime= t \label{Protection_STU: con2}
\end{align}
\label{Protection_Function_STU}
\end{subequations}
\endgroup

\subsubsection{MILP Formulation}
\label{subsubsec:MILP_Formulation}

By applying strong duality~\cite{floudas1995nonlinear} to the linear formulations in~\eqref{Protection_Function_obj} and~\eqref{Protection_Function_STU} (and similarly for equivalent linear formulations related to other uncertain constraints), and replacing their dual problems instead of the protection function in the objective function~\eqref{Uncertainty: Obj} and the protection functions in the constraints~\eqref{Uncertainty: STU}-\eqref{Uncertainty: Demand_Heat}, respectively, the final \ac{milp} is formulated as~\eqref{Single-level_MILP}. The first and second lines of the objective function~\eqref{MILP: Obj} are identical to the first and second lines of~\eqref{Uncertainty: Obj}. The third line captures the negative effects of uncertainties in electricity (energy and reserve) prices. Constraint~\eqref{MILP_asymetric1} is equivalent to~\eqref{Uncertainty: Asymetric_DAprice}. Constraints~\eqref{MILP: DAMprice}-\eqref{MILP: downSRMprice} are the dual constraints of the linear problem related to electricity prices uncertainties in~\eqref{Protection_Function_obj}. Constraint~\eqref{MILP: STU1} represents the upper limit of thermal production of \ac{sf} of \ac{csp} while accounting for uncertainty. Since the term $\Gamma_{\theta} - \sum_{t^\prime \in \mathscr{T}_{\theta}, t^\prime \neq t} z_{\theta,t^\prime}^{*}$ in~\eqref{Protection_STU: con1} can only take a value of zero or one, depending on the number of worst-case periods in its defined set, a new binary variable $\chi_{\theta,t}$ is introduced to represent these conditions. Additionally, a new positive auxiliary variable $y_{\theta,t}$ is defined in~\eqref{MILP: STU1} and constrained in~\eqref{MILP: STU2} using the big-M method~\cite{floudas1995nonlinear} to represent the dual term $\chi_{\theta,t} \phi_{\theta} + {\zeta}_{\theta,t}$. Constraint~\eqref{MILP: STU3} is the dual constraint of the linear problem associated with \ac{csp} thermal production uncertainty in~\eqref{Protection_Function_STU}. Constraint~\eqref{MILP: STU4} assigns the uncertainty budget for thermal production of \ac{sf} of \ac{csp}. Constraints~\eqref{MILP: NDRES1}-\eqref{MILP: NDRES4} represent the \acp{ndrs} electrical production uncertain constraints, constraints~\eqref{MILP: Demand1}-\eqref{MILP: Demand_Heat4} represent the \acp{ed}/\acp{td} uncertain constraints. These constraints are defined similarly to the thermal production uncertainty of \ac{sf} of \ac{csp} constraints~\eqref{MILP: STU1}-\eqref{MILP: STU4}. The main difference is that, for \acp{ed}/\acp{td}, the worst-case uncertainty results in an increase in these parameters, whereas, for thermal production of \ac{sf} of \ac{csp} and \ac{ndrs} electrical production, it results in a decrease. The deterministic constraints that remain unaffected by uncertainty are outlined in~\eqref{MILP: Other_Cons_Det}. Finally, the nature of the positive and binary variables is defined in~\eqref{MILP: Auxilliary_obj}-\eqref{MILP: Binary_cons}.

It is worth noting that the refined \ac{milp} formulation in~\eqref{Single-level_MILP} is developed to flexibly consider both source- and load-side uncertainty by adjusting the uncertainty budget defined for temporal constraints. The flexibility to address different levels of uncertainty and the computational efficiency of the proposed approach are extensively studied through various case studies in Section~\ref{sec:Case_Studies}.

%constraints~\eqref{MILP: Demand1}-\eqref{MILP: Demand4} represent the electrical demand uncertainty constraints, and constraints~\eqref{MILP: Demand_Heat1}-\eqref{MILP: Demand_Heat4} represent the heat demand uncertainty constraints.

{\color{black}
\vspace{-1em}
\begingroup
\begin{subequations}
\allowdisplaybreaks
\begin{align} \label{MILP: Obj}
&\mathop {\max}\limits_{{\Xi ^{DA+SR+HPA}}} \scalebox{1.1}{\Bigg\{} -\sum\limits_{t \in \mathscr{T}} {{\lambda _t^{HT}h_t^{HT}\Delta t} } - \sum\limits_{t \in \mathscr{T}} {\sum\limits_{r \in \mathscr{R}} {C_rp_{r,t}\Delta t} } - \sum\limits_{t \in \mathscr{T}} {\sum\limits_{\theta \in \Theta} {C_{\theta}(p_{\theta,t}+h_{\theta,t})\Delta t} } \nonumber \\& + \sum\limits_{t \in \mathscr{T}} {\left[ {\tilde{\lambda}_t^{DA}p_t^{DA}\Delta t +{\hat{\tilde{\lambda}}_t^{{SR, \uparrow}}r_t^{SR,\uparrow} } +{\hat{\tilde{\lambda}}_t^{{SR, \downarrow}}r_t^{SR,\downarrow} }  } \right]} \nonumber
\\
& - \Gamma^{DA} \phi^{DA} - \Gamma^{SR,\uparrow} \phi^{SR,\uparrow} - \Gamma^{SR,\downarrow} \phi^{SR,\downarrow} - \sum\limits_{t \in \mathscr{T}} \left[{\zeta}^{DA}_t + {\zeta}^{SR,\uparrow}_t+ {\zeta}^{SR, \downarrow}_t\right]    \scalebox{1.1}{\Bigg\}}
\\
    \nonumber\text{st.} \\
    &- \frac{\check{\lambda}_t^{DA}}{\hat{\lambda}_t^{DA}} {y_t}^{DA} \le {p_t}^{DA} \Delta t \le {y_t}^{DA}~; 
    \hspace{18.6em}   
    \forall t \in \mathscr{T} \label{MILP_asymetric1} \\
    %
%    &- \frac{\check{\lambda}_t^{HT}}{\hat{\lambda}_t^{DA}} {y_t}^{HT} \le {p_t}^{HT} \Delta t \le {y_t}^{HT}~; 
%    \hspace{16em}   
%    \forall t \label{MILP_asymetric2} \\
    %
    &{\phi^{DA}} + \zeta_t^{DA} \ge  \check{\lambda}_t^{DA}y_t^{DA}~; 
    \hspace{21.7em}    
    \forall t \in \mathscr{T} \label{MILP: DAMprice} \\
    %
%    &{\phi ^{HT}} + \zeta_t^{HT} \ge  \check{\lambda}_t^{HT}y_t^{HT}~; 
%    \hspace{19.3em}    
%    \forall t \label{MILP: Heatprice} \\
    %
    &{\phi^{SR,\uparrow}} + \zeta_t^{SR,\uparrow} \ge ~\check{\lambda}_t^{SR,\uparrow} r_t^{SR,\uparrow}~; 
    \hspace{19.8em}   
    \forall t \in \mathscr{T} \label{MILP: upSRMprice} \\
    &{\phi^{SR,\downarrow}} + \zeta_t^{SR,\downarrow} \ge ~\check{\lambda}_t^{SR,\downarrow} r_t^{SR,\downarrow}~; 
    \hspace{19.7em}   
    \forall t \in \mathscr{T} \label{MILP: downSRMprice} \\
    & p_{\theta,t}^{SF} \leq \hat{\Tilde{P}}_{\theta,t}^{SF} - \chi_{\theta,t} \, \phi_{\theta} -  {\zeta}_{\theta,t} = \hat{\Tilde{P}}_{\theta,t}^{SF} - y_{\theta,t}~;
    \hspace{12.9em} \forall \theta, t \in {\Theta},\mathscr{T} \label{MILP: STU1} \\  
    &\phi_{\theta} +  {\zeta}_{\theta,t} - M (1 - \chi_{\theta,t}) \leq y_{\theta,t} \leq M \chi_{\theta,t}~;
    \hspace{13.6em} \forall \theta, t \in {\Theta},\mathscr{T} \label{MILP: STU2} \\ 
    &\phi_{\theta} + \zeta_{\theta,t} \ge  \check{P}_{\theta,t}~; 
    \hspace{22.3em}   
    \forall \theta, t \in {\Theta},\mathscr{T} \label{MILP: STU3} \\
    & \sum_{t} \chi_{\theta,t} = \Gamma_{\theta}~; 
    \hspace{24.8em}
    \forall \theta \in {\Theta} \label{MILP: STU4} \\
    & p_{r,t}+r_{r,t}^{\uparrow} \leq \hat{\Tilde{P}}_{r,t} - \chi_{r,t} \, \phi_{r} - {\zeta}_{r,t} = \hat{\tilde{P}}_{r,t} - y_{r,t}~;  
    \hspace{11.95em} \forall r,t \in \mathscr{R},\mathscr{T} \label{MILP: NDRES1} \\ 
    &\phi_{r} +  {\zeta}_{r,t} - M (1 - \chi_{r,t}) \leq y_{r,t} \leq M \chi_{r,t}~;  
    \hspace{13.9em} \forall r,t \in \mathscr{R},\mathscr{T} \label{MILP: NDRES2} \\ 
    &\phi_{r} + \zeta_{r,t} \ge  \check{P}_{r,t}~; 
    \hspace{22.6em}   
    \forall r, t \in \mathscr{R},\mathscr{T} \label{MILP: NDRES3} \\
    & \sum_{t} \chi_{r,t} = \Gamma_{r}~; 
    \hspace{25.0em}
    \forall r \in \mathscr{R} \label{MILP: NDRES4} \\
    & p_{d,t} \geq \tilde{\check{P}}_{d,t} + \chi_{d,t} \, \phi_{d} + {\zeta}_{d,t} = \tilde{\check{P}}_{d,t} + y_{d,t}~; 
    \hspace{13.2em} \forall d,t \in \mathscr{D},\mathscr{T} \label{MILP: Demand1}\\
    &h_{d,t} \geq \tilde{\check{H}}_{d,t} + \chi_{d,t} \, \phi_{d} + {\zeta}_{d,t} = \tilde{\check{H}}_{d,t} + y_{d,t}~; 
    \hspace{13.1em} \forall d,t \in \mathscr{D},\mathscr{T} \label{MILP: Demand_Heat1}\\
    &\phi_{d} +  {\zeta}_{d,t} - M (1 - \chi_{d,t}) \leq y_{d,t} \leq M \chi_{d,t}~; 
    \hspace{13.2em} \forall d,t \in \mathscr{D},\mathscr{T}  \label{MILP: Demand2}\\
    &\phi_{d} + \zeta_{d,t} \ge  \hat{P}_{d,t}~; 
    \hspace{22.1em}   
    \forall d, t \in \mathscr{D},\mathscr{T} \label{MILP: Demand3} \\
     %
%    & \sum_{t} \chi_{d,t} = \Gamma_{d}~; 
%    \hspace{22.9em}
%    \forall d \label{MILP: Demand4} \\
    %
%    &\phi_{d} +  {\zeta}_{d,t} - M (1 - \chi_{d,t}) \leq y_{d,t} \leq M \chi_{d,t}~; 
%    \hspace{13.4em} \forall d,t   \label{MILP: Demand_Heat2}\\
%
    &\phi_{d} + \zeta_{d,t} \ge  \hat{H}_{d,t}~; 
    \hspace{22.1em}   
    \forall d, t \in \mathscr{D},\mathscr{T} \label{MILP: Demand_Heat3} \\
    & \sum_{t} \chi_{d,t} = \Gamma_{d}~; 
    \hspace{24.7em}
    \forall d \in \mathscr{D} \label{MILP: Demand_Heat4} \\
    &\eqref{RVPP: Supply-Demand}, \eqref{Deterministic: STU2}-\eqref{Deterministic: STU4}, \eqref{Deterministic: STU6}-\eqref{Deterministic: STU8}, \eqref{STU_SOS}, \eqref{Deterministic: TS}, \eqref{cons: NDRES2}-\eqref{cons: NDRES4}, \eqref{cons: Demand2}-\eqref{cons: Demand_Heat3}~; 
    &   
    \label{MILP: Other_Cons_Det} \\
    &\phi^{DA}, \phi^{SR,\uparrow}, \phi^{SR,\downarrow},  \zeta_t^{DA}, \zeta_t^{SR,\uparrow}, \zeta_t^{SR,\downarrow}, y_t^{DA} \ge 0~; 
    \hspace{13.5em} 
    \forall t \in \mathscr{T} \label{MILP: Auxilliary_obj} \\
    & \phi_{\theta}, \phi_{r}, \phi_d, \zeta_{\theta,t}, \zeta_{r,t}, \zeta_{d,t}, y_{\theta,t}, y_{r,t}, y_{d,t} \ge 0~; 
    \hspace{9.2em}   
    \forall \theta, r,d, t \in {\Theta},\mathscr{R},\mathscr{D},\mathscr{T} \label{MILP: Auxilliary_cons} \\
    & \chi_{\theta,t}, \chi_{r,t}, \chi_{d,t} \in \{0, 1\}~; 
    \hspace{16.2em}   
    \forall \theta, r,d, t \in {\Theta},\mathscr{R},\mathscr{D},\mathscr{T} \label{MILP: Binary_cons} 
\end{align}
\label{Single-level_MILP}
\end{subequations}
\endgroup
\vspace{-1em}  
}

{\color{black}
\subsubsection{Flexibility Metrics}
\label{subsubsec:Flexibility_Metrics}
In this paper, the flexibility of each \ac{rvpp} unit is quantified as the total upward or downward reserve it provides over the scheduling horizon. Additionally, normalized definitions are employed, where the upward or downward reserve is expressed as a ratio relative to the capacity of each \ac{rvpp} unit.

Specifically, the flexibility is quantified using the metrics of total upward reserve, as presented in equation~\eqref{cons: Flexibility1}, and total downward reserve, as calculated in equation~\eqref{cons: Flexibility2}, where $r_{u,t}^{\uparrow}$ and $r_{u,t}^{\downarrow}$ represent the upward and downward reserve provided by unit $u$ at time $t$. The index $u \in \mathscr{U}$ represents the type of \ac{rvpp} unit (\ac{csp}, \ac{ed}, and \ac{ndrs}).

\vspace{-1em}
\begingroup
\allowdisplaybreaks
\begin{subequations}
\begin{align}
    & \sum\limits_{t \in \mathscr{T}} {r_{u,t}^{\uparrow} }~;
    & 
    \forall u \in \mathscr{U}  \label{cons: Flexibility1}\\
    & \sum\limits_{t \in \mathscr{T}} {r_{u,t}^{\downarrow} }~;
    & 
    \forall u \in \mathscr{U}  \label{cons: Flexibility2}
\end{align}
\label{Cons: Flexibility}
\end{subequations}
\endgroup
\vspace{-1em}

Additionally, to normalize flexibility contributions by unit capacity and enable fair comparisons across heterogeneous \ac{rvpp} components, we define the reserve-to-capacity ratio for upward reserve and downward reserve according to equations~\eqref{cons: Flexibility_Capacity1} and~\eqref{cons: Flexibility_Capacity2}. Where $P_u$ denotes the installed capacity of unit $u$.

\vspace{-1em}
\begingroup
\allowdisplaybreaks
\begin{subequations}
\begin{align}
    & \frac {\sum\limits_{t \in \mathscr{T}} {r_{u,t}^{\uparrow} }} {P_u}~;
    & 
    \forall u \in \mathscr{U}  \label{cons: Flexibility_Capacity1}\\
    & \frac {\sum\limits_{t \in \mathscr{T}} {r_{u,t}^{\downarrow} }} {P_u}~;
    & 
    \forall u \in \mathscr{U}  \label{cons: Flexibility_Capacity2}
\end{align}
\label{Cons: Flexibility_Capacity}
\end{subequations}
\endgroup
\vspace{-1em}
}

\begin{comment}
In this paper, flexibility is quantified as the aggregate upward or downward reserve provided by \ac{rvpp} units across scheduling horizon. Also normalized definition are used, in which the ratio of upward or downward reserve to capacity of each \ac{rvpp} units is used.

Specifically, we quantify flexibility using the following two metrics:

1. Total upward reserve:

$$\sum\limits_{t \in \mathscr{T}} {r_{u,t}^{\uparrow} }~; \qquad \forall u \in \mathscr{U}$$

where $r_{u,t}^{\uparrow}$ is the upward reserve provided by unit $u$ at time $t$, and the index $u \in \mathscr{U}$ represents the type of \ac{rvpp} unit (\ac{csp}:$\theta$, \ac{ed}: $d$, \ac{ndrs}: $r$).

2. Total downward reserve:

$$\sum\limits_{t \in \mathscr{T}} {r_{u,t}^{\downarrow} }~; \qquad \forall u \in \mathscr{U}$$

which reflects the total downward reserve provided by unit $u$ over the scheduling horizon.

Additionally, to normalize flexibility contributions by unit capacity and enable fair comparisons across heterogeneous \ac{rvpp} components, we define the reserve-to-capacity ratio as follows:

3. Upward reserve to capacity:

$$\frac {\sum\limits_{t \in \mathscr{T}} {r_{u,t}^{\uparrow} }} {P_u}~; \qquad \forall u \in \mathscr{U}$$

4. Downward reserve to capacity:

$$\frac {\sum\limits_{t \in \mathscr{T}} {r_{u,t}^{\downarrow} }} {P_u}~; \qquad \forall u \in \mathscr{U}$$

where $P_u$ denotes the installed capacity of unit $u$.
    
\end{comment}

\subsubsection{Uncertainty Characterization}
\label{subsubsec:Uncertainty_Characterization}

\begin{figure} [t!]
    \centering  \includegraphics[width=.8\textwidth]{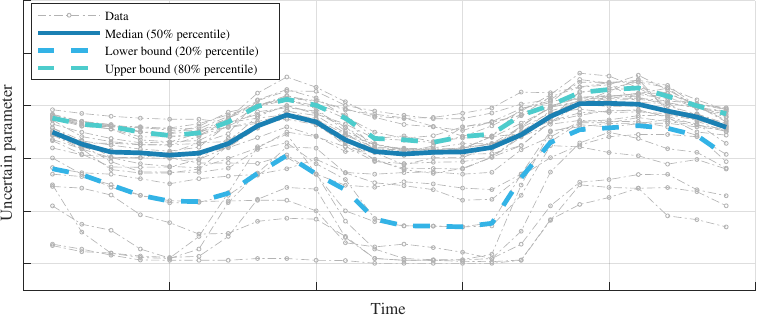}
    \vspace{-.5em}
    \caption{Assigning the bounds of uncertain parameters using historical data.}
    \label{fig:Uncertainty_bound}
    \vspace{1mm}
\end{figure}

The confidence bounds of uncertain parameters can be determined through various methods. These include leveraging historical records with observational and measurement data \citep{wang2019expansion}, fitting parametric distributions to model uncertainties \citep{srinivasan2023impact}, and employing bootstrapping techniques for resampling-based estimation \citep{wen2019performance}. Historical data is derived from past observations and measurements, which can be examined to identify underlying patterns and relationships. Figure~\ref{fig:Uncertainty_bound} illustrates the method used in this paper to assign bounds to uncertain parameters based on historical data. For this purpose, data for various uncertain parameters is collected over a 30-day period. The median, lower bound, and upper bound of these parameters are determined for each hour of the day. To prevent overly conservative solutions, the 20\% and 80\% percentiles of the data are used as the lower and upper bounds, respectively.

\section{Case Studies}
\label{sec:Case_Studies}

\begin{figure} [b!]
    \centering  \includegraphics[width=.91\textwidth]{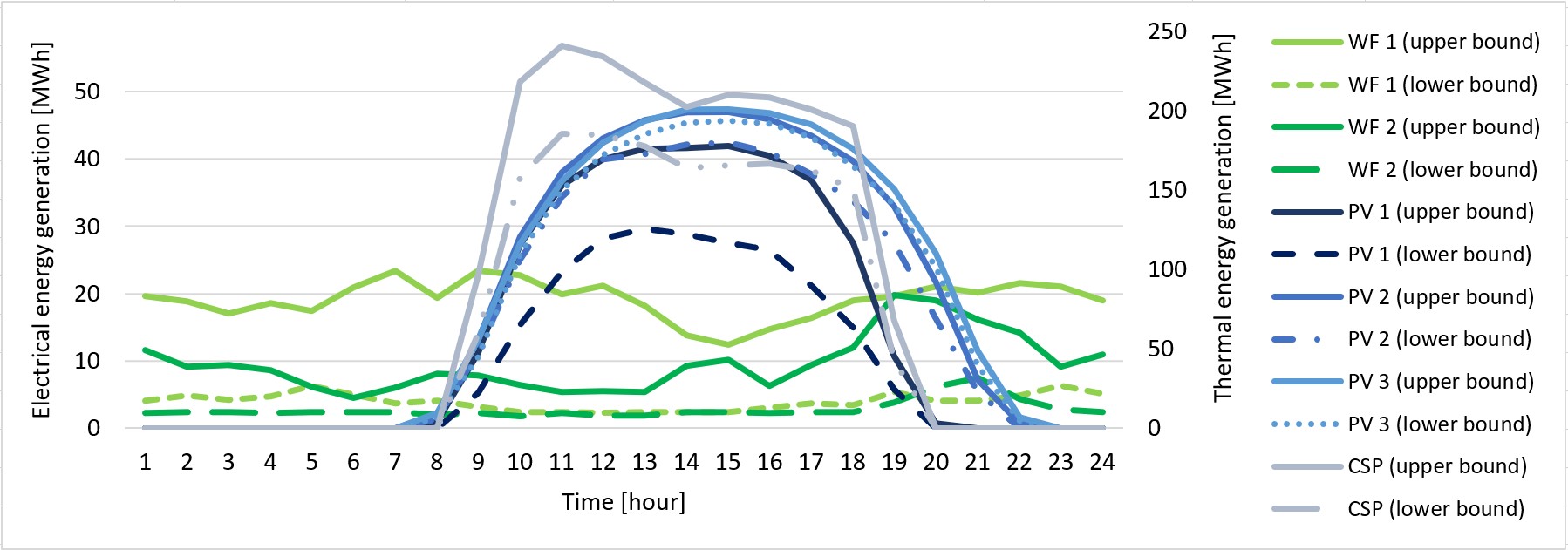}
    \vspace{-.5em}
    \caption{The forecast bounds of \acp{wf}, solar PVs, and \ac{csp}.}
    \label{fig:Data_Production}
    \vspace{2mm}
\end{figure}

\begin{figure} [b!]
    \centering  \includegraphics[width=.91\textwidth]{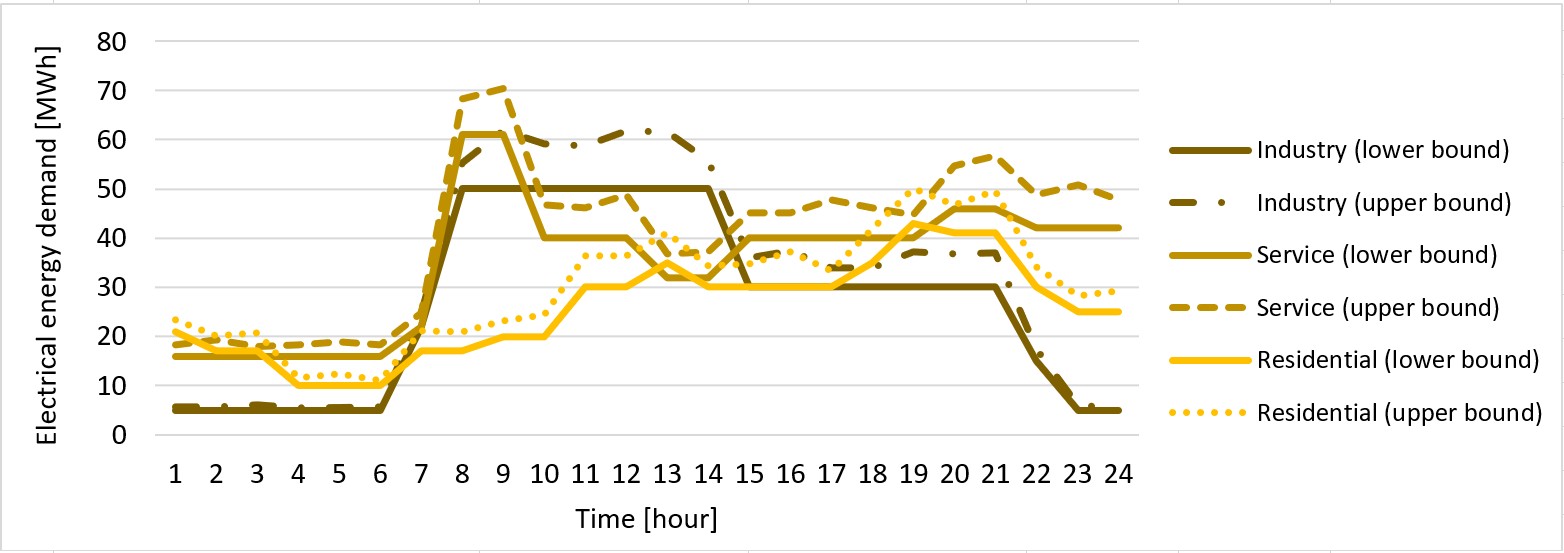}
    \vspace{-.5em}
    \caption{The forecast bounds of \acp{ed}.}
    \label{fig:Data_Demand}
   \vspace{2mm}
\end{figure}

\begin{figure} [b!]
    \centering  \includegraphics[width=.91\textwidth]{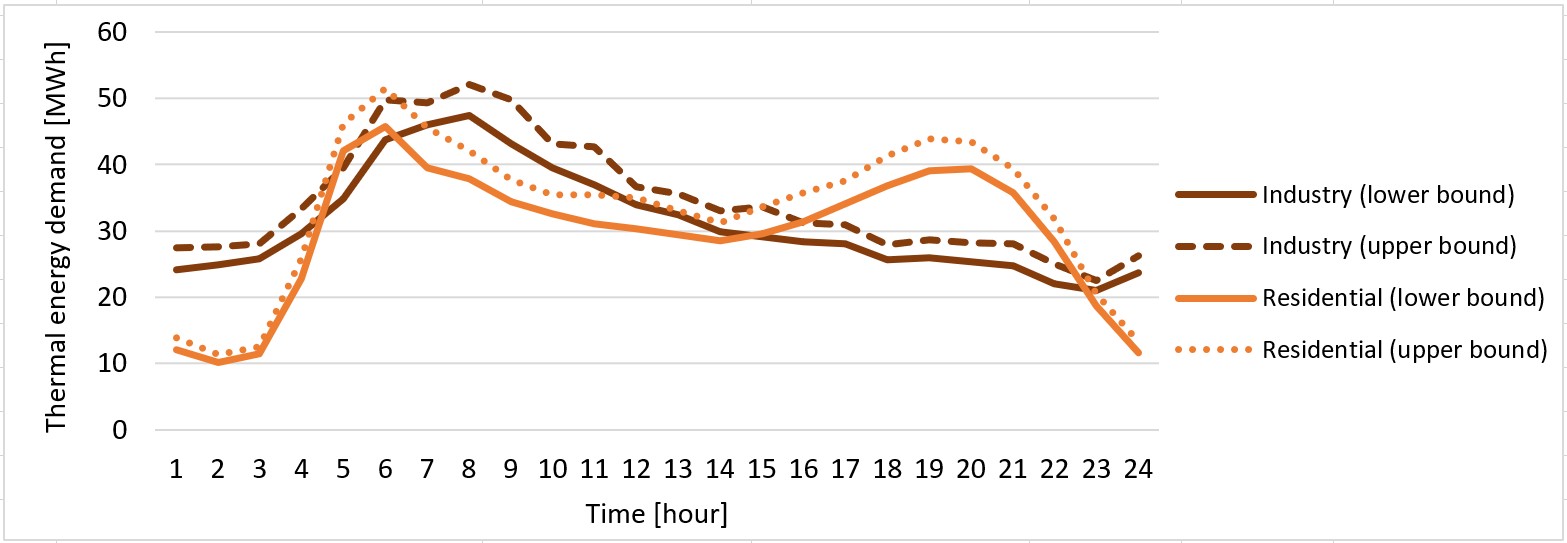}
    \vspace{-.5em}
    \caption{The forecast bounds of \acp{td}.}
    \label{fig:Data_TDemand}
    \vspace{2mm}
\end{figure}

\begin{figure}[t!]
    \centering
    \includegraphics[width=0.8\linewidth]{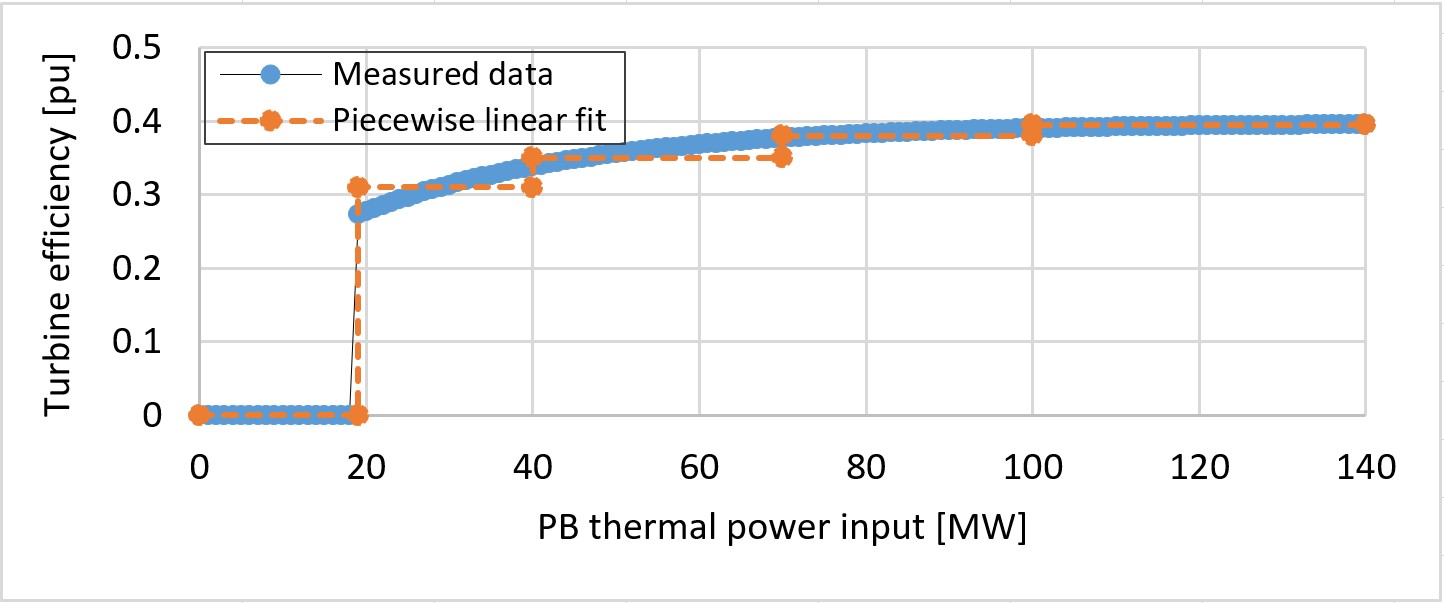}
    %\vspace{-5mm}
    \caption{Thermal to electrical conversion efficiency of \ac{pb}.}
     \label{fig:PB_conversion_efficiency}
\end{figure}

\begin{figure} [t!]
    \centering  \includegraphics[width=.9\textwidth]{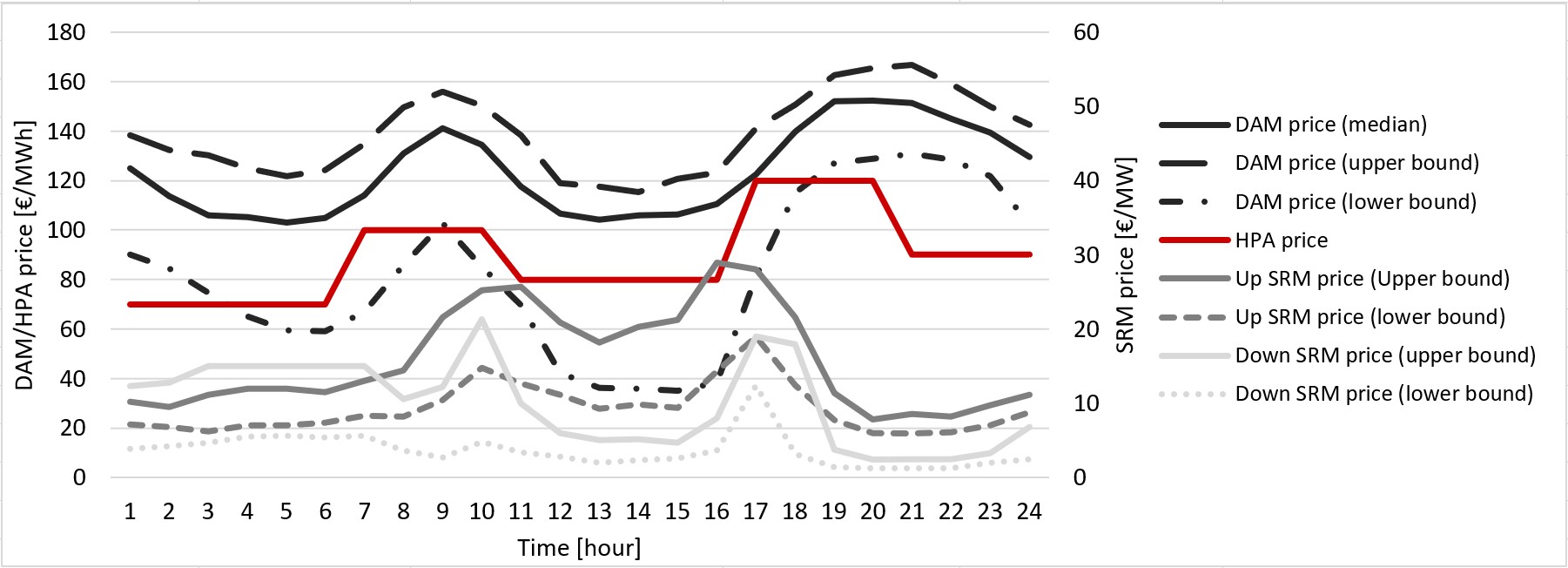}
    \vspace{-.5em}
    \caption{The DAM, SRM, and HPA price data.}
    \label{fig:Data_Price}
    \vspace{1mm}
\end{figure}

This section presents the simulation results of the proposed two-stage \ac{ro} model. The simulations are conducted on an \ac{rvpp} that includes one \ac{csp}, two \acp{wf}, three solar PV plants, three types of \acp{ed} (residential, industrial, and service), and two types of \acp{td} (residential and industrial). The forecast bounds for electrical and thermal energy generation units within the \ac{rvpp} are illustrated in Figure~\ref{fig:Data_Production}. These bounds are derived using the methodology described in Section~\ref{subsubsec:Uncertainty_Characterization}, based on historical data from CIEMAT Spain~\cite{web:ciemat_spain} for solar PVs and the \ac{csp}, and from Iberdrola Spain~\cite{web:iberdrola_spain} for \acp{wf}. {\color{black}The economic data, including the ranges of electricity and thermal energy prices, the operational costs of \ac{rvpp} units, and the forecast bounds for various uncertain parameters, are presented in Table~\ref{table:Data_Economic}. The electrical and thermal energy consumption of the demands is illustrated in Figures~\ref{fig:Data_Demand} and~\ref{fig:Data_TDemand}, based on the hourly demand profiles provided in~\citep{oladimeji2022modeling, JASM2019}. The main technical data related to the demands are provided in Table~\ref{table:Data_Demand}. The technical and design characteristics of the \ac{csp} are presented in Tables~\ref{table:Data_STU} and~\ref{table:Design_Data1_STU}, based on data from CIEMAT Spain~\cite{web:ciemat_spain} and~\cite{garcia2011performance}. The hourly thermal energy output profiles used in this study are adopted from~\cite{garcia2011performance}, which simulated \ac{csp} performance using TMY3-format meteorological data representative of high-\ac{dni} locations in Southern Europe and North Africa. These scenarios, produced for March 2018, inherently account for solar irradiance and other weather variables and are used here as inputs for simulation analysis. The technical information for the \ac{ndrs}, including \acp{wf} and solar \ac{pv} plants, is provided in Table~\ref{table:Data_NDRES}.}
%The nominal capacities of the \acp{wf} and solar PV plants are both 50 MW, with operation costs of 15 €/MWh and 7.5 €/MWh, respectively. The flexibility of \acp{ed} is assumed to be 10\%. The \ac{csp} has an operation cost of 25 €/MWh, and its technical characteristics are presented in Table~\ref{table:Data_STU} based on data from CIEMAT Spain~\cite{web:ciemat_spain}.
The efficiency of thermal-to-electrical energy conversion in the turbine varies depending on the thermal power of the \ac{pb}, following the piecewise linear relationship depicted in Figure~\ref{fig:PB_conversion_efficiency}. The forecast bounds for \ac{dam} and \ac{srm} electricity prices, derived from historical data in~\cite{REE2025}, are shown in Figure~\ref{fig:Data_Price}. The cost of purchasing thermal energy via a \ac{hpa} is determined by a time-of-use contract between the \ac{rvpp} and the thermal energy service provider. Figure~\ref{fig:Data_Price} illustrates the \ac{hpa} price at different hours of the day, estimated based on the costs associated with thermal energy production using various technologies in Spain~\cite{Statista2024}. Table~\ref{table:Data_Budget} provides information on the assumed uncertainty budget for different uncertain parameters across various case studies. Since the production of solar \acp{pv} and the \ac{sf} of \ac{csp} is zero during night hours, a lower uncertainty budget is considered for these units. In this way, the percentage of hours with deviations over the 24-hour period is defined within the same range for all uncertain parameters.

\begin{table*}[t!]
  \centering
    {\color{black}
  \caption{Economic data.}
    \label{table:Data_Economic}
  \small
  \setlength{\tabcolsep}{1pt}
  \renewcommand{\arraystretch}{1}
  \vspace{-.5em}
 %\small
  \begin{threeparttable}
  \begin{tabular}{lcc}
    \toprule

    \multicolumn{1}{c}{\textbf{Market data}}  
    && \multicolumn{1}{c}{\textbf{Value}} 
    \\

    \cmidrule{1-1} \cmidrule{3-3} 

   \multirow{1}{*}{\text{\ac{dam} price ranges [€/MWh]}}  &&  35-167 \\ [0.2em]

    \multirow{1}{*}{\text{Up \ac{srm} price ranges [€/MW] }}  &&  6-29 \\ [0.2em]

    \multirow{1}{*}{\text{Down \ac{srm} price ranges [€/MW]}}  &&   1-22 \\[0.2em]

    \multirow{1}{*}{\text{\ac{hpa} price ranges [€/MW]}}  &&  70-120 \\ [0.2em]

    \toprule
    \multicolumn{1}{c}{\textbf{Operation cost data}}  
    && \multicolumn{1}{c}{\textbf{Value}} 
    \\

    \cmidrule{1-1} \cmidrule{3-3}

    \multirow{1}{*}{\text{\ac{csp} operation cost [€/MWh]
}}    && 25 \\ [0.2em]

   \multirow{1}{*}{\text{\ac{wf} operation cost [€/MWh]}}   && 15    \\ [0.2em]

      \multirow{1}{*}{\text{\ac{pv} operation cost [€/MWh]}}  &&  7.5   \\ [0.2em] 

    \toprule
    \multicolumn{1}{c}{\textbf{Uncertain data}}  
    && \multicolumn{1}{c}{\textbf{Value}} 
    \\

    \cmidrule{1-1} \cmidrule{3-3}

%      \multirow{1}{*}{\text{Price date time}}  &&  December 2024   \\ [0.2em] 

%     \multirow{1}{*}{\text{\ac{pv} date time}}  &&  March, May 2021   \\ [0.2em] 

%     \multirow{1}{*}{\text{\ac{wf} date time}}  &&  March 2021   \\ [0.2em] 

%          \multirow{1}{*}{\text{\ac{csp} date time}}  &&  March 2018   \\ [0.2em] 

     \multirow{1}{*}{\text{Forecast percentile [\%]}}  &&  20-80    \\ [0.2em] 

\bottomrule
  \end{tabular}
\end{threeparttable}}
      %\vspace{-2mm}
\end{table*}

\begin{table*}[t!]
  \centering
    {\color{black}
  \caption{\ac{ed} and \ac{td} data.}
    \label{table:Data_Demand}
  \small
  \setlength{\tabcolsep}{2pt}
  \renewcommand{\arraystretch}{1}
  \vspace{-.5em}
 %\small
  \begin{threeparttable}
  \begin{tabular}{lcccccccc}
    \toprule

    \multicolumn{1}{c}{}  
    && \multicolumn{3}{c}{\textbf{\ac{ed}}}
    && \multicolumn{2}{c}{\textbf{\ac{td}}} 
    \\

 \cmidrule{3-5} \cmidrule{7-8}

    \multicolumn{1}{c}{\textbf{Parameter}}  
    && \multicolumn{1}{c}{\textbf{Industry}} & \multicolumn{1}{c}{\textbf{Service}} & \multicolumn{1}{c}{\textbf{Residential}}
    && \multicolumn{1}{c}{\textbf{Industry}}  & \multicolumn{1}{c}{\textbf{Residential}} 
    \\

  \cmidrule{1-1} \cmidrule{3-5} \cmidrule{7-8}

   \multirow{1}{*}{\text{Minimum energy consumption [MWh]}}    && 600 & 800 & 600
   && 700 & 700 \\ [0.2em]

   \multirow{1}{*}{\text{Minimum power output [MW]}}    && 2 & 10 & 5
   && 10 & 5 \\ [0.2em]

    \multirow{1}{*}{\text{Maximum power output [MW]}}    && 70 & 80 & 60
   && 80 & 70 \\ [0.2em]

    \multirow{1}{*}{\text{Initial power output [MW]}}    && 5 & 42 & 25
   && 23 & 11 \\ [0.2em]

     \multirow{1}{*}{\text{Secondary reserve ramp rate [MW/min]}}    && 30 & 25  & 10 && - & - \\ [0.2em]

    \multirow{1}{*}{\text{Secondary reserve relative to the power capacity [\%]}}    && 10 & 10  & 10 && - & - \\ [0.2em]

\bottomrule
  \end{tabular}
\end{threeparttable}}
      %\vspace{-2mm}
\end{table*}

\begin{table*}[t!]
  \centering
      {\color{black}
  \caption{\ac{csp} data.}
    \label{table:Data_STU}
  \vspace{-.5em}
 \small
  \begin{threeparttable}
  \begin{tabular}{lcc}
    \toprule

    \multicolumn{1}{c}{\textbf{\ac{sf}}}  
    & \multicolumn{1}{c}{\textbf{}} 
    \\

    \cmidrule{1-2}

%    \multirow{1}{*}{\text{Concentrating solar technology \& fluid}}    &  \makecell{Parabolic-trough\\with thermal oil} \\ [0.2em]

    \multirow{1}{*}{\text{\ac{sf} maximum thermal power output [MW]}}    & 300  \\ [0.2em]

    \toprule

    \multicolumn{1}{c}{\textbf{\ac{pb}/Turbine}}  
    & \multicolumn{1}{c}{\textbf{}} 
    \\

    \cmidrule{1-2}

    \multirow{1}{*}{\text{\ac{pb} maximum thermal power input [MW]}}    & 140 \\ [0.2em] 

    \multirow{1}{*}{\text{Turbine maximum electrical power output [MW]}}    & 55 \\ [0.2em] 

    \multirow{1}{*}{\text{Turbine minimum electrical power output [MW]}}    & 5 \\ [0.2em] 

    \multirow{1}{*}{\text{Turbine minimum up/down time [hour]}}    & 6 \\ [0.2em]

    \multirow{1}{*}{\text{\ac{csp} output thermal power efficiency [\%]
}}    & 90 \\ [0.2em]

    \multirow{1}{*}{\text{\ac{csp} secondary reserve ramp rate [MW/min]
}}    & 25 \\ [0.2em]

    \multirow{1}{*}{\text{\ac{csp} secondary reserve relative to power capacity  [\%]
}}    & 50 \\ [0.2em]

    \toprule

    \multicolumn{1}{c}{\textbf{\ac{ts}}}  
    & \multicolumn{1}{c}{\textbf{}} 
    \\

    \cmidrule{1-2}

%   \multirow{1}{*}\text{\ac{ts} technology \& fluid} & 2-tank with molten salt    \\ [0.2em]

   \multirow{1}{*}\text{Maximum thermal energy [MWh]}
   & 1100    \\ [0.2em]

    \multirow{1}{*}\text{Minimum thermal energy [MWh]}   & 110     \\ [0.2em] 

    \multirow{1}{*}Discharging thermal power [MW]    & 115   \\ [0.2em] 

    \multirow{1}{*}Charging thermal power [MW]   & 140   \\ [0.2em]

    \multirow{1}{*}\text{Dis/charging efficiency [\%]}  & 95   \\ [0.2em]

\bottomrule
  \end{tabular}
\end{threeparttable}}
      %\vspace{-2mm}
\end{table*}

\begin{table*}[t!]
  \centering
      {\color{black}
  \caption{\ac{csp} main design data.}
    \label{table:Design_Data1_STU}
  \vspace{-.5em}
 \small
  \begin{threeparttable}
  \begin{tabular}{lcc}
    \toprule

    \multicolumn{1}{c}{\textbf{\ac{sf} characteristics}}  
    & \multicolumn{1}{c}{\textbf{Value}} 
    \\

    \cmidrule{1-2}

    \multirow{1}{*}{\text{Concentrating solar technology}}    &  Parabolic-trough \\ [0.2em]

    \multirow{1}{*}{\text{Heat transfer fluid}}    &  Oil \\ [0.2em]

    \multirow{1}{*}{\text{Heat transfer fluid temperature range} [°C]}    &  296-390 \\ [0.2em]

    \multirow{1}{*}{\text{Receiver type}}    & Tube  \\ [0.2em]
    
    \multirow{1}{*}{\text{Mirror area} [$\mathrm{m^2}$]}    &  3270 \\ [0.2em]    

    \multirow{1}{*}{\text{\ac{ts} medium}}    &  2-tank with molten salt  \\ [0.2em]

    \multirow{1}{*}{\text{\ac{ts} capacity} [MWh]}    &  1100 \\ [0.2em]

    \multirow{1}{*}{\text{\ac{ts} duration} [h]}    &  7.5 \\ [0.2em]

    \multirow{1}{*}{\text{Turbine type or cycle}}    &  Rankine steam cycle \\ [0.2em]

    \multirow{1}{*}{\text{\ac{pb} efficiency} [\%]}    &  27-40 \\ [0.2em]

\bottomrule
  \end{tabular}
\end{threeparttable}}
\end{table*}

\begin{comment}
  /\ac{pb}/Turbine  

    \multicolumn{1}{c}{\textbf{\ac{sf}/\ac{pb}/Turbine}}  
    & \multicolumn{1}{c}{\textbf{value}} 
    \\

    && \multicolumn{1}{c}{\textbf{\ac{ts}}}
    & \multicolumn{1}{c}{\textbf{value}} 
\end{comment}

\begin{table*}[t!]
  \centering
    {\color{black}
  \caption{\ac{ndrs} data.}
    \label{table:Data_NDRES}
  \small
  \setlength{\tabcolsep}{1pt}
  \renewcommand{\arraystretch}{1}
  \vspace{-.5em}
 %\small
  \begin{threeparttable}
  \begin{tabular}{lccccc}
    \toprule

    \multicolumn{1}{c}{\textbf{Parameter}}  
    && \multicolumn{1}{c}{\textbf{\ac{pv}}}
    && \multicolumn{1}{c}{\textbf{\ac{wf}}} 
    \\

    \cmidrule{1-1} \cmidrule{3-3} \cmidrule{5-5}

   \multirow{1}{*}{\text{Maximum/minimum power output [MW]}}    && 50/0 
   && 50/0  \\ [0.2em]

     \multirow{1}{*}{\text{Secondary reserve ramp up rate [MW/min]}}    && 10 && 15    \\ [0.2em]

    \multirow{1}{*}{\text{Secondary reserve ramp down rate [MW/min]}}    && 25 && 25    \\ [0.2em]

    \multirow{1}{*}{\text{Secondary reserve relative to power capacity [\%]}}    && 10 && 10    \\ [0.2em]

\bottomrule
  \end{tabular}
\end{threeparttable}}
     % \vspace{-2mm}
\end{table*}

\begin{table*}[t!]
  \centering
  \caption{Uncertainty budgets for different uncertain parameters for different RVPP strategies (optimistic, balanced, and pessimistic).}
 \small
\vspace{-.5em}
  \begin{threeparttable}
  \begin{tabular}{lcccccc}
    \toprule

    \multicolumn{1}{c}{\textbf{}}   
    && \multicolumn{1}{c}{\textbf{\ac{dam}/\ac{srm}}} 
    & \multicolumn{1}{c}{\textbf{\acp{wf}}}
    & \multicolumn{1}{c}{\textbf{PVs}} 
    & \multicolumn{1}{c}{\textbf{\ac{sf} thermal}}
    & \multicolumn{1}{c}{\textbf{\acp{ed}/\acp{td}}}
    \\

    \multicolumn{1}{c}{\textbf{Strategy}}   
    && \multicolumn{1}{c}{\textbf{prices}} 
    & \multicolumn{1}{c}{\textbf{production}}
    & \multicolumn{1}{c}{\textbf{production}} 
    & \multicolumn{1}{c}{\textbf{production}}
    & \multicolumn{1}{c}{\textbf{consumption}}
    \\

 \cmidrule{1-1} \cmidrule{3-7}

    \multirow{1}{*}{\textbf{Optimistic}}    && 3 & 3  &  2 & 2 & 3  \\ [0.2em]

    \multirow{1}{*}{\textbf{Balanced}}    && 6 & 6   & 4 & 4 &  6 \\ [0.2em]

    \multirow{1}{*}{\textbf{Pessimistic}}    && 9 & 9   & 6 & 6 & 9 \\ [0.2em]

\bottomrule
  \end{tabular}
\end{threeparttable}
  \label{table:Data_Budget}
\end{table*}

Three case studies are performed to assess the effectiveness of the proposed model. In the first case study, the scheduling of \ac{rvpp} units and the traded electrical and thermal energy of \ac{rvpp} are obtained without considering uncertainty. The thermal-to-electrical energy conversion of \ac{csp}, the energy in its different elements, and the proposed approach to adjusting a share of \ac{ts} energy for reserve provision are studied. In the second case study, the performance of the proposed two-stage \ac{ro} approach in handling different uncertainties—such as electricity price, \ac{ndrs} production, thermal production of \ac{csp}, and consumption of \acp{ed}/\acp{td}—is studied for three different \ac{rvpp} operator strategies (optimistic, balanced, and pessimistic) against uncertainty. In the third case study, the profitability of \ac{rvpp} under different market (contract) participation strategies and various decision-making approaches for managing uncertainties is evaluated to demonstrate the effectiveness of the proposed coordinated approach compared to separate market participation strategies. Furthermore, the effect of different possible future prices of \ac{hpa} (scenarios with thermal energy production from renewable resources rather than fossil fuels)~\cite{Statista2024} is evaluated. {\color{black}In the fourth case study, the performance of the proposed two-stage \ac{ro} approach is compared to the \ac{sp} approach proposed in~\citep{zhao2021coordinated} using out-of-sample assessment.}

The characteristics of these three case studies are summarized as follows:

\begin{itemize}
    \item Case 1: Identify the deterministic optimal operation of \ac{rvpp} units, including \ac{csp}, for supplying both \acp{ed}/\acp{td}. Additionally, determine the \ac{rvpp}'s bidding strategy in the electrical energy market and its participation in thermal \ac{hpa} contract.

    \item Case 2: Assess the optimal bidding strategy of the \ac{rvpp} while accounting for uncertainties related to \ac{dam} and \ac{srm} prices, \ac{ndrs} production, and variations in \acp{ed}/\acp{td}.

      \item Case 3: Analyze the \ac{rvpp}'s profitability under different market (contract) participation strategies, considering different decision-making approaches to manage uncertainties.

      \item {\color{black}Case 4: Compare the performance of the proposed two-stage \ac{ro} approach with the \ac{sp} model in~\citep{zhao2021coordinated}, considering different decision-making approaches in the proposed model and varying numbers of scenarios in~\citep{zhao2021coordinated}.}

\end{itemize}

Simulations are conducted on a Dell XPS equipped with an i7-1165G7 2.8 GHz processor and 16 GB of RAM using the CPLEX solver in GAMS 39.1.1. The computation time for all simulations remains under 5 seconds, demonstrating the computational efficiency of the proposed two-stage \ac{ro} approach for \ac{rvpp} market participation.

\subsection{Case 1}
\label{subsec: Case 1}

\begin{figure} [b!]
    \centering  \includegraphics[width=.92\textwidth]{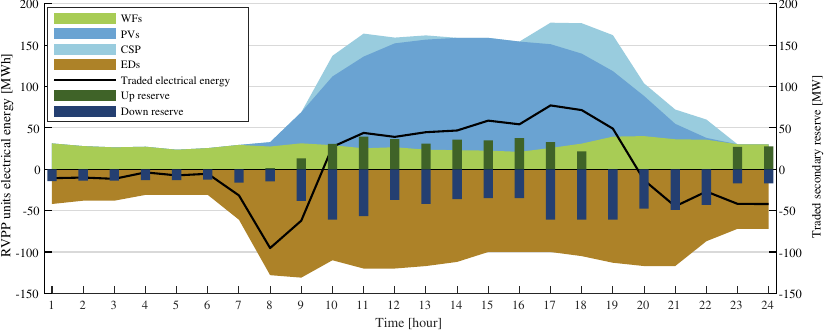}
    \vspace{-.5em}
    \caption{The RVPP traded electrical energy in DAM and traded reserve in SRM, and electrical energy of RVPP units.}
    \label{fig:RVPP_Units_energy}
    \vspace{-2mm}
\end{figure}

Figure~\ref{fig:RVPP_Units_energy} shows the traded electrical energy and reserve of the \ac{rvpp} in the \ac{dam} and \ac{srm}, as well as the cumulative energy of the \ac{rvpp} units for the deterministic case. The figure illustrates that the \ac{rvpp} is an energy seller in the \ac{dam} between hours 10–19, and an energy buyer during other hours. This is due to the high available power from solar \acp{pv} and \ac{csp} production during hours 10-19. In the night and early morning hours (1–6), the \acp{ed} are low and are mainly provided by the \acp{wf}. However, the \ac{rvpp} buys a small amount of energy during these hours to maintain supply and demand balance. Between hours 7–9, the demand increases, and solar \ac{pv} production is minimal or zero. As a result, the \ac{rvpp} buys a larger amount of energy during these hours to supply its demands. The \ac{csp} turbine starts up at hour 10 and produces electrical energy between hours 10–13 to supply the morning demand and between hours 17–22 to supply the evening demand (when solar \acp{pv} production is not at its maximum and demand is high). The \ac{rvpp} provides upward reserve in the \ac{srm} between hours 8–24 and downward reserve in all hours. The upward reserve is mainly provided by the \ac{csp} between hours 10–18 and 23–24, and by the demands between hours 8–18. The demands also provide downward reserve throughout all hours. Since the operational cost of the \ac{csp} is higher than that of solar \acp{pv} and \acp{wf}, these units only produce energy and do not provide reserve.

\begin{figure} [t!]
    \centering  \includegraphics[width=.92\textwidth]{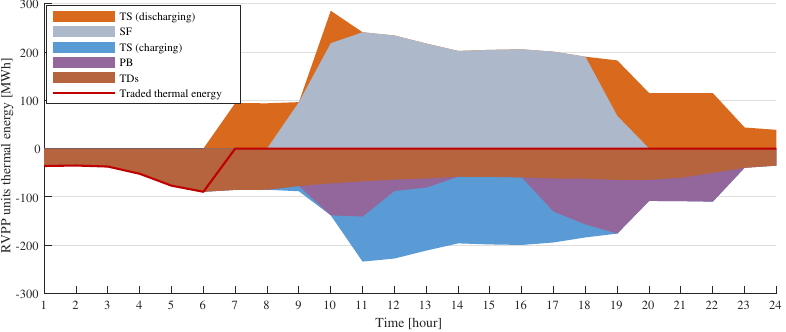}
    \vspace{-.5em}
    \caption{The RVPP thermal energy through HPA and thermal energy of RVPP units.}
    \label{fig:RVPP_Units_Tenergy}
    \vspace{-2mm}
\end{figure}

Figure~\ref{fig:RVPP_Units_Tenergy} shows the thermal energy traded through the \ac{hpa} contract, along with the thermal production and demands of the \ac{rvpp}. The \ac{csp} provides all the energy for \acp{td} between hours 7–24. This energy is supplied either by discharging the \ac{ts} during hours when the sun is not available, by using the \ac{sf} of the \ac{csp}, or by a combination of both. During the night hours (1–6), the thermal energy for demands is provided through the \ac{hpa} contract. These hours are selected for buying thermal energy since the price of thermal energy is lower compared to other hours (see Figure~\ref{fig:Data_Price}). Additionally, by purchasing thermal energy through the \ac{hpa} contract during these hours, the \ac{csp} can retain enough energy to supply both \acp{ed}/\acp{td} during the morning hours.

\begin{figure} [t!]
    \centering  \includegraphics[width=.95\textwidth]{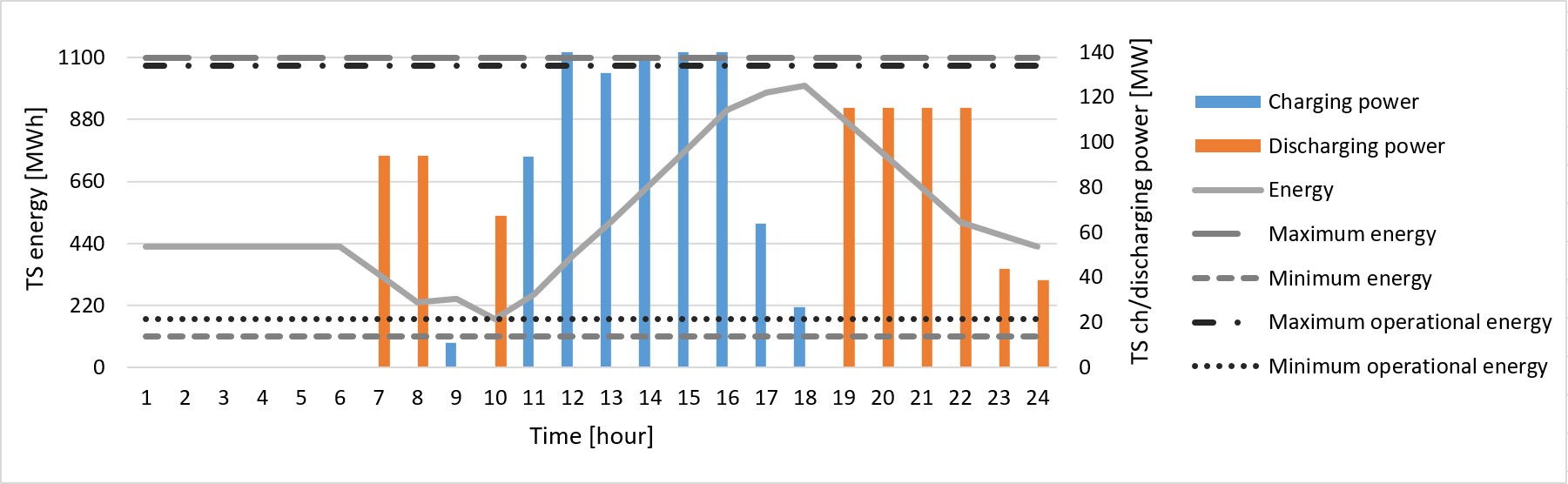}
    \vspace{-.5em}
    \caption{The TS energy and charging/discharging power.}
    \label{fig:Ts_energy_determinitic}
    \vspace{-2mm}
\end{figure}

Figure~\ref{fig:Ts_energy_determinitic} shows the energy and charging/discharging power of the \ac{ts} of the \ac{csp}. The figure illustrates that the \ac{ts} discharges during hours 7, 8, 10, and 19–24. In hours 7 and 8, as discussed in Figure~\ref{fig:RVPP_Units_Tenergy}, the discharged power of the \ac{ts} is used to supply the local \acp{td} of the \ac{rvpp}. In hour 10, the discharging power of the \ac{ts} assists the \ac{csp} turbine in starting up and producing electricity. Between hours 19 and 24, the discharged power of the \ac{ts} is used to support the turbine and/or supply \acp{td}. Figure~\ref{fig:Ts_energy_determinitic} also shows additional margins compared to the maximum and minimum energy levels of the \ac{ts}, which are allocated solely for providing reserve. These margins are determined by the optimization problem to avoid the complete depletion or operation at the maximum energy level of the \ac{ts}. In this way, when the \ac{tso} requests upward or downward reserve, there is enough capacity available to provide it. The algorithm adjusts these margins based on the reserve that needs to be provided by the \ac{csp}. For example, if providing reserve is not favorable for the \ac{rvpp}, the optimization problem relaxes these margins accordingly. It is important to note that if these margins are not considered, the \ac{rvpp} may face penalization or exemption from the market due to its inability to provide the requested reserve by the \ac{tso}. For instance, at hour 10, if the energy level of the \ac{ts} were at its minimum (represented by the dashed grey line), it would not be possible for the \ac{csp} to provide the requested upward reserve during this time period.

\subsection{Case 2}
\label{subsec: Case 2}

\begin{figure} [t!]
    \centering  \includegraphics[width=.95\textwidth]{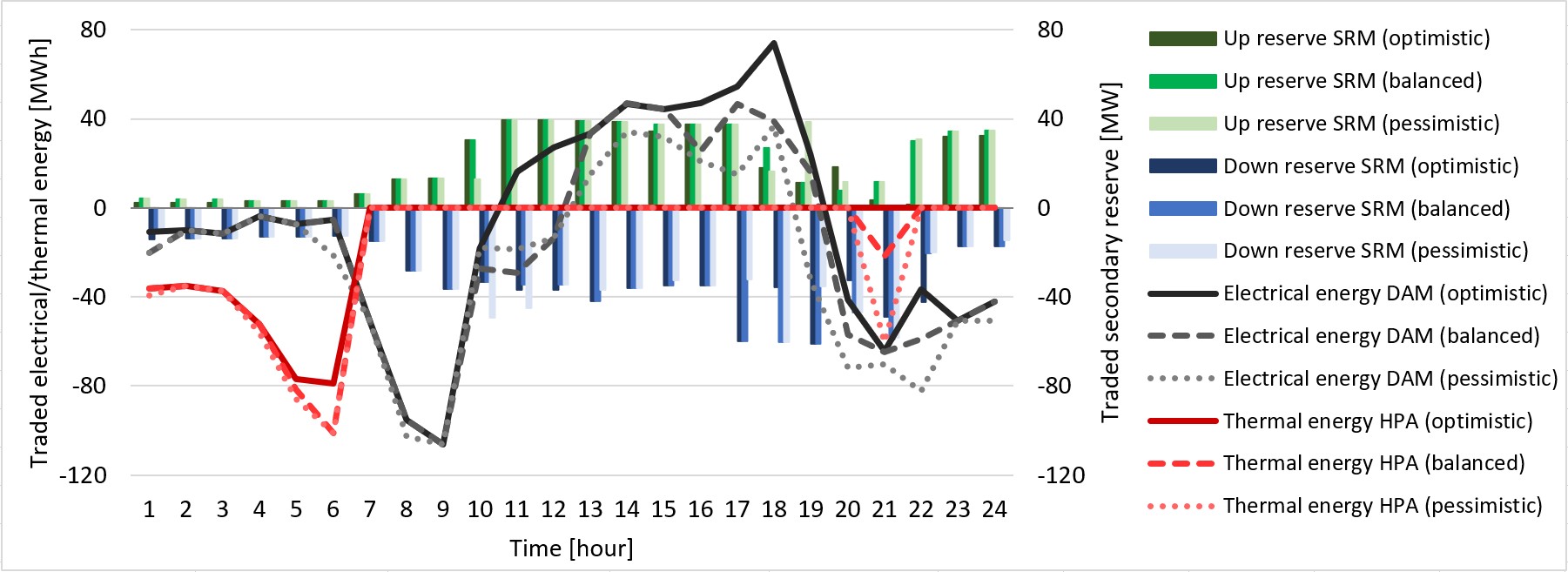}
    \vspace{-.5em}
    \caption{The RVPP traded electrical and thermal energy by considering different uncertainties.}
    \label{fig:RVPP_trade_case2}
    \vspace{-2mm}
\end{figure}

Figure~\ref{fig:RVPP_trade_case2} shows the traded electrical energy and reserve of the \ac{rvpp} in the \ac{dam} and \ac{srm}, as well as the purchased thermal energy through the \ac{hpa} from the thermal service provider, while considering different uncertainties. For the optimistic decision (only a small number of hours are considered as the worst case), the \ac{rvpp} purchases the lowest amount of electrical and thermal energy when it is an energy buyer in the market. Conversely, when the \ac{rvpp} is a seller in the market, it sells more electrical energy. 
%This is because only a small number of hours (2 or 3 out of 24, as shown in Table~\ref{table:Data_Budget}) are considered as the worst case in the optimization problem. 
For the balanced decision, the amounts of purchased electrical and thermal energy increase, while the amount of sold electrical energy decreases compared to the optimistic decision. Specifically, during hours 5, 6, and 21, the \ac{rvpp} buys more thermal energy to manage \acp{td} uncertainties. The traded electrical energy is affected in more hours compared to thermal energy, as it is influenced by multiple uncertainties, including variations in the production of \acp{wf}, solar \acp{pv}, and \ac{csp}, as well as fluctuations in \acp{ed}. Since the worst-case scenario for production or demand can occur at different hours, the hours 11, 12, 16, 18, 20, and 22 exhibit the highest deviations in traded electrical energy compared to the optimistic decision. As a significant portion of the \ac{rvpp}'s energy production comes from solar \acp{pv}, most of these hours correspond to periods of high volatility in solar production—morning hours (11 and 12) and evening hours (16, 18, and 20). Additionally, some hours are strongly affected by uncertainty, leading to a change in the direction of the \ac{rvpp}'s traded electrical energy. For example, in hours 11 and 12, the \ac{rvpp} becomes an energy buyer, whereas in the optimistic decision, it was an energy seller during these hours.

For the pessimistic decision, a greater number of hours (6 or 9 out of 24, according to Table~\ref{table:Data_Budget}) are selected as the worst case for each uncertain parameter. As a result, the \ac{rvpp} operator adopts a more conservative strategy. The thermal energy purchased by the \ac{rvpp} through the \ac{hpa} increases in hours 1, 4, 5, and 21, with a more significant increase in hour 21. In this hour, the \acp{td} of the \ac{rvpp} remain similar to those in the balanced decision. However, the \ac{rvpp} prefers to purchase more thermal energy at this time to compensate for a lack of thermal energy in the \ac{csp}. This shortage of thermal energy in the \ac{csp} results from greater variations in the production and demand of other units in previous hours, which the \ac{csp} would otherwise compensate for. The \ac{rvpp} is an electrical energy seller only between hours 13-18; during all other hours, it buys electrical energy from the market. %This is because, in most hours of the 24-hour period, there are variations in uncertain parameters. Therefore, the \ac{rvpp} prefers to purchase electricity from the market to ensure it can meet its demands. 
The variation of traded electrical energy from the balanced to the pessimistic decision is more significant in hours 6, 13-15, 17, 19, 20, and 22 than in other hours. These hours are typically those in which \ac{rvpp} units still experience moderate to high deviations in production and demand but have a subsequent effect on the objective function of the optimization problem compared to worst case hours in the balanced decisions.

\begin{figure} [t!]
    \centering  \includegraphics[width=.95\textwidth]{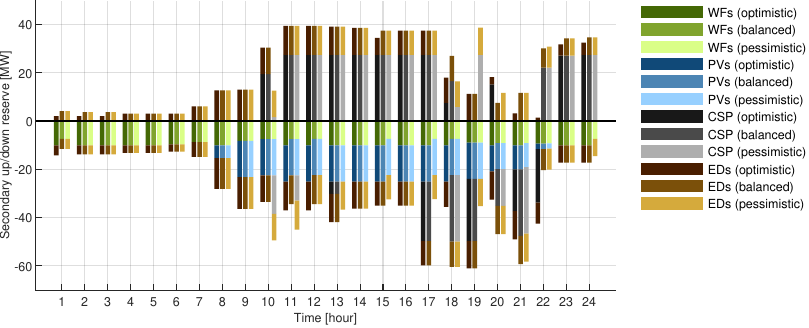}
    \vspace{-.5em}
    \caption{The RVPP units provided up and down reserve by considering different uncertainties.}
    \label{fig:RVPP_reserve_case2}
    \vspace{-2mm}
\end{figure}

Figure~\ref{fig:RVPP_trade_case2} also illustrates the changes in upward and downward reserves provided under the three decision strategies considered. In hour 10, under the pessimistic strategy, the upward reserve provided by the \ac{csp} is reduced compared to the optimistic and balanced strategies. Instead of providing upward reserve, the \ac{csp} generates more energy to meet higher levels of \acp{ed}/\acp{td} due to uncertainty. This higher production in hour 10 enables the \ac{csp} to offer more downward reserve compared to the optimistic and balanced strategies. In hour 19, the opposite situation occurs: the \ac{rvpp} provides a higher amount of upward reserve and a lower amount of downward reserve under the pessimistic strategy compared to the other two strategies. The \ac{csp} does not generate electrical energy in this hour, allowing it to allocate more capacity for upward reserve. Conversely, since the \ac{csp} is not producing energy, it cannot provide any downward reserve in this hour.

To provide a more in-depth analysis of the reserve provision by the \ac{rvpp}, Figure~\ref{fig:RVPP_reserve_case2} presents the reserve contribution of each technology under the three strategies. The figure shows that, for all strategies, both the \ac{csp} and demands provide upward reserve. Since the production cost of solar \acp{pv} and \acp{wf} is lower than that of the \ac{csp}, they do not contribute to upward reserve provision. The \ac{csp} has a higher share than demands in providing upward reserve due to the flexibility offered by its \ac{ts}. All technologies in the \ac{rvpp} contribute to downward reserve provision. A significant portion of the downward reserve is supplied by solar \acp{pv} between hours 9-19 due to their high production during these hours. Between hours 17-22, the \ac{csp} provides the majority of the downward reserve, as its production level is higher in the evening. \acp{wf} contribute to downward reserve provision throughout all time periods, as they can reduce their output from the maximum production level when needed.

{\color{black}
Table~\ref{table:RVPP_flexibility_Case2} compares the defined quantitative flexibility metrics for different \ac{rvpp} units (as described in Section~\ref{subsubsec:Flexibility_Metrics}) under various uncertainty handling strategies. The results show that the \ac{csp} unit provides the highest contribution—measured by the reserve-to-capacity ratio—in providing upward reserve compared to other \ac{rvpp} units. Furthermore, the upward reserve-to-capacity ratio is increased by adopting a more conservative strategy for both \ac{csp} and \acp{ed}, enabling them to more effectively handling uncertainty. These findings highlight the critical role of \ac{csp} in enhancing the overall flexibility of the \ac{rvpp}.}

\begin{table*}[t!]
  \centering
    {\color{black}
  \caption{The flexibility provided by RVPP units by considering different uncertainties.}
    \label{table:RVPP_flexibility_Case2}
 \small
   \setlength{\tabcolsep}{2pt} % Adjust 3pt to a smaller value if needed
  \renewcommand{\arraystretch}{1} % Adjust the factor (default is 1.0)
\vspace{-.5em}
  \begin{threeparttable}
  \begin{tabular}{llccccc}
    \toprule

     \multicolumn{2}{c}{\textbf{}}   
    && \multicolumn{4}{c}{\textbf{Flexibility meteric}} 
    \\

     \cmidrule{4-7}

    \multicolumn{1}{c}{\textbf{Strategy}}  
    & \multicolumn{1}{c}{\textbf{Unit}}
    && \multicolumn{1}{c}{\textbf{\makecell{Upward \\ reserve [MW]}}} 
    & \multicolumn{1}{c}{\textbf{\makecell{Downward \\ reserve [MW]}}}
    & \multicolumn{1}{c}{\textbf{\makecell{Upward reserve \\to capacity [\%]}}} 
    & \multicolumn{1}{c}{\textbf{\makecell{Downward reserve \\to capacity [\%]}}}
    \\

 \cmidrule{1-2} \cmidrule{4-7}

%    \multirow{4}{*}{\textbf{Deterministic}} & {\textbf{\acp{wf}}} && 0  & 239.5   & 0  & 2.39    \\ [0.2em] 

%    \multirow{1}{*}{\textbf{}} & {\textbf{\acp{pv}}} && 0 & 193.3   & 0  & 1.29    \\ [0.2em] 

%    \multirow{1}{*}{\textbf{}} & {\textbf{\ac{csp}}} && 276.5 & 182.7   & 5.03  & 3.32    \\ [0.2em] 

%    \multirow{1}{*}{\textbf{}} & {\textbf{\acp{ed}}} && 93.0 & 207.3   & 0.44  & 0.99    \\ [0.2em] 

%    \toprule

    \multirow{4}{*}{\textbf{Optimistic}} & {\textbf{\acp{wf}}} && 0 & 232.1   & 0  & 2.32    \\ [0.2em] 

    \multirow{1}{*}{\textbf{}} & {\textbf{\acp{pv}}} && 0 & 193.3   & 0  & 1.29    \\ [0.2em] 

    \multirow{1}{*}{\textbf{}} & {\textbf{\ac{csp}}} && 289.4  & 95.1   & 5.26  &  1.73   \\ [0.2em] 

    \multirow{1}{*}{\textbf{}} & {\textbf{\acp{ed}}} && 171.8 &  209.3  & 0.82  & 1.00    \\ [0.2em] 

    \toprule

    \multirow{4}{*}{\textbf{Balanced}} & {\textbf{\acp{wf}}} && 0 & 220.6  & 0  & 2.20    \\ [0.2em] 

    \multirow{1}{*}{\textbf{}} & {\textbf{\acp{pv}}} && 0 &  193.3  &  0 & 1.29    \\ [0.2em] 

    \multirow{1}{*}{\textbf{}} & {\textbf{\ac{csp}}} && 305.3 & 126.0   & 5.55  & 2.29    \\ [0.2em] 

    \multirow{1}{*}{\textbf{}} & {\textbf{\acp{ed}}} && 204.5 &  209.3  & 0.97  & 1.00    \\ [0.2em] 

    \toprule

    \multirow{4}{*}{\textbf{Pessimistic}} & {\textbf{\acp{wf}}} && 0 & 211.6  & 0  & 2.11    \\ [0.2em] 

    \multirow{1}{*}{\textbf{}} & {\textbf{\acp{pv}}} && 0 & 193.3   &  0 & 1.29    \\ [0.2em] 

    \multirow{1}{*}{\textbf{}} & {\textbf{\ac{csp}}} && 304.4 & 96.8   & 5.53  & 1.76    \\ [0.2em] 

    \multirow{1}{*}{\textbf{}} & {\textbf{\acp{ed}}} && 209.3 & 209.3   & 1.00  &  1.00   \\ [0.2em] 

\bottomrule
  \end{tabular}
\end{threeparttable}}
\end{table*}

\begin{figure} [t!]
    \centering  \includegraphics[width=.95\textwidth]{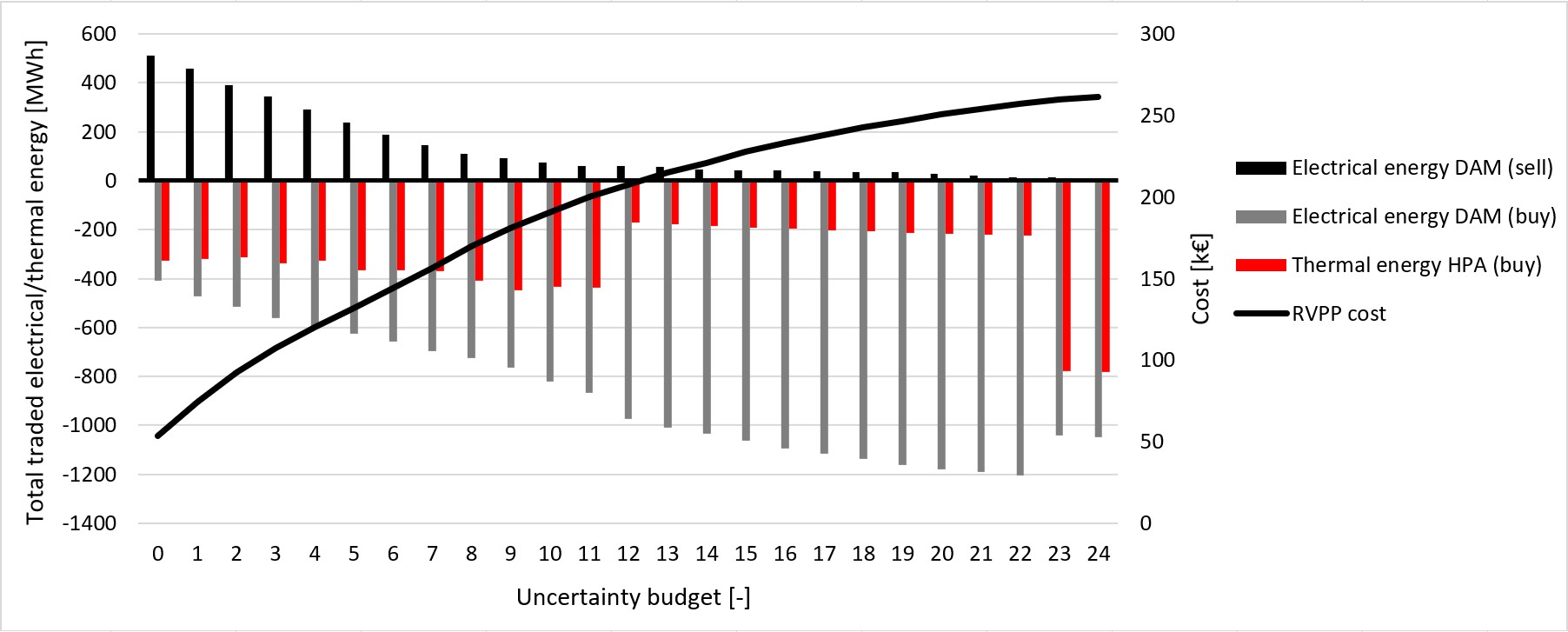}
    \vspace{-.5em}
    \caption{  {\color{black}Sensitivity analysis for the RVPP cost as well as total traded electrical and thermal energy versus uncertainty budget.}}
    \label{fig:RVPP_total_trade_case2}
    \vspace{-2mm}
\end{figure}

\begin{figure} [t!]
    \centering  \includegraphics[width=.95\textwidth]{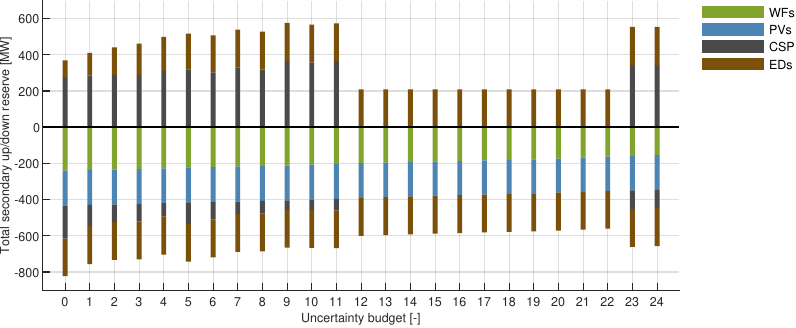}
    \vspace{-.5em}
    \caption{  {\color{black}Sensitivity analysis for the RVPP units provided up and down reserve versus uncertainty budget.}}
    \label{fig:RVPP_total_reserve_case2}
    \vspace{-2mm}
\end{figure}

{\color{black}A sensitivity analysis is performed to examine how varying the uncertainty budget from 0 to 24 affects the bidding strategy, reserve allocation, and overall profitability of the \ac{rvpp}. The results of this analysis are presented in Figures~\ref{fig:RVPP_total_trade_case2} and~\ref{fig:RVPP_total_reserve_case2}. These figures illustrate that adopting a higher uncertainty budget leads the \ac{rvpp} to follow a more conservative strategy—selling less electricity when acting as a seller and purchasing more when acting as a buyer in the market. This shift results in lower profitability (i.e., higher total cost) under more conservative scenarios. Additionally, the total upward and downward reserve provisions are closely linked to the volume of traded electrical and thermal energy, demonstrating the interdependence between energy trading and reserve provision. For example, under a low uncertainty budget, the upward reserve is generally lower, while the downward reserve is higher, since the \ac{rvpp} tends to sell more and buy less electricity compared to scenarios with higher uncertainty budgets. Notably, for uncertainty budget values between 12 and 22, the \ac{csp} unit is turned off and does not produce electricity; as a result, it provides no reserve. In these cases, the \ac{csp} allocates most of its capacity to thermal energy provision, thereby reducing \ac{rvpp} thermal energy procurement through the \ac{hpa} contract.}

\subsection{Case 3}
\label{subsec: Case 3}

Table~\ref{table:RVPP_profit_Case3} presents the cost of the \ac{rvpp} under different market (contract) participation strategies and different strategies for handling uncertainties. The table shows that if only \ac{dam} participation is considered, the \ac{rvpp} achieves the highest cost compared to other market (contract) participation strategies. Adopting more conservative strategies against uncertainties leads to an increase in the \ac{rvpp}'s cost for market participation. For example, compared to the deterministic strategy, the cost increases by 67.6\%, 113.5\%, and 152.1\% for the optimistic, balanced, and pessimistic strategies, respectively. These cost increase occur because, as more conservative strategies are implemented, the \ac{rvpp} submits lower bids for selling electricity and higher bids for purchasing electricity from the \ac{dam} to ensure it can supply its demands. 

When the \ac{rvpp} participates in the \ac{dam} for electricity trading and also enters into a \ac{hpa} contract to purchase thermal energy, %(strategy \ac{dam} + \ac{hpa}), 
its cost decreases in the deterministic, optimistic, and balanced strategies by 2.0\%, 1.8\%, and 1.3\%, respectively, compared to the \ac{dam}-only participation strategy. Since the majority of the \acp{td} energy are supplied by the \ac{csp} of the \ac{rvpp}, the cost decrease from considering the \ac{hpa} is relatively marginal. However, the \ac{hpa} can have a more substantial impact on costs when the \ac{hpa} price is lower, when \acp{td} require more energy, or when the \ac{csp} is unable to fully supply the energy needed for \acp{td}. When the \ac{rvpp} is allowed to participate in both the \ac{dam} and \ac{srm}, 
%(strategy \ac{dam} + \ac{srm}),
its cost decreases in the deterministic, optimistic, balanced, and pessimistic strategies by 21.7\%, 10.3\%, 4.4\%, and 3.3\%, respectively, compared to the \ac{dam}-only participation strategy. As more conservative strategies are adopted, the percentage of cost decrease inclines. Uncertainty negatively impacts both the reserve provision of the \ac{rvpp} and the \ac{srm} price, thereby limiting the additional profit that can be gained from participating in the \ac{srm}. Finally, when participation in all \ac{dam}, \ac{srm}, and \ac{hpa} contract is considered, 
%(strategy \ac{dam} + \ac{srm} + \ac{hpa}), 
the \ac{rvpp} achieves its minimum cost. The cost of the \ac{rvpp} decreases in the deterministic, optimistic, balanced, and pessimistic strategies by 23.9\%, 12.0\%, 8.3\%, and 4.1\%, respectively, compared to the \ac{dam}-only participation strategy.

{\color{black}
Table~\ref{table:RVPP_profit_w/o_CSP_Case3} presents the economic analysis of the proposed \ac{rvpp} without its \ac{csp} component, to more thoroughly evaluate the added value of \ac{csp} integration. The results indicate that the total cost of the \ac{rvpp} with \ac{csp}—participating only in the \ac{dam}—is 58.1\%, 44.4\%, 38.3\%, and 35.0\% lower than that of the \ac{rvpp} without \ac{csp} participating in all market and contract types (\ac{dam}+\ac{srm}+\ac{hpa}), under the deterministic, optimistic, balanced, and pessimistic strategies, respectively. For the case where both configurations (\ac{rvpp} with and without \ac{csp}) participate in all market and contract types, the cost reductions achieved with \ac{csp} integration are 68.1\%, 51.1\%, 43.4\%, and 37.6\% under the same respective strategies. These findings highlight the significant incremental value of incorporating \ac{csp} into the \ac{rvpp}, particularly when it is allowed to participate in multiple energy and reserve markets.
}

%Table~\ref{table:RVPP_profit_Case3.2} shows the effect of different \ac{hpa} prices on the cost of the \ac{rvpp}.

To analyze the impact of various potential future prices of \ac{hpa} (considering scenarios where thermal energy is primarily produced from renewable sources instead of fossil fuels~\cite{Statista2024}), Table~\ref{table:RVPP_profit_Case3.2} is presented. The table indicates that as the \ac{hpa} price decreases, the cost of the \ac{rvpp} also decreases. For example, when considering constant \ac{hpa} prices of 70, 60, and 50 €/MWh—compared to the time-of-use \ac{hpa} price in Figure~\ref{fig:Data_Price}—the cost of the \ac{rvpp} decreases by 6.6\%, 24.8\%, and 48.9\% for the deterministic strategy; 4.5\%, 15.1\%, and 28.8\% for the optimistic strategy; 4.2\%, 12.5\%, and 25.2\% for the balanced strategy; and 3.9\%, 11.9\%, and 20.7\% for the pessimistic strategy. A similar trend is observed when adopting more conservative strategies, where cost reduction becomes more challenging. The results in Table~\ref{table:RVPP_profit_Case3} and Table~\ref{table:RVPP_profit_Case3.2} demonstrate that by implementing the proposed coordinated approach in the \ac{dam}, \ac{srm}, and \ac{hpa}, the \ac{rvpp} achieves the lowest cost, highlighting the effectiveness and practicality of the proposed approach in this paper.

\begin{table*}[t!]
  \centering
  \caption{The RVPP cost [k€] by considering different market (contract) participation strategies.}
 \small
\vspace{-.5em}
  \begin{threeparttable}
  \begin{tabular}{lccccc}
    \toprule

     \multicolumn{1}{c}{\textbf{}}   
    && \multicolumn{4}{c}{\textbf{Market (contract)}} 
    \\

     \cmidrule{3-6}

    \multicolumn{1}{c}{\textbf{Strategy}}   
    && \multicolumn{1}{c}{\textbf{DAM}} 
    & \multicolumn{1}{c}{\textbf{DAM+HPA}}
    & \multicolumn{1}{c}{\textbf{DAM+SRM}} 
    & \multicolumn{1}{c}{\textbf{DAM+SRM+HPA}}
    \\

 \cmidrule{1-1} \cmidrule{3-6}

    \multirow{1}{*}{\textbf{Deterministic}}    && 69.9 & 68.5   & 54.7  & 53.2    \\ [0.2em]

    \multirow{1}{*}{\textbf{Optimistic}}    && 117.2 & 115.1  &  105.1 & 103.1   \\ [0.2em]

    \multirow{1}{*}{\textbf{Balanced}}    && 149.3 & 147.3   & 142.7 & 136.9  \\ [0.2em]

    \multirow{1}{*}{\textbf{Pessimistic}}    && 176.2 & 176.2   & 170.4 & 169.0  \\ [0.2em]

\bottomrule
  \end{tabular}
\end{threeparttable}
  \label{table:RVPP_profit_Case3}
\end{table*}

\begin{table*}[t!]
  \centering
      {\color{black}
  \caption{The RVPP cost [k€] (without CSP) by considering different market (contract) participation strategies.}
    \label{table:RVPP_profit_w/o_CSP_Case3}
 \small
\vspace{-.5em}
  \begin{threeparttable}
  \begin{tabular}{lccc}
    \toprule

     \multicolumn{1}{c}{\textbf{}}   
    && \multicolumn{2}{c}{\textbf{Market (contract)}} 
    \\

     \cmidrule{3-4}

    \multicolumn{1}{c}{\textbf{Strategy}}   
    && \multicolumn{1}{c}{\textbf{DAM+HPA}} 
    & \multicolumn{1}{c}{\textbf{DAM+SRM+HPA}}
    \\

 \cmidrule{1-1} \cmidrule{3-4}

    \multirow{1}{*}{\textbf{Deterministic}}    &&  175.3   &  167.0   \\ [0.2em]

    \multirow{1}{*}{\textbf{Optimistic}}    &&  218.3  & 210.8   \\ [0.2em]

    \multirow{1}{*}{\textbf{Balanced}}    && 248.7   & 242.1  \\ [0.2em]

    \multirow{1}{*}{\textbf{Pessimistic}}    &&  276.9  & 271.0  \\ [0.2em]

\bottomrule
  \end{tabular}
\end{threeparttable}}
\end{table*}

\begin{table*}[t!]
  \centering
  \caption{The RVPP cost [k€] by considering different \ac{hpa} price.}
 \small
\vspace{-.5em}
  \begin{threeparttable}
  \begin{tabular}{lccccc}
    \toprule

     \multicolumn{1}{c}{\textbf{}}   
    && \multicolumn{4}{c}{\textbf{\ac{hpa} price [€/MWh]}} 
    \\

     \cmidrule{3-6}

    \multicolumn{1}{c}{\textbf{Strategy}}   
    && \multicolumn{1}{c}{\textbf{Figure~\ref{fig:Data_Price}}} 
    & \multicolumn{1}{c}{\textbf{70}}
    & \multicolumn{1}{c}{\textbf{60}} 
    & \multicolumn{1}{c}{\textbf{50}}
    \\

 \cmidrule{1-1} \cmidrule{3-6}

    \multirow{1}{*}{\textbf{Deterministic}}    && 53.2 & 49.7   & 40.0  &  27.2   \\ [0.2em]

    \multirow{1}{*}{\textbf{Optimistic}}    && 103.1 & 98.5  &  87.6 & 73.4   \\ [0.2em]

    \multirow{1}{*}{\textbf{Balanced}}    && 136.9 & 131.1   & 119.8 & 105.4  \\ [0.2em]

    \multirow{1}{*}{\textbf{Pessimistic}}    && 169.0 & 160.9   & 148.9 & 134.0  \\ [0.2em]

\bottomrule
  \end{tabular}
\end{threeparttable}
  \label{table:RVPP_profit_Case3.2}
\end{table*}

\begin{figure} [ht!]
    \centering  \includegraphics[width=.95\textwidth]{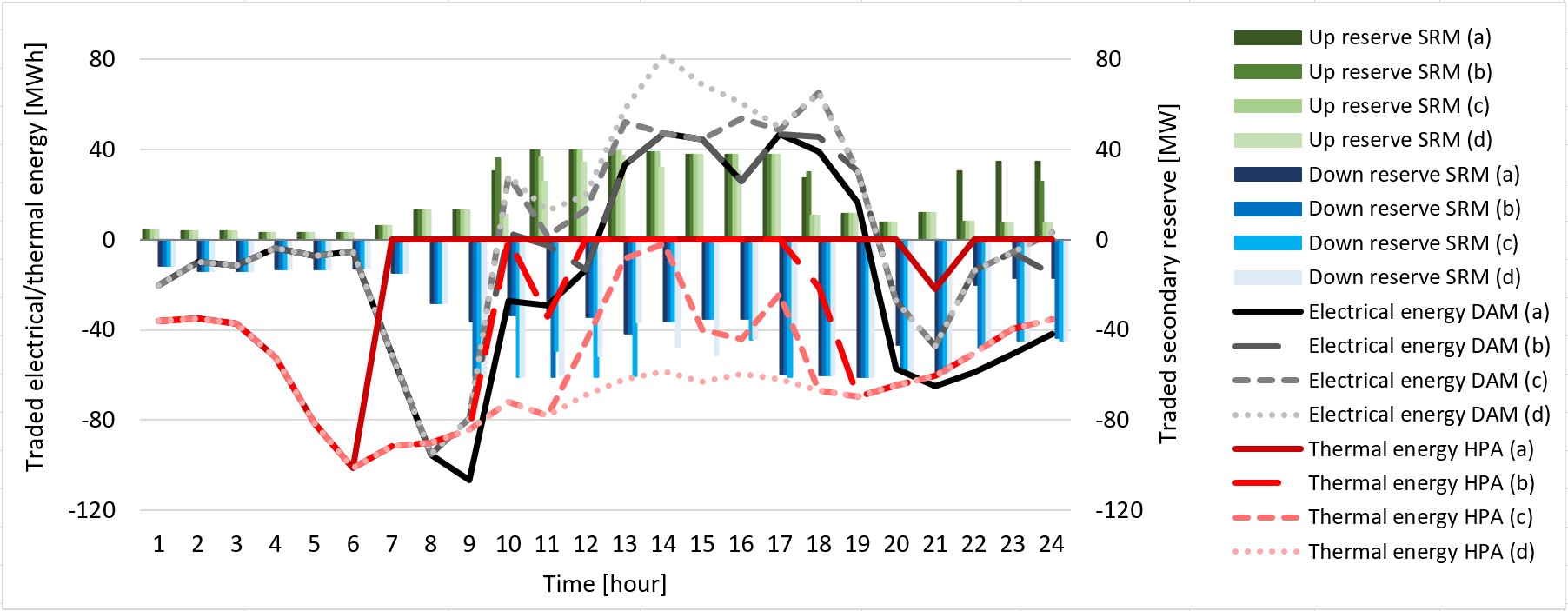}
    \vspace{-.5em}
    \caption{The RVPP traded electrical and thermal energy by considering different \ac{hpa} price [€/MWh]. (a) according to Figure~\ref{fig:Data_Price};  (b) 70; (c) 60; and (d) 50 €/MWh, $\forall t$.  Balanced strategy is considered in all cases.}
    \label{fig:RVPP_trade_case3}
    \vspace{-2mm}
\end{figure}

Figure~\ref{fig:RVPP_trade_case3} shows the traded electrical and thermal energy of the \ac{rvpp} considering different \ac{hpa} prices. The results indicate that when a constant price of 70 €/MWh is assumed, the \ac{rvpp} purchases thermal energy through the \ac{hpa} during the morning, evening, and night hours (1–9, 11, and 18–24). Compared to the strategy with a time-of-use \ac{hpa} price (Figure~\ref{fig:Data_Price}), the \ac{rvpp} sells more and buys less electrical energy in the \ac{dam}, as the \ac{csp} has more thermal energy available for conversion to electrical energy rather than supplying \acp{td}. Notably, the up secondary reserve provided by the \ac{rvpp} decreases in hours 22–24, as the \ac{rvpp} opts to purchase less electrical energy in the \ac{dam}. Additionally, the down secondary reserve provided by the \ac{rvpp} increases in hours 9, 11, 20, and 22–24 due to higher electrical production from \ac{rvpp} units. The results for lower \ac{hpa} prices (60 and 50 €/MWh) show that as the \ac{hpa} price decreases further, a greater portion of the \ac{rvpp}'s \acp{td} is met through \ac{hpa} purchases. Moreover, electrical production and down reserves provided by the \ac{rvpp} increase, while the up reserve decreases. This highlights the significant impact of different thermal \ac{hpa} contract prices on the electrical energy and reserve traded by the \ac{rvpp}.

\hspace{1cm}
\subsection{Case 4}
\label{subsec: Case 4}

{\color{black}The effectiveness of the proposed approach is evaluated in comparison with the \ac{sp} model from~\citep{zhao2021coordinated}, using the out-of-sample method proposed in~\citep{baringo2018day}. The scenario generation method used for the \ac{sp} model is based on Monte Carlo simulations applied to the same historical data used to define the uncertainty bounds in the proposed approach. Moreover, separate datasets are employed for training the models and conducting the out-of-sample evaluation. Importantly, both models are assessed using an identical set of scenarios during testing, ensuring that they are exposed to the same uncertainty. This setup is designed to isolate the effects of the modeling techniques themselves, so that any observed performance differences can be attributed to methodological differences rather than inconsistencies in the testing conditions.}

{\color{black}Table~\ref{table:OutofSample_This_Case4} summarizes the out-of-sample performance of the proposed model under various uncertainty handling strategies. Following the definitions in~\citep{baringo2018day}, three metrics are reported: the average sampled cost (representing market expenses excluding penalty costs), the average sampled penalty cost (due to constraint violations), and the average sampled net cost (i.e., the sum of the average cost and the penalty cost). The simulation results indicate that adopting more conservative strategies significantly reduces the penalty costs associated with the \ac{rvpp}, thereby lowering the overall net cost. Specifically, compared to the deterministic strategy, the optimistic, balanced, and pessimistic strategies reduce penalty costs by 62.9\%, 72.2\%, and 79.0\%, respectively. However, this increased robustness comes at the expense of higher average costs, which rise by 76.6\%, 128.1\%, and 181.9\% for the same strategies. Among the strategies evaluated, the optimistic approach yields the lowest net cost, achieving a 33.8\% reduction compared to the deterministic case. Additionally, the proposed model demonstrates strong computational efficiency, with solution times remaining under 2 seconds across all strategies.}

{\color{black}Table~\ref{table:OutofSample_[2]_Case4} presents the out-of-sample results for the \ac{sp} model proposed in~\citep{zhao2021coordinated}, evaluated under varying numbers of scenarios. The \ac{sp} model demonstrates fair performance in keeping the average cost of the \ac{rvpp} close to that of the deterministic strategy, which yields the lowest average market participation cost. For instance, when using 5, 10, 15, 20, and 25 scenarios, the average cost increases by 29.2\%, 24.0\%, 38.0\%, 54.4\%, and 47.9\%, respectively, compared to the deterministic strategy. While increasing the number of scenarios reduces the penalization cost, the extent of this reduction is less significant than that achieved by the proposed approach. Specifically, compared to the deterministic strategy, the average penalization cost is reduced by 31.2\%, 35.0\%, 40.6\%, 48.1\%, and 52.6\% for 5, 10, 15, 20, and 25 scenarios, respectively. Although these reductions are notable, they remain lower than the maximum 79.0\% reduction achieved with the proposed method. Overall, while the \ac{sp} model effectively maintains a low average market cost, it performs less well in capturing worst-case scenarios that lead to high penalty costs. As a result, a higher net cost is obtained with the \ac{sp} approach compared to the proposed approach. Additionally, as the number of scenarios increases, the average net cost of the \ac{rvpp} decreases, emphasizing the importance of selecting an appropriate number of scenarios in the \ac{sp} model. The lowest net cost in the \ac{sp} model is observed with 25 scenarios, amounting to 56.0~k€. However, this comes at a high computational cost, with the solution time reaching 6900 seconds.}

\begin{table*}[t!]
  \centering
   {\color{black}
  \caption{The out-of-sample results for the proposed model.}
    \label{table:OutofSample_This_Case4}
 \small
\vspace{-.5em}
  \begin{threeparttable}
  \begin{tabular}{lccccc}
    \toprule

    \multicolumn{1}{c}{\textbf{Strategy}}   
    && \multicolumn{1}{c}{\textbf{Average sampled}} 
    & \multicolumn{1}{c}{\textbf{Average sampled}}
    & \multicolumn{1}{c}{\textbf{Average sampled}} 
    & \multicolumn{1}{c}{\textbf{Computational time}}
    \\

    \multicolumn{1}{c}{\textbf{}} && \textbf{cost [k€]} & \textbf{penalization cost [k€]} & \textbf{net cost [k€]}  & \text{[s]}
    \\

 \cmidrule{1-1} \cmidrule{3-6}

    \multirow{1}{*}{\textbf{Deterministic}}    && 17.1 & 64.8   & 81.9  & 1.5    \\ [0.2em]

    \multirow{1}{*}{\textbf{Optimistic}}    && 30.2 & 24.0  &  54.2 & 1.8   \\ [0.2em]

    \multirow{1}{*}{\textbf{Balanced}}    && 39.0 & 18.0   & 57.0 & 1.0  \\ [0.2em]

    \multirow{1}{*}{\textbf{Pessimistic}}    && 48.2 & 13.6   & 61.8 & 1.4  \\ [0.2em]

\bottomrule
  \end{tabular}
\end{threeparttable}}
\end{table*}

\begin{table*}[t!]
  \centering
   {\color{black}
  \caption{The out-of-sample results for the model in~\cite{zhao2021coordinated}.}
    \label{table:OutofSample_[2]_Case4}
 \small
\vspace{-.5em}
  \begin{threeparttable}
  \begin{tabular}{cccccc}
    \toprule

    \multicolumn{1}{c}{\textbf{Number of}}   
    && \multicolumn{1}{c}{\textbf{Average sampled}} 
    & \multicolumn{1}{c}{\textbf{Average sampled}}
    & \multicolumn{1}{c}{\textbf{Average sampled}} 
    & \multicolumn{1}{c}{\textbf{Computational time}}
    \\

    \multicolumn{1}{c}{\textbf{scenarios [-]}} && \textbf{cost [k€]} & \textbf{penalization cost [k€]} & \textbf{net cost [k€]}  & \text{[s]}
    \\

 \cmidrule{1-1} \cmidrule{3-6}

    \multirow{1}{*}{\textbf{5}}    && 22.1 & 44.6  &  66.7 & 73.9   \\ [0.2em]

    \multirow{1}{*}{\textbf{10}}    && 21.2 & 42.1   & 63.3 & 547.2  \\ [0.2em]

    \multirow{1}{*}{\textbf{15}}    && 23.6 & 38.5  & 62.1 & 836.0  \\ [0.2em]

    \multirow{1}{*}{\textbf{20}}    && 26.4 & 33.6  & 60.0 & 1447.2  \\ [0.2em]

    \multirow{1}{*}{\textbf{25}}    && 25.3  & 30.7   & 56.0 & 6900.1 \\ [0.2em]

\bottomrule
  \end{tabular}
\end{threeparttable}}
\end{table*}

\section{Conclusion}
\label{sec:Conclusion}

In this paper, a new two-stage robust model is proposed for the integration of \acp{csp} in the \ac{rvpp}. The \ac{rvpp} includes \acp{csp}, \acp{wf}, solar PVs, \acp{ed}, and \acp{td}, and trades electrical energy and reserves in the \ac{dam} and \ac{srm}, while also purchasing thermal energy to supply its local \acp{td}. Multiple uncertainties related to electricity prices, as well as the source and load sides of \ac{rvpp} production and consumption, are taken into account by incorporating temporal constraints to determine the flexible worst case in \ac{ro}. The energy and reserve provision of \ac{csp} is modeled by assigning an adjustable level of energy from the \ac{ts} of \ac{csp} for reserve provision. Several case studies are conducted to demonstrate the applicability and computational efficiency of the proposed approach. The simulation results show that the \ac{csp} provides electrical energy to meet demand during early morning and late evening hours when the production of solar PV plants in the \ac{rvpp} is low. Additionally, the majority of the up reserve provided by the \ac{rvpp} is scheduled to be supplied by the \ac{csp} due to its flexibility. Furthermore, the \ac{csp} can supply \acp{td} during daylight hours, primarily with the help of its \ac{sf} production, and supply \ac{td} during other hours when sunlight is unavailable by discharging the \ac{ts}. The results show that adopting an optimistic decision by the \ac{rvpp} against uncertain parameters leads to selling more electrical energy and buying less electrical and thermal energy compared to balanced and pessimistic strategies. The traded reserve in the \ac{srm} is also affected by the electrical and thermal energy traded. For example, in the pessimistic case, during some hours, since the \ac{csp} has higher electrical energy production, its provided up and down reserves decrease and increase, respectively. The results also show that by adopting the proposed coordinated strategy (\ac{dam} + \ac{srm} + \ac{hpa}), the cost of the \ac{rvpp} decreases by 4.1\%–23.9\% compared to the \ac{dam}-only participation strategy. Additionally, the effect of different \ac{hpa} prices on traded thermal energy and the profitability of the \ac{rvpp} is analyzed. The results indicate that for lower \ac{hpa} prices, the purchased thermal energy through \ac{hpa} increases. The cost of the \ac{rvpp} for an \ac{hpa} price of 50 €/MWh decreases by 20.7\%–48.9\% for different \ac{rvpp} strategies against uncertainties compared to the assumed time-of-use price for \ac{hpa}. 

{\color{black}Future work would focus on modeling and analyzing the potential spatiotemporal correlations among uncertain parameters, and how they can influence \ac{rvpp} performance. }

%\clearpage
\appendix

\section*{Appendix A. \ac{csp} Turn On/Off Constraints}
\label{appendix:A}

\setcounter{equation}{0} % Reset equation counter
\renewcommand{\theequation}{A.\arabic{equation}} % Custom format A1, A2,...

The minimum up and down time constraints for turbine of \ac{csp} are formulated in~\eqref{Deterministic:STU_minimum_up_down_time} and are adopted from~\cite{carrion2006computationally}. Constraints~\eqref{Deterministic: STU10} and~\eqref{Deterministic: STU11} define the commitment status of turbine. The parameters $N_{\theta}^{ON}$ and $N_{\theta}^{OFF}$ are the number of initial periods during which turbine must be online/offline, respectively. Equation~\eqref{Deterministic: STU12} defines the initial status of turbine as defined by $N_{\theta}^{ON}$. Constraint~\eqref{Deterministic: STU13} restricts the minimum up time during all combination of the subsequent periods ${UT}_{\theta}$. Constraint~\eqref{Deterministic: STU14} defines the minimum up time for the final periods ${UT}_{\theta}-1$. Constraints~\eqref{Deterministic: STU15}-\eqref{Deterministic: STU17} are the minimum down time version of~\eqref{Deterministic: STU12}-\eqref{Deterministic: STU14}. The nature of binary variables is shown in~\eqref{Deterministic: STU18}.

{\color{black}
\begingroup
\allowdisplaybreaks
\begin{subequations}
\begin{align}
    &u_{\theta,t} - u_{\theta,t-1} = v_{\theta,t}^{SU} - v_{\theta,t}^{SD}~; 
    &\forall \theta, t \in {\Theta},\mathscr{T} \label{Deterministic: STU10} \\
    & v_{\theta,t}^{SU} + v_{\theta,t}^{SD} \leq 1~; 
    &\forall \theta, t \in {\Theta},\mathscr{T} \label{Deterministic: STU11} \\
    & \sum_{t=1}^{N_\theta^{ON}} \left[ 1 - u_{\theta,t} \right] = 0~; 
    &\forall \theta \in {\Theta} \label{Deterministic: STU12} \\
    & {UT}_\theta \left( u_{\theta,t} - u_{\theta,t-1} \right) \leq \sum_{{t}^\prime =t}^{t+UT_\theta-1} u_{\theta, t^{\prime} }~; 
    &\forall \theta \in {\Theta}, t = N_\theta^{ON}+1, ..., |\mathscr{T}|-UT_\theta+1 \label{Deterministic: STU13} \\
    & 0 \leq \sum_{{t}^\prime=t}^{|\mathscr{T}|} \left[ u_{\theta,{t}^\prime} - \left( u_{\theta,t} - u_{\theta,t-1} \right) \right]~; 
    &\forall \theta \in {\Theta}, t = |\mathscr{T}|-UT_\theta+2, ..., |\mathscr{T}| \label{Deterministic: STU14} \\
    & \sum_{t=1}^{N_\theta^{OFF}} u_{\theta,t} = 0~; 
    &\forall \theta \in {\Theta} \label{Deterministic: STU15} \\
    & DT_\theta \left( u_{\theta,t-1} - u_{\theta,t} \right) \leq \sum_{{t}^\prime=t}^{t+DT_\theta-1} \left[ 1 - u_{\theta,{t}^\prime} \right]~; 
    &\forall \theta \in {\Theta}, t = N_\theta^{OFF}+1, ..., |\mathscr{T}|-DT_\theta+1 \label{Deterministic: STU16} \\
    & 0 \leq \sum_{{t}^\prime=t}^{|\mathscr{T}|} \left[ 1 - u_{\theta,{t}^\prime} - \left( u_{\theta,t-1} - u_{\theta,t} \right) \right]~; 
    &\forall \theta \in {\Theta}, t = |\mathscr{T}|-DT_\theta+2, ..., |\mathscr{T}| \label{Deterministic: STU17} \\
    & u_{\theta,t}, v_{\theta,t}^{SU}, v_{\theta,t}^{SD} \in \{0,1\}~; 
    &\forall \theta, t \in {\Theta},\mathscr{T} \label{Deterministic: STU18}     
    \end{align}    
\label{Deterministic:STU_minimum_up_down_time}
\end{subequations}
\endgroup
}

\section*{Acknowledgments}
The authors wish to thank Comunidad de Madrid for the financial support to PREDFLEX project (TEC-2024/ECO-287), through the R\&D activity programme Tecnologías 2024.

The authors wish to thank Mr. Ione Lopez with Iberdrola, Spain, and the people with CIEMAT, Spain, especially Dr. Mario Biencinto and Dr. Loreto Valenzuela, for the discussions on the modeling of concentrated solar power plants and flexible demands, and the provision of data used in the case study.

\clearpage

\bibliography{refs.bib}

\begin{thebibliography}{10}
\expandafter\ifx\csname url\endcsname\relax
  \def\url#1{\texttt{#1}}\fi
\expandafter\ifx\csname urlprefix\endcsname\relax\def\urlprefix{URL }\fi
\expandafter\ifx\csname href\endcsname\relax
  \def\href#1#2{#2} \def\path#1{#1}\fi

\bibitem{dominguez2012optimal}
R.~Dominguez, L.~Baringo, A.~Conejo, Optimal offering strategy for a concentrating solar power plant, Applied Energy 98 (2012) 316--325.

\bibitem{zhao2021coordinated}
Y.~Zhao, S.~Liu, Z.~Lin, F.~Wen, Y.~Ding, Coordinated scheduling strategy for an integrated system with concentrating solar power plants and solar prosumers considering thermal interactions and demand flexibilities, Applied Energy 304 (2021) 117646.

\bibitem{USDOE_CSP}
{U.S. Department of Energy}, \href{https://www.energy.gov/eere/solar/concentrating-solar-power}{Concentrating solar power} (2023).
\newline\urlprefix\url{https://www.energy.gov/eere/solar/concentrating-solar-power}

\bibitem{garcia2011performance}
I.~L. Garc{\'\i}a, J.~L. {\'A}lvarez, D.~Blanco, Performance model for parabolic trough solar thermal power plants with thermal storage: Comparison to operating plant data, Solar Energy 85~(10) (2011) 2443--2460.

\bibitem{miron2023cost}
D.~Miron, A.~Navon, Y.~Levron, J.~Belikov, C.~Rotschild, The cost-competitiveness of concentrated solar power with thermal energy storage in power systems with high solar penetration levels, Journal of Energy Storage 72 (2023) 108464.

\bibitem{IRENA2025}
I.~R. E.~A. (IRENA), \href{https://www.irena.org/-/media/Files/IRENA/Agency/Publication/2025/Jan/IRENA_Renewable_energy_benefits_leveraging_capacity_CSP_2025.pdf}{Renewable energy benefits: Leveraging local capacity for concentrated solar power}, Tech. rep., International Renewable Energy Agency (IRENA), Abu Dhabi (2025).
\newline\urlprefix\url{https://www.irena.org/-/media/Files/IRENA/Agency/Publication/2025/Jan/IRENA_Renewable_energy_benefits_leveraging_capacity_CSP_2025.pdf}

\bibitem{khan2022progress}
M.~I. Khan, F.~Asfand, S.~G. Al-Ghamdi, Progress in research and technological advancements of thermal energy storage systems for concentrated solar power, Journal of Energy Storage 55 (2022) 105860.

\bibitem{yao2023concentrated}
L.~Yao, Y.~Wang, X.~Xiao, Concentrated solar power plant modeling for power system studies, IEEE Transactions on Power Systems (2023).

\bibitem{alvaroPOSYFT}
{\'A}.~Ortega, O.~Oladimeji, H.~Nemati, L.~Sigrist, P.~Sánchez-Martín, L.~Rouco, E.~Lobato, M.~Biencinto, I.~López, Modeling of {VPPs} for their optimal operation and configuration, Tech. rep., POSYTYF Consortium, deliverable 5.1. (2021).

\bibitem{conejo2010decision}
A.~J. Conejo, M.~Carri{\'o}n, J.~M. Morales, et~al., Decision making under uncertainty in electricity markets, Vol.~1, Springer, 2010.

\bibitem{naval2021virtual}
N.~Naval, J.~M. Yusta, Virtual power plant models and electricity markets-{A} review, Renewable and Sustainable Energy Reviews 149 (2021) 111393.

\bibitem{fernandes2016participation}
C.~Fernandes, P.~Fr{\'\i}as, J.~Reneses, Participation of intermittent renewable generators in balancing mechanisms: A closer look into the spanish market design, Renewable Energy 89 (2016) 305--316.

\bibitem{sun2022day}
S.~Sun, S.~M. Kazemi-Razi, L.~G. Kaigutha, M.~Marzband, H.~Nafisi, A.~S. Al-Sumaiti, Day-ahead offering strategy in the market for concentrating solar power considering thermoelectric decoupling by a compressed air energy storage, Applied Energy 305 (2022) 117804.

\bibitem{kong2020robust}
X.~Kong, J.~Xiao, D.~Liu, J.~Wu, C.~Wang, Y.~Shen, Robust stochastic optimal dispatching method of multi-energy virtual power plant considering multiple uncertainties, Applied Energy 279 (2020) 115707.

\bibitem{ghasemi2022coordinated}
F.~Ghasemi~Olanlari, T.~Amraee, M.~Moradi-Sepahvand, A.~Ahmadian, Coordinated multi-objective scheduling of a multi-energy virtual power plant considering storages and demand response, IET Generation, Transmission \& Distribution 16~(17) (2022) 3539--3562.

\bibitem{foroughi2021bi}
M.~Foroughi, A.~Pasban, M.~Moeini-Aghtaie, A.~Fayaz-Heidari, A bi-level model for optimal bidding of a multi-carrier technical virtual power plant in energy markets, International Journal of Electrical Power \& Energy Systems 125 (2021) 106397.

\bibitem{wang2020non}
L.~Wang, W.~Gu, Z.~Wu, H.~Qiu, G.~Pan, Non-cooperative game-based multilateral contract transactions in power-heating integrated systems, Applied Energy 268 (2020) 114930.

\bibitem{hasni2023case}
S.~Hasni, W.~J. Platzer, Case study on decarbonization strategies for {LNG} export terminals using heat and power from {CSP/PV} hybrid plants, Solar Energy Advances 3 (2023) 100041.

\bibitem{kircher2021heat}
K.~J. Kircher, K.~M. Zhang, Heat purchase agreements could lower barriers to heat pump adoption, Applied Energy 286 (2021) 116489.

\bibitem{roald2023power}
L.~A. Roald, D.~Pozo, A.~Papavasiliou, D.~K. Molzahn, J.~Kazempour, A.~Conejo, Power systems optimization under uncertainty: A review of methods and applications, Electric Power Systems Research 214 (2023) 108725.

\bibitem{singh2022uncertainty}
V.~Singh, T.~Moger, D.~Jena, Uncertainty handling techniques in power systems: A critical review, Electric Power Systems Research 203 (2022) 107633.

\bibitem{venegas2022review}
J.~F. Venegas-Zarama, J.~I. Mu{\~n}oz-Hernandez, L.~Baringo, P.~Diaz-Cachinero, I.~De~Domingo-Mondejar, A review of the evolution and main roles of virtual power plants as key stakeholders in power systems, IEEE Access 10 (2022) 47937--47964.

\bibitem{xiong2024distributionally}
H.~Xiong, F.~Luo, M.~Yan, L.~Yan, C.~Guo, G.~Ranzi, Distributionally robust and transactive energy management scheme for integrated wind-concentrated solar virtual power plants, Applied Energy 368 (2024) 123148.

\bibitem{yu2019uncertainties}
S.~Yu, F.~Fang, Y.~Liu, J.~Liu, Uncertainties of virtual power plant: Problems and countermeasures, Applied energy 239 (2019) 454--470.

\bibitem{nemati2025segan}
H.~Nemati, P.~S{\'a}nchez-Mart{\'\i}n, {\'A}.~Ortega, L.~Sigrist, E.~Lobato, L.~Rouco, Flexible robust optimal bidding of renewable virtual power plants in sequential markets under asymmetric uncertainties, Sustainable Energy, Grids and Networks (2025) 101801.

\bibitem{rahimi2021optimal}
M.~Rahimi, F.~J. Ardakani, A.~J. Ardakani, Optimal stochastic scheduling of electrical and thermal renewable and non-renewable resources in virtual power plant, International Journal of Electrical Power \& Energy Systems 127 (2021) 106658.

\bibitem{xiao2024windfall}
D.~Xiao, Z.~Lin, H.~Chen, W.~Hua, J.~Yan, Windfall profit-aware stochastic scheduling strategy for industrial virtual power plant with integrated risk-seeking/averse preferences, Applied Energy 357 (2024) 122460.

\bibitem{kalantari2023strategic}
N.~T. Kalantari, A.~Abdolahi, S.~H. Mousavi, S.~C. Khavar, F.~S. Gazijahani, Strategic decision making of energy storage owned virtual power plant in day-ahead and intra-day markets, Journal of Energy Storage 73 (2023) 108839.

\bibitem{li2023robust}
Y.~Li, Y.~Deng, Y.~Wang, L.~Jiang, M.~Shahidehpour, Robust bidding strategy for multi-energy virtual power plant in peak-regulation ancillary service market considering uncertainties, International Journal of Electrical Power \& Energy Systems 151 (2023) 109101.

\bibitem{gough2023bi}
M.~Gough, S.~F. Santos, M.~S. Javadi, J.~M. Home-Ortiz, R.~Castro, J.~P. Catal{\~a}o, Bi-level stochastic energy trading model for technical virtual power plants considering various renewable energy sources, energy storage systems and electric vehicles, Journal of Energy Storage 68 (2023) 107742.

\bibitem{NEMATI2025136421}
H.~Nemati, P.~S{\'a}nchez-Mart{\'\i}n, A.~Baringo, {\'A}.~Ortega, Single-level flexible robust optimal bidding of renewable-only virtual power plant in energy and secondary reserve markets, Energy 328 (2025) 136421.

\bibitem{nemati2025flexible}
H.~Nemati, P.~S{\'a}nchez-Mart{\'\i}n, L.~Sigrist, L.~Rouco, {\'A}.~Ortega, Flexible robust optimization for renewable-only {VPP} bidding on electricity markets with economic risk analysis, International Journal of Electrical Power \& Energy Systems 167 (2025) 110594.

\bibitem{yan2022two}
Q.~Yan, M.~Zhang, H.~Lin, W.~Li, Two-stage adjustable robust optimal dispatching model for multi-energy virtual power plant considering multiple uncertainties and carbon trading, Journal of cleaner production 336 (2022) 130400.

\bibitem{zhao2020mixed}
Y.~Zhao, S.~Liu, Z.~Lin, F.~Wen, L.~Yang, Q.~Wang, A mixed {CVaR}-based stochastic information gap approach for building optimal offering strategies of a {CSP} plant in electricity markets, IEEE Access 8 (2020) 85772--85783.

\bibitem{he2016optimal}
G.~He, Q.~Chen, C.~Kang, Q.~Xia, Optimal offering strategy for concentrating solar power plants in joint energy, reserve and regulation markets, IEEE Transactions on sustainable Energy 7~(3) (2016) 1245--1254.

\bibitem{xu2016coordinated}
T.~Xu, N.~Zhang, Coordinated operation of concentrated solar power and wind resources for the provision of energy and reserve services, IEEE Transactions on Power Systems 32~(2) (2016) 1260--1271.

\bibitem{xiong2023dp}
H.~Xiong, M.~Yan, C.~Guo, Y.~Ding, Y.~Zhou, {DP} based multi-stage {ARO} for coordinated scheduling of {CSP} and wind energy with tractable storage scheme: Tight formulation and solution technique, Applied Energy 333 (2023) 120578.

\bibitem{pousinho2014self}
H.~M.~I. Pousinho, H.~Silva, V.~Mendes, M.~Collares-Pereira, C.~P. Cabrita, Self-scheduling for energy and spinning reserve of wind/{CSP} plants by a {MILP} approach, Energy 78 (2014) 524--534.

\bibitem{fang2020look}
Y.~Fang, S.~Zhao, Look-ahead bidding strategy for concentrating solar power plants with wind farms, Energy 203 (2020) 117895.

\bibitem{khaloie2021day}
H.~Khaloie, F.~Vall{\'e}e, C.~S. Lai, J.-F. Toubeau, N.~D. Hatziargyriou, Day-ahead and intraday dispatch of an integrated biomass-concentrated solar system: A multi-objective risk-controlling approach, IEEE Transactions on Power Systems 37~(1) (2021) 701--714.

\bibitem{oladimeji2022optimal}
O.~Oladimeji, {\'A}.~Ortega, L.~Sigrist, L.~Rouco, P.~S{\'a}nchez-Mart{\'\i}n, E.~Lobato, Optimal participation of heterogeneous, {RES}-based virtual power plants in energy markets, Energies 15~(9) (2022) 3207.

\bibitem{fang2023optimization}
Q.~Fang, N.~Liang, Z.~Liu, M.~Miao, Optimization scheduling of virtual power plant with concentrated solar power plant considering carbon trading and demand response, in: 2023 IEEE International Conference on Power Science and Technology (ICPST), IEEE, 2023, pp. 717--722.

\bibitem{bertsimas04}
D.~Bertsimas, M.~Sim, The price of robustness, Operations Research 52~(1) (2004) 35--53.

\bibitem{zhang2022frequency}
Z.~Zhang, M.~Zhou, Z.~Wu, S.~Liu, Z.~Guo, G.~Li, A frequency security constrained scheduling approach considering wind farm providing frequency support and reserve, IEEE Transactions on Sustainable Energy 13~(2) (2022) 1086--1100.

\bibitem{yin2021state}
S.~Yin, J.~Wang, Z.~Li, X.~Fang, State-of-the-art short-term electricity market operation with solar generation: A review, Renewable and Sustainable Energy Reviews 138 (2021) 110647.

\bibitem{oladimeji2022modeling}
O.~Oladimeji, A.~Ortega, L.~Sigrist, P.~S{\'a}nchez-Mart{\'\i}n, E.~Lobato, L.~Rouco, Modeling demand flexibility of {RES}-based virtual power plants, in: 2022 IEEE Power \& Energy Society General Meeting (PESGM), IEEE, 2022, pp. 1--5.

\bibitem{chen2007robust}
X.~Chen, M.~Sim, P.~Sun, A robust optimization perspective on stochastic programming, Operations research 55~(6) (2007) 1058--1071.

\bibitem{floudas1995nonlinear}
C.~A. Floudas, Nonlinear and mixed-integer optimization: fundamentals and applications, Oxford University Press, 1995.

\bibitem{wang2019expansion}
J.~Wang, Z.~Hu, S.~Xie, Expansion planning model of multi-energy system with the integration of active distribution network, Applied Energy 253 (2019) 113517.

\bibitem{srinivasan2023impact}
A.~Srinivasan, R.~Wu, P.~Heer, G.~Sansavini, Impact of forecast uncertainty and electricity markets on the flexibility provision and economic performance of highly-decarbonized multi-energy systems, Applied Energy 338 (2023) 120825.

\bibitem{wen2019performance}
Y.~Wen, D.~AlHakeem, P.~Mandal, S.~Chakraborty, Y.-K. Wu, T.~Senjyu, S.~Paudyal, T.-L. Tseng, Performance evaluation of probabilistic methods based on bootstrap and quantile regression to quantify {PV} power point forecast uncertainty, IEEE transactions on neural networks and learning systems 31~(4) (2019) 1134--1144.

\bibitem{web:ciemat_spain}
{Ciemat Spain}, \href{{https://www.ciemat.es/}}{{PV-STU production forecast}}.
\newline\urlprefix\url{{https://www.ciemat.es/}}

\bibitem{web:iberdrola_spain}
{Iberdrola Spain}, \href{{https://www.iberdrola.es/.}}{{Wind production forecast}}.
\newline\urlprefix\url{{https://www.iberdrola.es/.}}

\bibitem{JASM2019}
{SCCER JASM Data Platform}, \href{https://data.sccer-jasm.ch/demand-hourly-profile/2019-02-27/}{{Demand Hourly Profile}} (2019).
\newline\urlprefix\url{https://data.sccer-jasm.ch/demand-hourly-profile/2019-02-27/}

\bibitem{REE2025}
{Red Eléctrica de España (REE)}, \href{https://www.esios.ree.es/}{{Electricity Market Data}}.
\newline\urlprefix\url{https://www.esios.ree.es/}

\bibitem{Statista2024}
{Statista}, \href{https://www.statista.com/statistics/1534706/levelized-cost-of-heat-spain\\-by-technology/}{Levelized cost of heat in spain by technology} (2024).
\newline\urlprefix\url{https://www.statista.com/statistics/1534706/levelized-cost-of-heat-spain\\-by-technology/}

\bibitem{baringo2018day}
A.~Baringo, L.~Baringo, J.~M. Arroyo, Day-ahead self-scheduling of a virtual power plant in energy and reserve electricity markets under uncertainty, IEEE Transactions on Power Systems 34~(3) (2018) 1881--1894.

\bibitem{carrion2006computationally}
M.~Carri{\'o}n, J.~M. Arroyo, A computationally efficient mixed-integer linear formulation for the thermal unit commitment problem, IEEE Transactions on power systems 21~(3) (2006) 1371--1378.

\end{thebibliography}
\bibliographystyle{elsarticle-num.bst} 
%\printbibliography

%\clearpage

\end{document}